\documentclass{aa}  
\usepackage{graphicx}
\usepackage{txfonts}
\usepackage{pdflscape}

\usepackage{natbib,twoopt}
\usepackage{amsmath} 
\usepackage[breaklinks=true]{hyperref} 
\bibpunct{(}{)}{;}{a}{}{,} 
\makeatletter
\newcommandtwoopt{\citeads}[3][][]{\href{http://adsabs.harvard.edu/abs/#3}%
{\def\hyper@linkstart##1##2{}%
\let\hyper@linkend\@empty\citealp[#1][#2]{#3}}}
\newcommandtwoopt{\citepads}[3][][]{\href{http://adsabs.harvard.edu/abs/#3}%
{\def\hyper@linkstart##1##2{}%
\let\hyper@linkend\@empty\citep[#1][#2]{#3}}}
\newcommandtwoopt{\citetads}[3][][]{\href{http://adsabs.harvard.edu/abs/#3}%
{\def\hyper@linkstart##1##2{}%
\let\hyper@linkend\@empty\citet[#1][#2]{#3}}}
\newcommandtwoopt{\citeyearads}[3][][]%
{\href{http://adsabs.harvard.edu/abs/#3}
{\def\hyper@linkstart##1##2{}%
\let\hyper@linkend\@empty\citeyear[#1][#2]{#3}}}
\makeatother

\usepackage{color}
\definecolor{mygreen}{RGB}{0,128,0}

\hypersetup{colorlinks=true,linkcolor=blue,citecolor=blue,urlcolor=blue}

\newcommand{\gacs}[1]{{\footnotesize\texttt{#1}}}
\usepackage{wasysym}

\begin{document}

   \title{\textit{Gaia} Early Data Release 3}
   \subtitle{Parallax bias versus magnitude, colour, and position}

\author{
L.~Lindegren\inst{\ref{inst:lund}}\fnmsep\thanks{Corresponding author: L. Lindegren\newline
e-mail: \href{mailto:lennart@astro.lu.se}{\tt lennart@astro.lu.se}}
\and U.~Bastian\inst{\ref{inst:ari}}
\and M.~Biermann\inst{\ref{inst:ari}}
\and A.~Bombrun\inst{\ref{inst:hespaceesac}}
\and A.~de~Torres\inst{\ref{inst:hespaceesac}}
\and E.~Gerlach\inst{\ref{inst:tud}}
\and R.~Geyer\inst{\ref{inst:tud}}
\and J.~Hern{\'a}ndez\inst{\ref{inst:esac}}
\and T.~Hilger\inst{\ref{inst:tud}}
\and D.~Hobbs\inst{\ref{inst:lund}}
\and S.A.~Klioner\inst{\ref{inst:tud}}
\and U.~Lammers\inst{\ref{inst:esac}}
\and P.J.~McMillan\inst{\ref{inst:lund}}
\and M.~Ramos-Lerate\inst{\ref{inst:vitrocisetesac}}
\and H.~Steidelm\"{u}ller\inst{\ref{inst:tud}}
\and C.A.~Stephenson\inst{\ref{inst:vegaesac}}
\and F.~van~Leeuwen\inst{\ref{inst:ioa}}
}  


\institute{
     Lund Observatory, Department of Astronomy and Theoretical Physics, Lund University, Box 43, 22100 Lund, Sweden\label{inst:lund}
\and
Astronomisches Rechen-Institut, Zentrum f\"{ u}r Astronomie der Universit\"{ a}t Heidelberg, M\"{ o}nchhofstr. 12-14, 69120 Heidelberg, Germany
\label{inst:ari}
\and
HE Space Operations BV for European Space Agency (ESA), Camino bajo del Castillo, s/n, Urbanizacion Villafranca del Castillo, Villanueva de la Ca\~{n}ada, 28692 Madrid, Spain
\label{inst:hespaceesac}
\and
Lohrmann Observatory, Technische Universit\"{ a}t Dresden, Mommsenstra{\ss}e 13, 01062 Dresden, Germany
\label{inst:tud}
\and 
European Space Agency (ESA), European Space Astronomy Centre (ESAC), Camino bajo del Castillo, s/n, Urbanizacion Villafranca del Castillo, Villanueva de la Ca\~{n}ada, 28692 Madrid, Spain
\label{inst:esac}
\and
Vitrociset Belgium for European Space Agency (ESA), Camino bajo del Castillo, s/n, Urbanizacion Villafranca del Castillo, Villanueva de la Ca\~{n}ada, 28692 Madrid, Spain
\label{inst:vitrocisetesac}
\and
Telespazio Vega UK Ltd for European Space Agency (ESA), Camino bajo del Castillo, s/n, Urbanizacion Villafranca del Castillo, Villanueva de la Ca\~{n}ada, 28692 Madrid, Spain
\label{inst:vegaesac}
\and
Institute of Astronomy, University of Cambridge, Madingley Road, Cambridge CB3~0HA, United Kingdom
\label{inst:ioa}
} 

   \date{ }

 
\abstract
  {\textit{Gaia} Early Data Release 3 (\textit{Gaia} EDR3) gives trigonometric parallaxes 
  for nearly 1.5~billion sources. Inspection of the EDR3 data for sources identified as 
  quasars reveals that their parallaxes are biased, that is systematically offset from the 
  expected distribution around zero, by a few tens of microarcsec.}  
  {We attempt to map the main dependencies of the parallax bias in EDR3.
  In principle this could provide a recipe for correcting the EDR3 parallaxes.}  
  {For faint sources the quasars provide the most direct way to estimate parallax bias. 
  In order to extend this to brighter sources and a broader range of colours, we use
  differential methods based on physical pairs (binaries) and sources in the Large Magellanic
  Cloud. The functional forms of the dependencies are explored by mapping the systematic
  differences between EDR3 and DR2 parallaxes.}  
  {The parallax bias is found to depend in a non-trivial way on (at least) the magnitude, 
  colour, and ecliptic latitude of the source. Different dependencies apply to the five- and
  six-parameter solutions in EDR3. While it is not possible to derive a definitive recipe for the 
  parallax correction, we give tentative expressions to be used at the researcher's discretion
  and point out some possible paths towards future improvements.}  
  {}


   \titlerunning{\textit{Gaia} EDR3 -- Parallax bias versus magnitude, colour, and position} 
   \authorrunning{L.~Lindegren et al.}

   \maketitle

%

\section{Introduction} 
\label{sec:intro}

The (early) Third \textit{Gaia} Data Release (hereafter EDR3; \citealt{EDR3-DPACP-130}) provides 
trigonometrically determined parallaxes for nearly 1478~million sources in the 
magnitude range $G\simeq 6$ to 21. The sources include stars, unresolved binaries,
compact extragalactic objects such as active galactic nuclei (AGNs), and other 
objects that appear roughly pointlike at the angular resolution of \textit{Gaia}
($\sim$0.1~arcsec). Although \textit{Gaia} in principle determines absolute parallaxes 
(e.g.\ \citeads{2011EAS....45..109L}), without relying on distant background objects,
imperfections in the instrument and data processing methods inevitably result in
systematic errors in the published astrometric data. For example, it is well known 
that the parallax solution is degenerate with respect to certain variations
of the `basic angle' between the viewing directions of \textit{Gaia}'s two telescopes
\citepads{2017A&A...603A..45B}. This means that one cannot, purely from \textit{Gaia}'s
own astrometric observations, simultaneously determine absolute parallaxes \textit{and} 
calibrate this particular perturbation of the instrument. Conversely, if the actual 
instrument perturbations contain a component of this form, it will produce
biased parallax values in the astrometric solution.

Since the second release of \textit{Gaia} data in April 2018 
(DR2; \citeads{2018A&A...616A...1G}), a number of investigations have been published, 
using a variety of astrophysical objects, that have resulted in estimates of the parallax 
systematics in DR2 
\citep[see, for example,][and references therein]{2020MNRAS.493.4367C}. 
Reported values of the \textit{Gaia} DR2 `parallax zero point' (i.e., the quantity to be
subtracted from the DR2 parallaxes) range from about $-30~\mu$as to $-80~\mu$as.
While quasars in DR2 yield a median parallax of $-29~\mu$as \citep{2018A&A...616A...2L},
the cited studies, which typically use much brighter objects, tend to give more negative 
values. There is consequently a strong suspicion
that the parallax offset in DR2 depends on the magnitude, and possibly also on the colour of
the sources \citep[e.g.,][]{2019ApJ...878..136Z}. Furthermore, it is known that 
\textit{Gaia} DR2 has position-dependent offsets of the parallaxes on angular scales 
down to $\sim$1~deg \citep{2018A&A...616A..17A,2018A&A...616A...2L}.
The systematic offset of parallaxes in DR2 could therefore be a complicated function
of several variables $\vec{x}$, including at least the magnitude, colour, 
and position of the object; in the following we write this $Z_\text{DR2}(\vec{x})$.

In \textit{Gaia} EDR3 quasars have a median parallax of about $-17~\mu$as, and already
a simple plot of the parallaxes versus the $G$ magnitude or colour index reveals 
systematic variations at a level of $\sim$10~$\mu$as (see Sect.~\ref{sec:qso}).
Position-dependent variations are also seen on all angular scales, although with
smaller amplitudes than in DR2 \citep{EDR3-DPACP-128}. Thus it is possible to define 
an offset function $Z_\text{EDR3}(\vec{x})$ that is in general different from 
$Z_\text{DR2}(\vec{x})$, although it may depend on the same variables $\vec{x}$.

The letter $Z$ adopted for these functions is a mnemonic for `zero point'. What is meant
is an estimate of the bias (or systematics) of the parallax estimate as a function of certain 
known variables $\vec{x}$. The practical determination of this function is fraught 
with difficulties and uncertainties, and even if we knew the true parallax of every 
object in the catalogue, it would not be possible to define a unique function $Z(\vec{x})$ 
short of tabulating the bias for every source. The function will necessarily depend 
on, for example, the choice of arguments in $\vec{x}$, their resolution and numerical
representation. At best, what can be 
achieved is a prescription such that the use of $\varpi_i-Z(\vec{x}_i)$ for the parallax of
source $i$, instead of the catalogue value $\varpi_i$, will in most cases give more 
accurate and consistent results in astrophysical applications.

The aim of this paper is to map some of the main dependencies of the \textit{Gaia} EDR3 
parallax bias (zero point), as found in the course of the internal validations carried out 
by the \textit{Gaia} astrometry team prior to the publication of the data. Because the 
parallax determinations in \textit{Gaia} EDR3 are of two kinds, known as five- and
six-parameter solutions, with distinctly different properties, the results are
given in the form of two functions $Z_5(\vec{x})$ and $Z_6(\vec{x})$ describing the
bias as a function of magnitude, colour, and position for each kind of solution.
In the following, the functions $Z_5$ and $Z_6$ always refer to EDR3, while for DR2 there
is only one kind of solution, with bias function $Z_\text{DR2}$.

Section~\ref{sec:instr} is a brief overview of some aspects of the \textit{Gaia} instrument
and data processing that are particularly relevant for the bias functions, such as the different 
observing modes depending on the magnitude of the source, and the use of colour information
in the five- and
six-parameter solutions. In Sect.~\ref{sec:dr2} we discuss the systematic differences 
between DR2 and EDR3 parallaxes. While the DR2 parallaxes are completely superseded 
by the later release, the differences could give important clues to the systematic 
dependencies. In Sect.~\ref{sec:p5} we estimate $Z_5$, the bias function for the
five-parameter solutions in EDR3. In Sect.~\ref{sec:p6} we use a differential procedure
to estimate $Z_6$, the bias function for the six-parameter solutions. A limited
test of the derived bias functions is made in Sect.~\ref{sec:val}. Some possible future
improvements are discussed in Sect.~\ref{sec:future} before the findings are 
summarised in Sect.~\ref{sec:sum}. Certain technical details are put in Appendices.

\begin{figure}
\centering
  \includegraphics[width=\hsize]{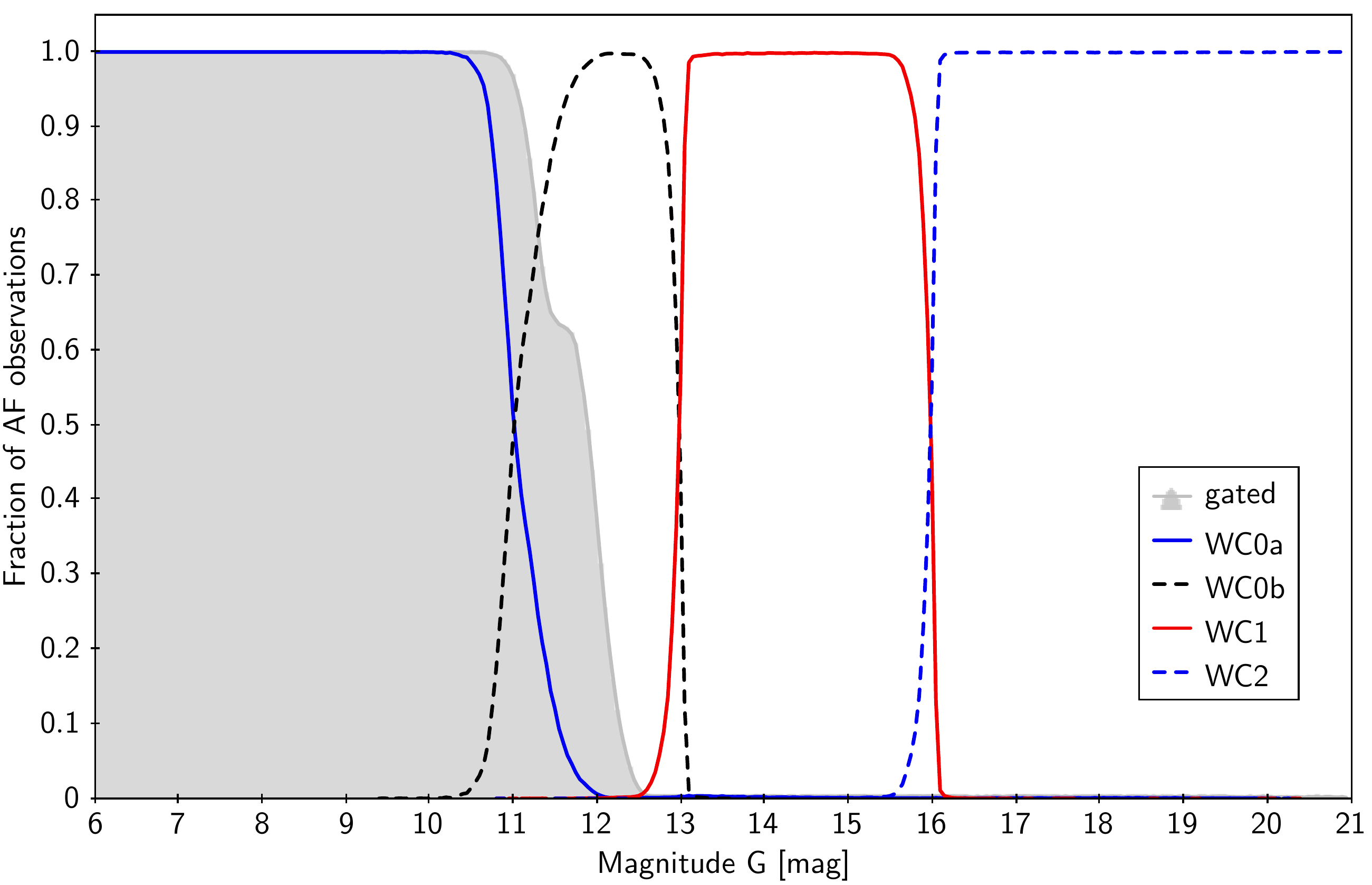}
    \caption{Fraction of observations in different modes. The shaded area represents
    gated observations; the curves show the fraction of AF observations made in 
    window classes WC0a, WC0b, WC1, and WC2.}
    \label{fig:gClass}
\end{figure}

\begin{figure}
\centering
  \includegraphics[width=\hsize]{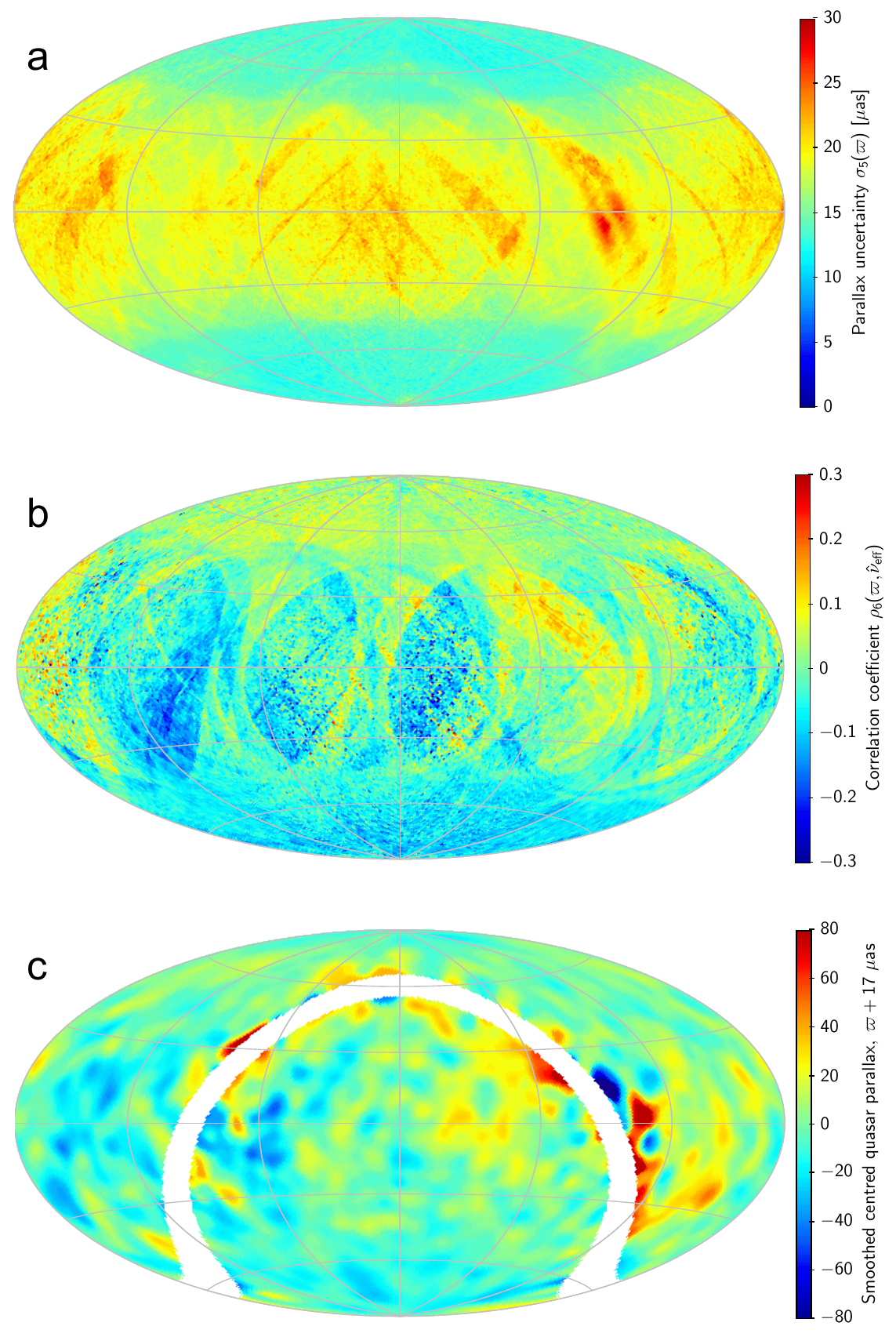}   
    \caption{Celestial maps in ecliptic coordinates of some selected statistics in \textit{Gaia} EDR3:
    (\textit{a}) mean parallax uncertainty for sources with five-parameter solutions and $G<14$~mag.
    (\textit{b}) mean correlation coefficient between parallax and pseudocolour for sources with six-parameter solutions.
    (\textit{c}) smoothed median parallax of quasars, corrected for the global median offset of $-17~\mu$as.
    Owing to the small number of quasars identified at small Galactic latitudes, no data are shown for $|\sin b\,|<0.1$.  
    The maps use a Hammer--Aitoff projection in ecliptic coordinates with $\lambda=\beta=0$ at the centre, 
    ecliptic north up, and ecliptic longitude $\lambda$ increasing from right to left. All three statistics exhibit
    some degree of systematic dependence on ecliptic latitude, which may be symmetric (as in \textit{a})
    or antisymmetric (as in \textit{b} and \textit{c}), but also even larger variations as functions of both coordinates.}
    \label{figA20}
\end{figure}

\section{Observing and processing modes for the EDR3 astrometry}
\label{sec:instr}

The determination of biases in \textit{Gaia} astrometry must rely on a comparison with 
external data. In principle, this process requires no prior knowledge on how the astrometry
was generated, but the mapping of possible dependencies is surely facilitated by some 
acquaintance with certain aspects of the instrument and data processing. This section 
provides a brief overview of relevant topics.

\subsection{Window classes and gates}
\label{sec:wcgates}

The \textit{Gaia} instrument and its routine operations are described in Sects.~3 and 5
of \citet{2016A&A...595A...1G}. Further details on the initial treatment of the data and
the subsequent  astrometric processing can be found in \citet{EDR3-DPACP-73} and 
\citet{EDR3-DPACP-128}.
The astrometric results are derived almost exclusively from observations made with 
the 62 CCDs occupying the central part of the focal plane assembly, the
so-called astrometric field (AF; see Fig.~4 in \citeads{2016A&A...595A...1G}).
Of relevance here is that only small patches (`windows') of the CCD images around 
detected point sources are transmitted to the ground, and that different sampling
schemes (window size, pixel binning, etc.) are used depending on the on-board 
estimate of the brightness of the source. 
Furthermore, to avoid pixel saturation for sources brighter
than $G\simeq 12$ 
(where $G$ is the magnitude in the \textit{Gaia} passband; \citeads{2018A&A...616A...4E}), 
the CCD integration time may be reduced by means of gates \citepads{2016A&A...595A...6C}.

In principle, the different sampling schemes (known as window classes, WC), and their
different uses together with the gates, may require separate calibrations both in the
initial treatment (image parameter determination, IPD) and in the astrometric solution.
This need arises both because of subtle differences in the way the CCDs and associated
electronics function depending on their mode of operation \citepads{2018A&A...616A..15H}, 
and because different data processing models are used, for example depending on 
whether the window contains a one- or two-dimensional image. Further complications
are caused by unavoidable conflicts in the on-board resource management, for example
overlapping and truncated windows. 

In the EDR3 astrometric solution \citep{EDR3-DPACP-128}, distinct calibration models were 
used for observations in the four window classes designated WC0a, WC0b, WC1, and WC2.
The first two window classes have two-dimensional images, but different gate usages, 
while WC1 and WC2 have one-dimensional images (i.e., the pixels have been binned in 
the other dimension), but of different sizes. For the IPD a similar division was made,  
but without the distinction between WC1 and WC2. Figure~\ref{fig:gClass} shows the
fraction of AF observations in each window class as a function of $G$. Clearly, the
window classes map out distinct intervals in $G$, but the transitions around $G=11$, 
13, and 16 are slightly fuzzy because the decision on the sampling scheme is based on
the on-board real-time estimate of $G$, which differs from the on-ground estimated mean 
magnitude used in the plot. The transition between WC0a and WC0b occurs
gradually over a whole magnitude, because it involves gating thresholds that are
set individually for the CCDs. 

In EDR3, the astrometric parameters for a given source are typically obtained by
combining some 200--500 individual AF observations. Depending on the mean
magnitude of the source, the observations will be distributed among the window
classes and gates in proportion to the fractions shown in Fig.~\ref{fig:gClass}. This
explains some of the magnitude-dependent effects described in later sections. 
As it turns out, it is also relevant whether the observations are gated or not.
This creates another transition around $G=12$, indicated by the shaded area in the
diagram.

\subsection{Five- and six-parameter solutions, and the use of colour information}
\label{sec:nueff}

For a source in EDR3 that has a parallax, the astrometric parameters were determined either 
in a five-parameter solution or in a six-parameter solution. The choice of solution depends on
the availability of colour information, in the form of an effective wavenumber $\nu_\text{eff}$, 
at the time when the image parameter determination 
(IPD) is performed. The effective wavenumber comes from the processing of BP and RP spectra 
in the photometric pipeline \citep{EDR3-DPACP-117,EDR3-DPACP-118}.%
\footnote{Because the IPD is one of the first processes in the overall cyclic processing scheme 
of DPAC \citepads{2016A&A...595A...3F}, while the photometric processing is further downstream, 
the values of $\nu_\text{eff}$ used in EDR3 actually come from the spectrophotometric processing 
of the previous cycle, corresponding to the DR2 photometry.}
In IPD a calibrated point-spread function (PSF) or line-spread function (LSF) is fitted to 
the CCD samples in a window, yielding the accurate location and flux (in instrumental units) 
of the image in the 
pixel stream. In a previous step, the PSF and LSF have been calibrated as functions of 
several parameters, including window class and effective wavenumber.

A five-parameter solution is computed if the source has a reliable value of $\nu_\text{eff}$ 
that can be used to select the appropriate PSF or LSF for the IPD. This means that most 
of the colour-dependent image shifts, caused by diffraction phenomena in the telescopes
and electronic effects in the CCDs
have been eliminated already in the data input to the astrometric solution (AGIS). In AGIS, 
the image locations for a given source are fitted by the standard astrometric model with
five unknowns $\alpha$, $\delta$, $\varpi$, $\mu_{\alpha*}$, and $\mu_\delta$
\citepads{2012A&A...538A..78L}. These 
parameters (plus, for some nearby stars, the spectroscopic radial velocity) are sufficient to 
describe the astrometric observations of well-behaved sources such as single stars and quasars.

If a source does not have a reliable $\nu_\text{eff}$ at the time of the IPD, it will instead
obtain a six-parameter solution in AGIS, where the extra (sixth) parameter is an astrometric 
estimate of $\nu_\text{eff}$ known as the pseudocolour, denoted $\hat{\nu}_\text{eff}$.
At the IPD stage, image locations and fluxes for the source are determined by fitting the 
PSF or LSF at the default effective wavenumber $\nu_\text{eff}^\text{\,def}=1.43~\mu$m$^{-1}$.
This value is close to the mean $\nu_\text{eff}$ for the faint sources that make up the majority 
of sources without a photometric $\nu_\text{eff}$. The use of the default colour in IPD results 
in image locations that are biased in proportion to $\nu_\text{eff}-\nu_\text{eff}^\text{\,def}$,
where $\nu_\text{eff}$ is the actual (but unknown) colour. The coefficient of proportionality is 
a property of the instrument, known as the chromaticity; it varies with time and position in the 
field, but can be calibrated in the astrometric solution by means of stars of known colours
(for details, see \citealt{EDR3-DPACP-128}). Using the calibrated chromaticity, corrections to the default colour
can be estimated for the individual sources, which gives their pseudocolours. The 
determination of pseudocolour thus takes advantage of the (extremely small) along-scan 
shifts of the image centre versus wavelength, caused by optical wavefront errors and 
other imperfections in the astrometric instrument.

The instrument chromaticity is calibrated in AGIS by means of a special solution, using a
subset of about eight million primary sources for which IPD was executed twice: once using 
the actual effective wavenumber $\nu_\text{eff}$, known from the spectrophotometric processing,
and once using the default value $\nu_\text{eff}^\text{\,def}$. For these sources it is
possible to compute both five-parameter solutions (which are the published ones) 
and six-parameter solutions; the latter are not published but used in Sect.~\ref{sec:p6}
to derive the systematic differences between the five- and six-parameter solutions.

Because the parallax bias is colour-dependent, the functions $Z_5(\vec{x})$ and 
$Z_6(\vec{x})$ need to include a colour parameter among its arguments in $\vec{x}$.
Rather than using, for example, the colour index $G_\text{BP}-G_\text{RP}$, which is not
available for all sources, the colour parameter used here is the (photometric) effective 
wavenumber $\nu_\text{eff}$ (\gacs{nu\_eff\_used\_in\_astrometry}) in $Z_5$, and the 
(astrometric) pseudocolour $\hat{\nu}_\text{eff}$ (\gacs{pseudocolour}) in $Z_6$. 
By definition, this colour information is available for all sources with an astrometrically 
determined parallax. 

The use of colour information in IPD and AGIS has one additional feature that 
significantly affects the five-parameter solutions with 
$\nu_\text{eff}>1.72~\mu$m$^{-1}$ ($G_\text{BP}{-}G_\text{RP}\lesssim 0.14$)
or $\nu_\text{eff}<1.24~\mu$m$^{-1}$ ($G_\text{BP}{-}G_\text{RP}\gtrsim 3.0$).
The calibration of the PSF and LSF versus colour is only done for the well-populated 
interval $1.24<\nu_\text{eff}< 1.72$~$\mu$m$^{-1}$, where a quadratic variation of the 
displacement with $\nu_\text{eff}$ is assumed; if the IPD requests a PSF or LSF for 
a wavenumber outside of this range, then the calibration at the nearest boundary is 
used. This `clamping' of the wavenumbers guarantees that the LSF/PSF model used
by the IPD is always sensible, but on the downside it 
introduces some biases for sources of extreme colours. 
In the astrometric calibration model, the chromaticity is assumed to be linear over the 
entire range of wavenumbers. The combination of the two models may produce 
astrometric biases that are strongly non-linear in wavenumber, and there could in 
particular be discontinuities in $\partial Z_5/\partial\nu_\text{eff}$ at $\nu_\text{eff}=1.24$ 
and 1.72~$\mu$m$^{-1}$. As no clamping is used for sources with six-parameter 
solutions, such abrupt changes in slope are not expected for $Z_6$ versus pseudocolour, 
but the relation can still be non-linear from other effects.

Because $\nu_\text{eff}$ is computed from the detailed BP and RP spectra, while
$G_\text{BP}-G_\text{RP}$ depends on the ratio of the integrated fluxes in the two band,
there is no strict one-to-one relation between the two quantities. Nevertheless, for
$-0.5\le G_\text{BP}-G_\text{RP}\le 7$ the following simple formulae
\begin{gather}\label{eq:nuEff1}
\nu_\text{eff} \simeq 1.76 - \frac{1.61}{\pi}\,\text{atan}\,
\Bigl(0.531(G_\text{BP}-G_\text{RP})\Bigr)\quad \mu\text{m}^{-1}\, , \\
G_\text{BP}-G_\text{RP} \simeq \frac{1}{0.531}\,\text{tan}\,
\Biggl(\frac{\pi}{1.61}\,(1.76-\nu_\text{eff})\Biggr)\quad \text{mag} \, , \label{eq:nuEff2}
\end{gather}
represent the mean relation for stellar objects to within $\pm 0.007~\mu$m$^{-1}$ 
in the effective wavenumber \citep{EDR3-DPACP-128}.

\subsection{Scanning law}
\label{sec:scanlaw}

The two telescopes in \textit{Gaia} are continuously scanning the celestial sphere 
according to a pre-defined schedule known as the scanning law. Details are
given in Sect.~5.2. of \citet{2016A&A...595A...1G}. For thermal stability, it is
necessary that the spin axis is kept at a fixed angle ($45^\circ$) to the Sun at
all times, and as a consequence the pattern of scanning is roughly symmetric
about the ecliptic, at least in a statistical sense and after several years of 
observations. Thus, although equatorial (ICRS) coordinates are used throughout 
the processing and for the astrometric end products, many characteristics of the 
data are (approximately) aligned with ecliptic coordinates rather than equatorial. 

Three examples of the ecliptic alignment are shown in Fig.~\ref{figA20}.
In the top panel (\textit{a}), the precision of the parallaxes is shown to depend
systematically on the ecliptic latitude ($\beta$), with 50--60\% higher
uncertainties along the ecliptic than near the ecliptic poles. The middle panel
(\textit{b}) shows the mean correlation coefficient between parallax and
pseudocolour in the six-parameter solutions. This correlation is systematically
positive for $\beta\gtrsim 45^\circ$ and negative for $\beta\lesssim -45^\circ$,
which is relevant for the parallax bias, because an offset in the assumed 
colours of the sources translates into a parallax bias that is proportional to the
correlation coefficient. Although this correlation coefficient is only available for
the six-parameter solutions, the correlation between the errors in 
colour and parallax exists also for sources with five-parameter solutions. For
example, if $\nu_\text{eff}$ is systematically too high in the five-parameter
solutions, the correlations will produce a more positive parallax bias in the (ecliptic)
northern sky than in the south. At intermediate latitudes the correlation coefficient
exhibits more complex (and larger) variations related to the scanning law.
The bottom panel (\textit{c}) is a smoothed map 
of quasar parallaxes, increased by a constant $17~\mu$as to compensate for 
the global offset. Large regional variations are seen at middle latitudes, but for 
$|\,\beta\,|\gtrsim45^\circ$ there is clearly a systematic difference between north 
and south. Although such a systematic could be produced by the correlation
mechanism just described, several other explanations can be envisaged.

The maps in panels (\textit{a}) and (\textit{b}) of Fig.~\ref{figA20} show substantial 
regional and local variations, especially for $|\,\beta\,|\lesssim45^\circ$. These features 
are related to the scanning law and the (still) relatively poor coverage of the ecliptic zone 
obtained in the 33~months of data used for the EDR3 astrometry (global coverage is 
optimised for a mission length of 60~months). It can be expected that the parallax 
bias has regional and local variations of a similar character, and the map of quasar 
parallaxes (panel \textit{c}) seems to confirm this, although the finer details are made
invisible by the smoothing, and small-number statistics contribute to the variations 
most clearly along the Galactic zone of avoidance. In practice it is however very difficult
to determine local or even regional variations in $Z_5$ and $Z_6$ with any degree of 
confidence.
Even if we trust (some of) the variations seen in the quasar map, they are probably
only representative for the faint sources with colours similar to those of the quasars, 
that is bluer than the typical faint Galactic stars.

Recognising that a detailed mapping of $Z_5$ and $Z_6$ versus position is
not possible, but that a global variation with ecliptic latitude is expected on
theoretical grounds and also seen empirically, the only positional argument
in $Z_5$ and $Z_6$ is taken to be ecliptic latitude, $\beta$. In the \textit{Gaia}
Archive this coordinate is given (in degrees) as \gacs{ecl\_lat}. Alternatively,
it can be computed to within $\pm 43$~mas using the formula
\begin{equation}\label{eq:sinBeta}
\sin\beta=0.9174820621\sin\delta - 0.3977771559\cos\delta\sin\alpha \, .
\end{equation}
In this paper, all expressions involving the ecliptic latitude are written in terms
of $\sin\beta$.

\begin{figure}
\centering
  \includegraphics[height=195mm]{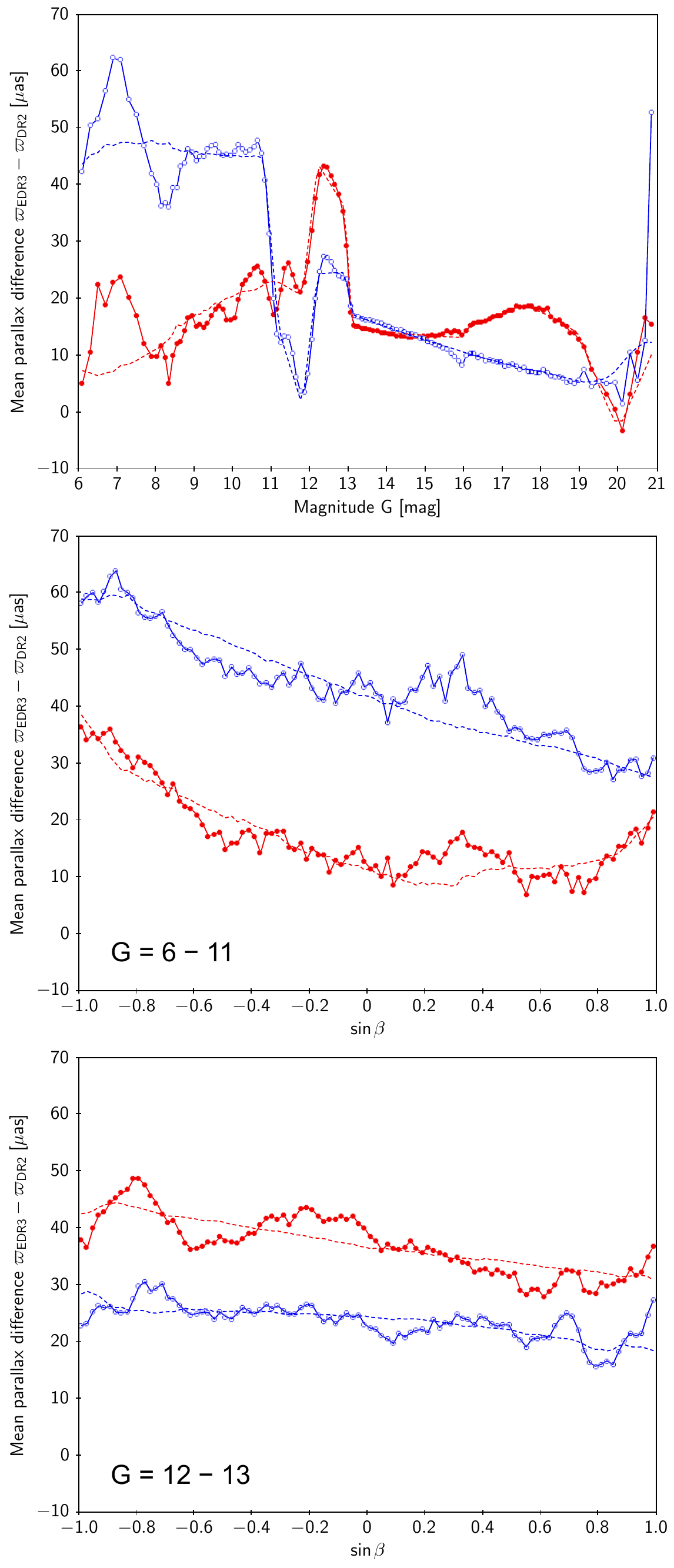}    
    \caption{Mean difference in parallax between EDR3 and DR2 versus magnitude (top)
    and, in two magnitude intervals, versus ecliptic latitude (middle and bottom). Circles 
    connected by solid lines are weighted mean values computed in bins of variable size, 
    with at least 1000~sources per bin; dashed lines are mean values of the fitted parametrized 
    function $\Delta Z$ (Eq.~\ref{eq:dZ1}), binned as for the sources.
    Red filled circles are for $\nu_\text{eff}<1.48~\mu$m$^{-1}$, blue open circles
    for $\nu_\text{eff}>1.48~\mu$m$^{-1}$.}
    \label{fig:dPlx3m2vsGB}
\end{figure}

\begin{figure*}
\centering
  \includegraphics[height=195mm]{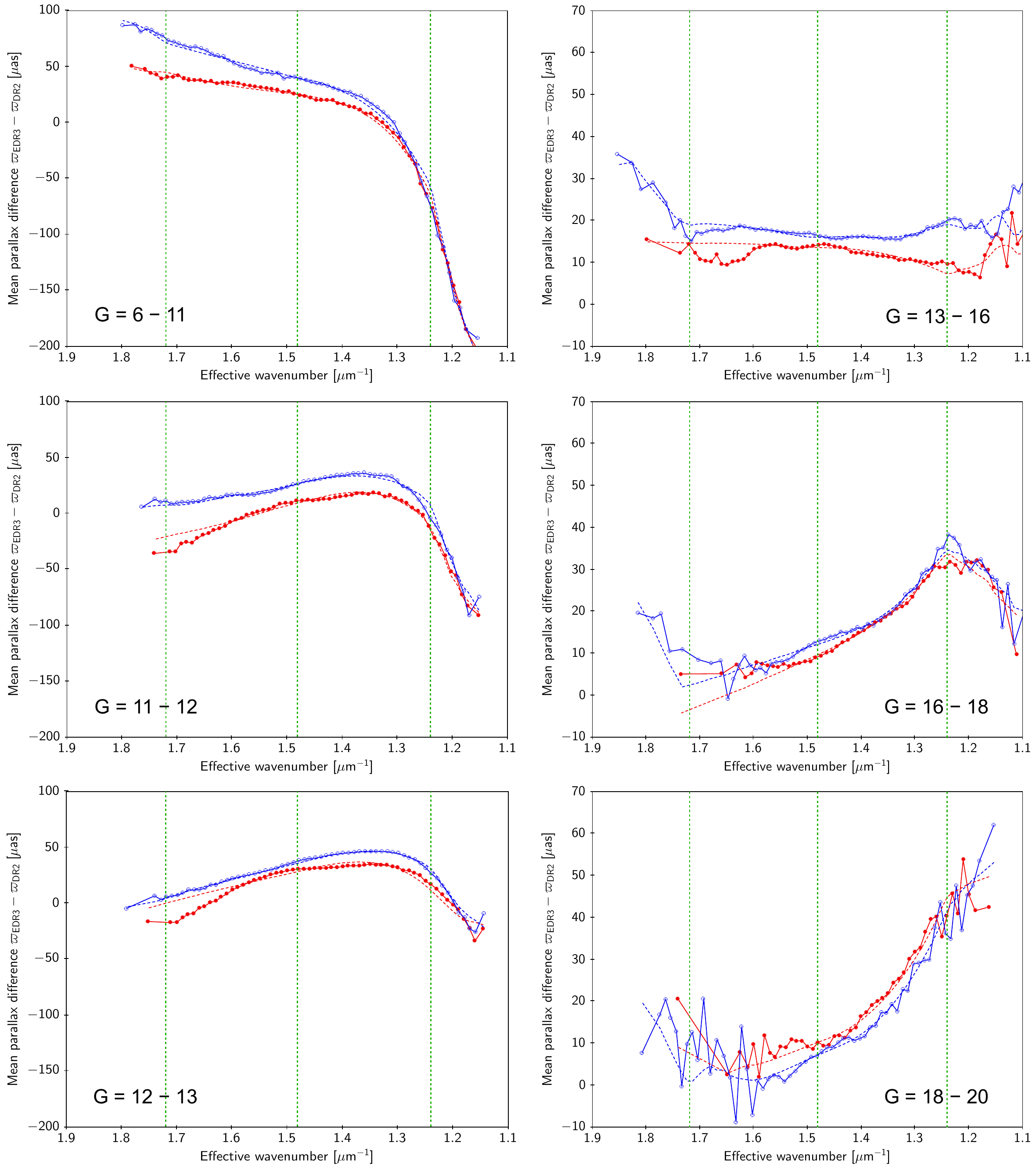}   
    \caption{Mean difference in parallax between EDR3 and DR2 as a function of effective 
    wavenumber for several ranges of the $G$ magnitude. Circles 
    connected by solid lines are weighted mean values computed in bins of variable size,
    with at least 1000~sources per bin;
    dashed lines are mean values of the fitted parametrized 
    function $\Delta Z$ (Eq.~\ref{eq:dZ1}), binned as for the sources.
    Red filled circles are for sources with $\beta>0$, blue open circles for $\beta<0$.
    The vertical dashed lines mark the breakpoints for the basis functions $c_j(\nu_\text{eff})$ 
    in Eq.~(\ref{eq:c}), namely the clamping limits at 1.24 and 1.72~$\mu$m$^{-1}$
    and the midpoint at 1.48~$\mu$m$^{-1}$.}
    \label{fig:dPlx3m2vsNuEff}
\end{figure*}

\begin{table*}
\caption{Coefficients for the function $\Delta Z(G,\,\nu_\text{eff},\,\beta)$ fitted to the
difference in parallax between EDR3 and DR2.  
\label{tab:dZ}}
\footnotesize\setlength{\tabcolsep}{4pt}
\begin{tabular}{crrrrrrrrrrrrrrr}
\hline\hline
\noalign{\smallskip}
 $G$ & $q_{00}$ & $q_{01}$ & $q_{02}$ & $q_{10}$ & $q_{11}$ & $q_{12}$ & $q_{20}$ & $q_{21}$ & $q_{22}$ & $q_{30}$ & $q_{31}$ & $q_{32}$ & $q_{40}$ & $q_{41}$ & $q_{42}$ \\
\noalign{\smallskip}\hline\noalign{\smallskip}
 6.0 & $+25.61$  & $ +3.66$  & $ -4.55$  & $+173.9$  & $ -37.1$  & $ -40.8$  & $ -3662$  & $ -1451$  & $ +3070$  & $+1783.0$  & $-$ & $-1299.0$  & $+408.8$  & $+121.7$\rlap{:} & $-609.6$  \\
10.8 & $+33.31$  & $-14.26$  & $ +9.32$  & $+101.8$  & $ -41.8$  & $ -52.5$  & $ -5556$  & $-$ & $ +2780$  & $+2477.6$  & $ -99.4$  & $-1314.0$  & $ +82.7$\rlap{:} & $-512.0$  & $-312.1$  \\
11.2 & $+19.91$  & $-18.28$  & $ +6.92$  & $ -73.5$  & $ -35.8$  & $ -43.4$  & $ -4844$  & $  -955$  & $ +1363$  & $+1827.7$  & $-219.8$  & $-1146.8$  & $-110.5$\rlap{:} & $-$ & $-$ \\
11.8 & $+12.45$  & $-11.50$  & $ +2.53$  & $-138.5$  & $ -25.9$  & $ -14.0$  & $ -2706$  & $  -549$  & $ +1712$  & $+1078.5$  & $-192.5$  & $-469.3$  & $-$ & $-$ & $-322.9$  \\
12.2 & $+37.20$  & $ -1.49$  & $ -2.63$  & $-151.2$  & $  +8.0$  & $ +10.8$  & $ -3548$  & $  -904$  & $ +1642$  & $+489.6$  & $-142.6$  & $-386.1$  & $-258.7$  & $ -96.6$\rlap{:} & $-229.1$\rlap{:} \\
12.9 & $+31.36$  & $ -9.14$  & $-$ & $ -99.1$  & $ +12.2$  & $-$ & $ -2730$  & $  -492$  & $  +922$  & $+705.6$  & $-265.8$  & $-290.1$  & $ -63.6$\rlap{:} & $-$ & $-$ \\
13.1 & $+15.20$  & $ -2.45$  & $ +3.99$  & $ +17.1$  & $  -9.5$  & $ -16.6$  & $  -159$  & $ -1087$  & $  +949$  & $ -51.9$  & $ -67.6$  & $-$ & $-$ & $-$ & $+281.3$  \\
15.9 & $+10.94$  & $ -0.60$  & $ +2.80$  & $ -11.0$  & $ +13.3$  & $ -21.6$  & $  +655$  & $  -116$\rlap{:} & $  +737$  & $-$ & $-$ & $-301.4$  & $-$ & $-176.6$  & $-$ \\
16.1 & $+10.86$  & $ -3.26$  & $ +2.11$  & $ -32.6$  & $ -19.5$  & $ -22.3$  & $  +688$  & $  -345$  & $ +1107$  & $-$ & $-$ & $-321.2$  & $-$ & $-188.5$  & $-$ \\
17.5 & $+10.27$  & $-$ & $ +3.20$  & $ -79.0$  & $-$ & $ -21.2$\rlap{:} & $ +1000$  & $-$ & $ +1441$  & $-$ & $+177.5$\rlap{:} & $-287.8$\rlap{:} & $-$ & $-$ & $+518.3$  \\
19.0 & $ +7.00$  & $ +4.40$  & $ +8.64$  & $ -45.6$  & $-$ & $-$ & $ +1851$  & $-$ & $-$ & $-271.1$\rlap{:} & $-$ & $-$ & $+356.4$  & $-$ & $-$ \\
20.0 & $ +0.43$\rlap{:} & $ +5.55$\rlap{:} & $+14.53$\rlap{:} & $+115.3$  & $-$ & $-$ & $ +3827$  & $-$ & $ -8769$\rlap{:} & $-$ & $-$ & $-$ & $-$ & $-$ & $-$ \\
21.0 & $+12.34$\rlap{:} & $-$ & $-$ & $-$ & $-$ & $-$ & $-$ & $-$ & $-$ & $-$ & $-$ & $-$ & $-$ & $-$ & $-$ \\
\noalign{\smallskip}
\hline
\end{tabular}
\tablefoot{The table gives $q_{jk}(G)$ in Eq.~(\ref{eq:dZ1}) at the values of $G$ in the first column. 
For other values of $G$, 
linear interpolation should be used. A dash (--) indicates that the coefficient is not significant at the 
$2\sigma$ level and should be taken to be zero. A colon (:) after the coefficient indicates that it is 
not significant at the $3\sigma$ level. Units are: $\mu$as (for $q_{0k}$), $\mu$as~$\mu$m 
($q_{1k}$, $q_{3k}$, and $q_{4k}$), and $\mu$as~$\mu$m$^2$ ($q_{20}$).} 
\end{table*}

\section{Systematic differences between EDR3 and DR2 parallaxes}
\label{sec:dr2}

Although it is not our goal to study the parallax bias for \textit{Gaia} DR2, 
that is the function $Z_\text{DR2}(\vec{x})$, it is instructive to start this investigation 
by looking at the systematic parallax difference between EDR3 and DR2, that is the 
differential bias function
\begin{equation}\label{eq:dr3m2}
\Delta Z(G,\,\nu_\text{eff},\,\beta) = Z_5(G,\,\nu_\text{eff},\,\beta) - Z_\text{DR2}(G,\,\nu_\text{eff},\,\beta) \, .
\end{equation}
In contrast to either $Z_5$ or $Z_\text{DR2}$, this difference can easily be mapped in 
considerable detail simply by computing mean differences in parallax for sources having 
the same identifier (\gacs{source\_id}) in the two releases.%
\footnote{For a small fraction of the sources in EDR3, no source with the same \gacs{source\_id}
is found in DR2 (and vice versa), owing to the evolution of the source list between 
releases \citep{EDR3-DPACP-130}. That complication is presently ignored, as the EDR3--DR2 
differences are not a main topic of the paper.}
Obviously, the difference 
contains no direct quantitative information on $Z_5$, but it is reasonable to expect that 
it gives a good qualitative indication of the relevant dependences on $G$, $\nu_\text{eff}$, and 
$\beta$. In fact, the definition of
the general parametrised function $Z(G,\,\nu_\text{eff},\,\beta)$ in Appendix~\ref{sec:num} 
is largely guided by the shape of $\Delta Z$.

In Eq.~(\ref{eq:dr3m2}) we use $Z_5$ rather than $Z_6$, because there are rather few sources 
with six-parameter solutions for $G<13$ that appear also in DR2. The colour argument is 
$\nu_\text{eff}$ given by \gacs{nu\_eff\_used\_in\_astrometry} in EDR3, which is available 
for all five-parameter solutions. 

Figure~\ref{fig:dPlx3m2vsGB} shows the mean parallax difference 
$\varpi_\text{EDR3}-\varpi_\text{DR2}$ for sources with five-parameter solutions in EDR3, 
subdivided by colour. 
The top panel shows the dependence on magnitude, while the middle and bottom panels show 
examples of the dependence on $\beta$ in two different magnitude intervals. 
In Fig.~\ref{fig:dPlx3m2vsNuEff} we similarly show the mean difference as a function of colour, 
but subdivided by magnitude and ecliptic latitude.%
\footnote{In this figure, and in all other diagrams with effective wavenumber or pseudocolour 
on the horizontal axis, the direction of the axis has been reversed to follow the usual convention for 
colour-magnitude diagrams, that is, blue objects to the left and red to the right.\label{fn:reverse}}
These figures, and all subsequent results on $\Delta Z$, were computed using all common 
sources for $G<14$ and a geometrically decreasing random fraction of the fainter sources.
In total about 26.7~million sources were used. Mean values of the parallax differences
were computed using weights proportional to 
$1/(\sigma_{\!\varpi,\,\text{EDR3}}^2+\sigma_{\!\varpi,\,\text{DR2}}^2)$.

The plots in Figs.~\ref{fig:dPlx3m2vsGB} and \ref{fig:dPlx3m2vsNuEff} show complex dependencies
on $G$, $\nu_\text{eff}$, and $\beta$, with interactions among all three arguments (that is, 
the colour-dependence is different depending on both $G$ and $\beta$, etc.). We use these
results to design a continuous, multi-dimensional function, relevant parts of which are used 
in Sects.~\ref{sec:p5} and \ref{sec:p6} to represent the functions $Z_5$ and $Z_6$ fitted 
to the EDR3 data. Owing to the scarcity of data available for those fits, the general function 
has to be quite schematic and cannot take into account many of the details seen in 
$\Delta Z$, especially for the brightest sources. With that purpose in mind, the following 
features in $\Delta Z$ are noteworthy:
\begin{itemize}
\item As a function of $G$, there are jumps at $G\simeq 11.0$, 12.0, 13.0, and 16.0, corresponding to
the boundaries shown in Fig.~\ref{fig:gClass}. The transitions are not abrupt, but occur over an
interval of 0.2--0.4~mag. Features at $G\lesssim 9$ will in the following be ignored because the
number of sources is too small for a reliable mapping of the parallax bias.\smallskip
\item As a function of $\nu_\text{eff}$, the effect of the clamping at 1.24 and 1.72~$\mu$m$^{-1}$ 
(Sect.~\ref{sec:nueff}) is visible in some plots as an abrupt change of slope; within these limits the 
relation is approximately linear for $1.48<\nu_\text{eff}<1.72~\mu$m$^{-1}$ and curved (cubic)
for $1.24<\nu_\text{eff}<1.48~\mu$m$^{-1}$, with no visible break at 1.48~$\mu$m$^{-1}$.
The `hook' for the reddest stars ($\nu_\text{eff}<1.18~\mu$m$^{-1}$) seen for $G<16$~mag
is not related to any known feature of the instrument or data processing, and since it concerns
very few objects it will be ignored in the following.\smallskip
\item As a function of $\beta$, the differences typically show an approximately linear or quadratic
dependence on $\sin\beta$.
\end{itemize}     
The parametrised function described in Appendix~\ref{sec:num} is designed to take into account
these features. Using that form, the differential bias is approximated by
\begin{equation}\label{eq:dZ1}
\Delta Z(G,\,\nu_\text{eff},\,\beta)=\sum_j \sum_k q_{jk}(G)\,c_j(\nu_\text{eff})\,b_k(\beta) \, ,
\end{equation}
where $c_j$ and $b_k$ are basis functions in $\nu_\text{eff}$ and $\beta$, specified by
Eqs.~(\ref{eq:c}) and (\ref{eq:b}), respectively, and $q_{jk}(G)$ are piecewise linear functions 
of $G$ given by Eqs.~(\ref{eq:g}) and (\ref{eq:Z2}) in terms of the fitted coefficients 
$z_{i\!jk}$ in Eq.~(\ref{eq:Z}).

The approximation in Eq.~(\ref{eq:dZ1}) has $13\times 5\times 3=195$ free parameters, 
namely all possible combinations of $i=0\dots 12$, $j=0\dots 4$, and $k=0\dots 2$ in 
the fitted coefficients $z_{i\!jk}$.
A simultaneous least-squares estimation of all 195 parameters shows that many of them
are poorly determined and contribute little to the overall fit. To avoid overfitting the
following procedure is used. First, all 195 parameters are estimated ($\hat{z}_{i\!jk}$), along 
with their formal uncertainties ($\sigma_{i\!jk}$). The parameter with the smallest standard score
$S_{i\!jk}=|\hat{z}_{i\!jk}|/\sigma_{i\!jk}$ is then removed (set to zero), and a new fit calculated 
with updated uncertainties. This procedure is repeated until $S_{i\!jk}>2$ for all the retained 
parameters. However, the parameters $z_{i\!jk}$ with $j=k=0$ are always retained 
at their estimated values independent of their scores. This is done in order to avoid that 
the sum in Eq.~(\ref{eq:dZ1}) defaults to zero at some $G$ independently of $\nu_\text{eff}$ 
and $\beta$.

Applying the above procedure to the EDR3--DR2 parallax differences yields the
137 non-zero coefficients shown in Table~\ref{tab:dZ}. For compactness, no uncertainties
are given; all values are significant at least on the $2\sigma$ level, although
some, indicated by a colon in the table, are below $3\sigma$. The precise values of the
coefficients, as well as the subset of significant coefficients, will depend on the selection of
sources used in the fit, which was increasingly non-exhaustive for $G>14$. Using more of
the faint sources in EDR3 and DR2 would undoubtedly give a better determination of the 
coefficients towards the faint end, and increase the number of significant coefficients.

As a check of the fit, the function $\Delta Z$ defined by the coefficients in Table~\ref{tab:dZ}
was evaluated for each of the 26.7~million sources used in the fit, and mean values in
bins of magnitude, colour, and ecliptic latitude were computed in exactly the same way as
was done for the parallax differences. The results, shown by the dashed lines in 
Figs.~\ref{fig:dPlx3m2vsGB} and \ref{fig:dPlx3m2vsNuEff}, thus represent our multidimensional
fit to the mean parallax differences shown as circles in the same diagrams. The overall fit is
reasonable, but there are a few systematic deviations that should be commented on:
\begin{itemize}
\item
Around $G\simeq 7.0$ and $\simeq 8.0$ there are large (10--15~$\mu$as) positive and 
negative excursions in the mean parallax differences that are not modelled by the fit
(Fig.~\ref{fig:dPlx3m2vsGB}, top). This
discrepancy is a direct consequence of our choice to use a linear dependence on $G$ 
in the interval 6.0 to 10.8, which in turn is motivated by the scarcity of data for estimating 
$Z_5$ in this interval (see Sect.~\ref{sec:pairs6}).
\item
As a function of $\beta$ (Fig.~\ref{fig:dPlx3m2vsGB}, middle and bottom) there are local 
excursions from the general quadratic trend on a level of $\pm 5~\mu$as. These are 
related to localised features on the sky (cf.\ Fig.~\ref{figA20}), whereas the fitted model 
is only meant to represent the position dependences on the largest scales.
\item
As a function of $\nu_\text{eff}$ (Fig.~\ref{fig:dPlx3m2vsNuEff}), the variation between
1.48 and 1.72~$\mu$m$^{-1}$ is not linear, as assumed in Eq.~(\ref{eq:c}), which sometimes
gives deviations of 5--10~$\mu$as for moderately blue sources 
($\nu_\text{eff}\simeq 1.65~\mu$m$^{-1}$). In the red end the curvature of the
cubic segment is not always sufficient to give a good fit around 
$\nu_\text{eff}\simeq 1.25~\mu$m$^{-1}$. 
\end{itemize}
These discrepancies could be caused by specific features in DR2, EDR3, or both.
It is of course also possible that the two data sets have similar systematics that cancel
in the parallax difference, and which therefore have not been identified and included in 
the adopted model (Appendix~\ref{sec:num}). Nevertheless, as shown in the next sections,
the model seems to be general and flexible enough to describe $Z_5$ and $Z_6$ to
the level of detail permitted by the data.

\begin{figure}
\centering
  \includegraphics[height=195mm]{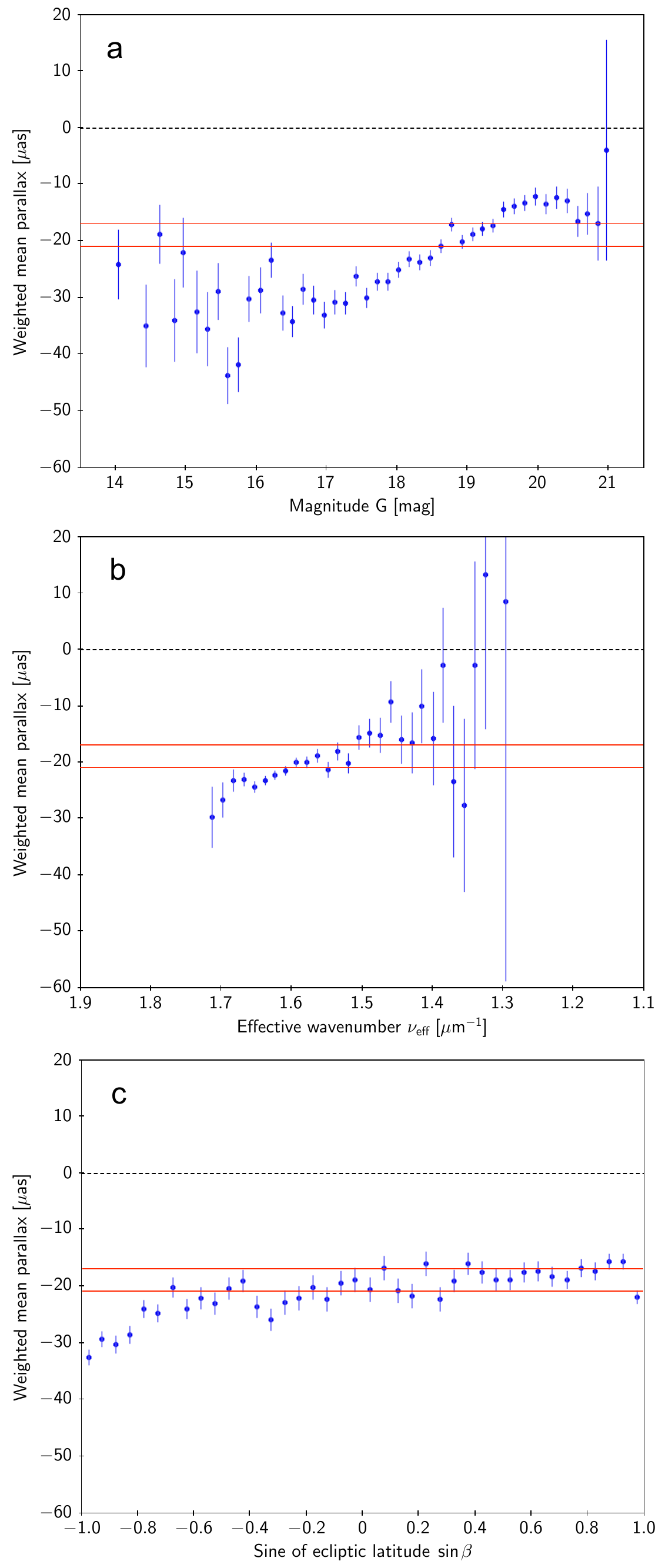}   
    \caption{Mean parallax of quasars binned by magnitude (\textit{a}), 
    effective wavenumber (\textit{b}), and sine of ecliptic latitude (\textit{c}).
    In each bin the dot is the mean parallax in EDR3 weighted by $\sigma_\varpi^{-2}$, with
    error bars indicating the estimated standard deviation of the weighted mean.
    The two red lines indicate the weighted mean value ($-21~\mu$as) and 
    median ($-17~\mu$as) of the full sample.}
    \label{figA10}
\end{figure}

\section{Five-parameter solutions}
\label{sec:p5}

The main methods for estimating $Z_5(G,\nu_\text{eff},\beta)$ in this paper are 
(i) quasars (Sect.~\ref{sec:qso}), which provide a direct estimate of the parallax bias
for $G\gtrsim 14$, although only in a rather narrow interval of $\nu_\text{eff}$; 
(ii) stars in the Large Magellanic Cloud (LMC; Sect.~\ref{sec:LMC}), which map differential 
variations of the bias over a range of magnitudes and colours, although only in a
specific location near the south ecliptic pole; and
(iii) physical pairs or binaries (Sect.~\ref{sec:pairs}), which can be used to map the differential 
variations for the bright stars. Additionally, we use red clump (RC) stars for a differential study 
of the bias in the red, where the number of quasars is too small (Appendix~\ref{sec:rc}). 
In subsequent sections, these methods will be developed in full detail.

In this process no assumption is made concerning the distance to the LMC, or the absolute 
magnitude of the RC stars. These objects are used purely differentially, and our estimates 
of the parallax biases are completely anchored in the quasars, that is, put on an 
absolute scale by means of their parallaxes. 

\subsection{Quasars}
\label{sec:qso}

As part of EDR3, the \textit{Gaia} Archive%
\footnote{\url{https://gea.esac.esa.int/archive/}}
contains the table \gacs{agn\_cross\_id}, listing a total of 1\,614\,173 sources constituting a very clean 
sample of quasar-like objects, whose positions and proper motions in EDR3 formally define the 
reference frame of the catalogue, \textit{Gaia}-CRF3. The list was constructed by cross-matching the 
full EDR3 catalogue with 17 external catalogues of active galactic nuclei (AGN) followed by a filtering 
based on the quality of the solutions and the astrometric parameters: the proper motions are consistent with 
zero, and the parallaxes with a constant offset of $-17~\mu$as, to within five times their respective
formal uncertainties. Full details of the selection procedure, and further characterisation of the sample, 
are given in \citep{EDR3-DPACP-133}. In spite of the strict selection criteria it is likely that the list contains
a small number of stellar contaminants. As described further down, we take this into account in a 
statistical way. 

The quasar sample used here is a subset of \gacs{agn\_cross\_id}, consisting of 1\,107\,770~sources%
\footnote{This investigation used a preliminary version of the table,
resulting in a slightly smaller sample than the corresponding selection from the 
published table. The final version was used for the check in Sect.~\ref{sec:val}.} 
with five-parameter solutions and effective wavenumbers 
($\nu_\text{eff}=\gacs{nu\_eff\_used\_in\_astrometry}$) 
in the range 1.24 to 1.72~$\mu$m$^{-1}$. The median $\nu_\text{eff}$ is 
1.59~$\mu$m$^{-1}$, with 1st and 99th percentiles at 1.44 and 1.69~$\mu$m$^{-1}$,
which makes this sample significantly bluer than typical stars of similar magnitudes, and
covering a smaller range of colours. The magnitudes range from $G\simeq 13.4$ to 21.0;
only 97 are brighter than $G=15$ and 541 are brighter than $G=16$.

\begin{figure}
\centering
  \includegraphics[height=65mm]{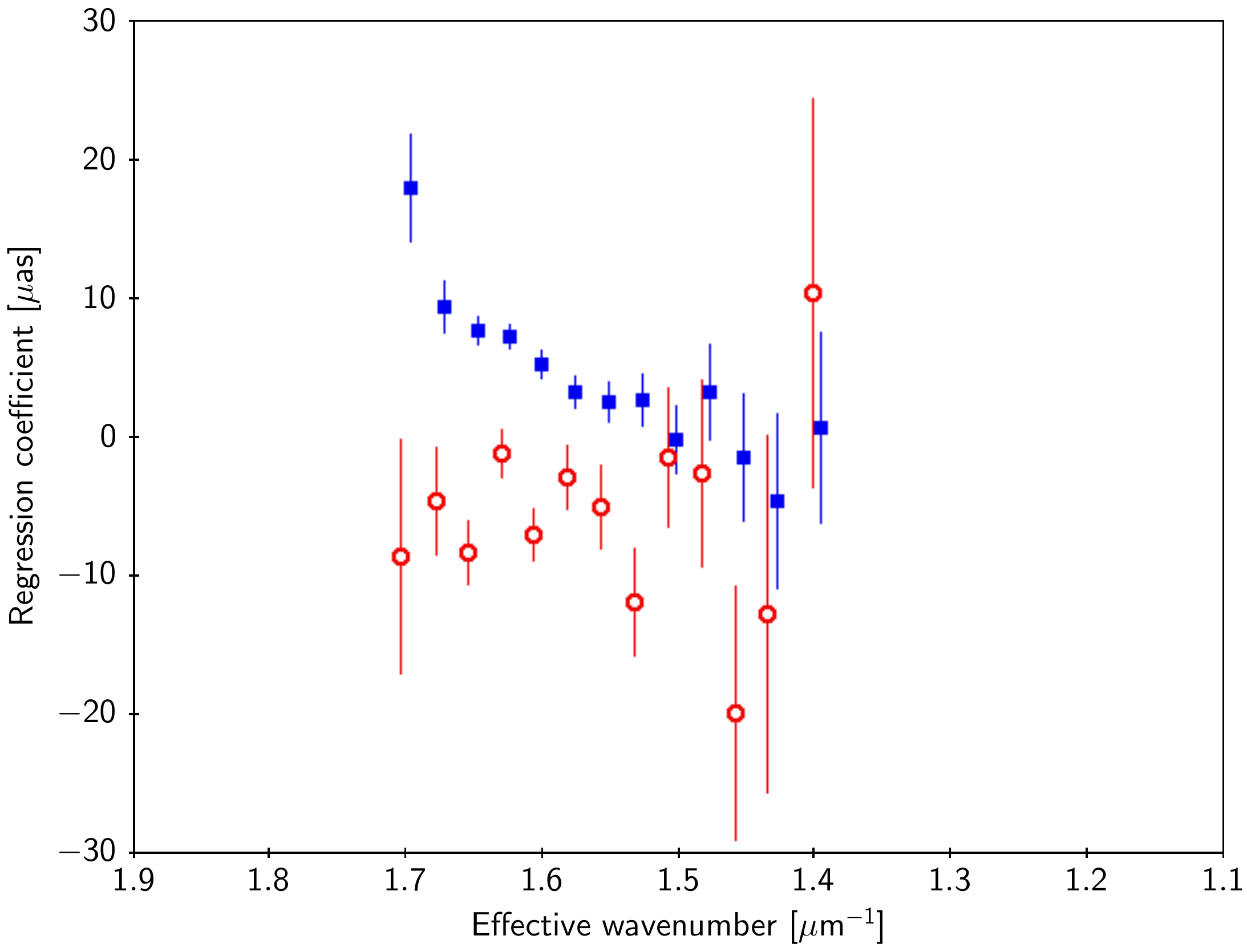}
  \caption{An example of interactions among the dependences of quasar parallaxes 
  on colour and ecliptic latitude. In a quadratic regression of parallax versus $\sin\beta$,
  the linear coefficient (filled blue squares) exhibits a strong variation with effective 
  wavenumber, while no such trend is shown by the quadratic coefficient (open red circles).
  The points have been slightly displaced sideways to avoid that error bars overlap.}
  \label{fig:CB}
\end{figure}

Figure~\ref{figA10} shows the weighted mean parallax plotted against 
$G$, $\nu_\text{eff}$, and $\beta$. The main trends are as follows. 
\begin{itemize}
\item On the whole, there is a negative parallax bias: the weighted mean parallax of the
sample used here is approximately $-21~\mu$as and the median is about $-17~\mu$as. 
For reference, these values are indicated by the red lines in Fig.~\ref{figA10}.
\item As a function of $G$, there is a clearly non-linear variation with an approximately 
linear increase from $G\simeq 17$ to 20, with plateaus on either side 
of this interval or perhaps a decreasing trend for $G>20$.
\item As a function of $\nu_\text{eff}$, the variation is approximately linear in the 
well-populated range of colours. If there is a curvature similar to what is seen in the 
EDR3--DR2 differences (Fig.~\ref{fig:dPlx3m2vsNuEff}) or the LMC data (Fig.~\ref{figA03}), 
the interval in $\nu_\text{eff}$ covered by the quasars is too narrow to reveal it.
\item As a function of $\beta$, the variation may be described by a quadratic polynomial 
in $\sin\beta$.
\end{itemize}
Similarly to what was observed in the EDR3--DR2 differences in Sect.~\ref{sec:dr2},
interactions among the three main variables are seen also in the quasar parallaxes.
For example, if the quasars are binned by effective wavenumber and a quadratic polynomial 
$a_0+a_1\sin\beta+a_2\sin^2\beta$ is fitted to the parallaxes in each bin, it is found that
$a_1$ has a strong dependence on $\nu_\text{eff}$, whereas no clear trend is seen 
for $a_2$ (Fig.~\ref{fig:CB}).

The trends described above, including interactions, are well approximated by the 
parametrised function $Z(G,\,\nu_\text{eff},\,\beta)$
defined in Appendix~\ref{sec:num}. However, the limited supports in $G$ and $\nu_\text{eff}$ 
make it necessary to constrain the general expression in Eq.~(\ref{eq:Z}) in several ways.
Specifically, for the basis functions in magnitude, $g_i(G)$ in Eq.~(\ref{eq:g}), we only use 
$i=6\dots 12$ (that is, $z_{i\!jk}$ is not fitted for $i<6$), and for the basis function in colour, 
$c_j(\nu_\text{eff})$ in Eq.~(\ref{eq:c}), we only use $j=0$ and 1. For $G<17.5$ there are not
enough quasars to determine reliably a linear variation with $G$, as permitted by the 
basis functions; instead we assume that the bias is independent of $G$ in the
intervals $13.1<G<15.9$ and $16.1<G<17.5$, but allow a step around $G=16$
representing the transition from window class WC1 to WC2 (Sect.~\ref{sec:wcgates}).%
\footnote{Algorithmically, this is achieved by using the combined basis functions 
$g_6(G)+g_7(G)$ and $g_8(G)+g_9(G)$ instead of the original four functions, or, 
equivalently, by adding the constraints $z_{6,jk}=z_{7,jk}$ and $z_{8,jk}=z_{9,jk}$ for 
all combinations of indices $j$ and $k$.\label{fn2}}
Additionally, the terms with $j=1$ and $k=2$ were all found to be insignificant, and
therefore constrained to zero. The resulting fit is given in Table~\ref{tab:qsz0}.

\begin{figure}[t]
\centering
  \includegraphics[height=65mm]{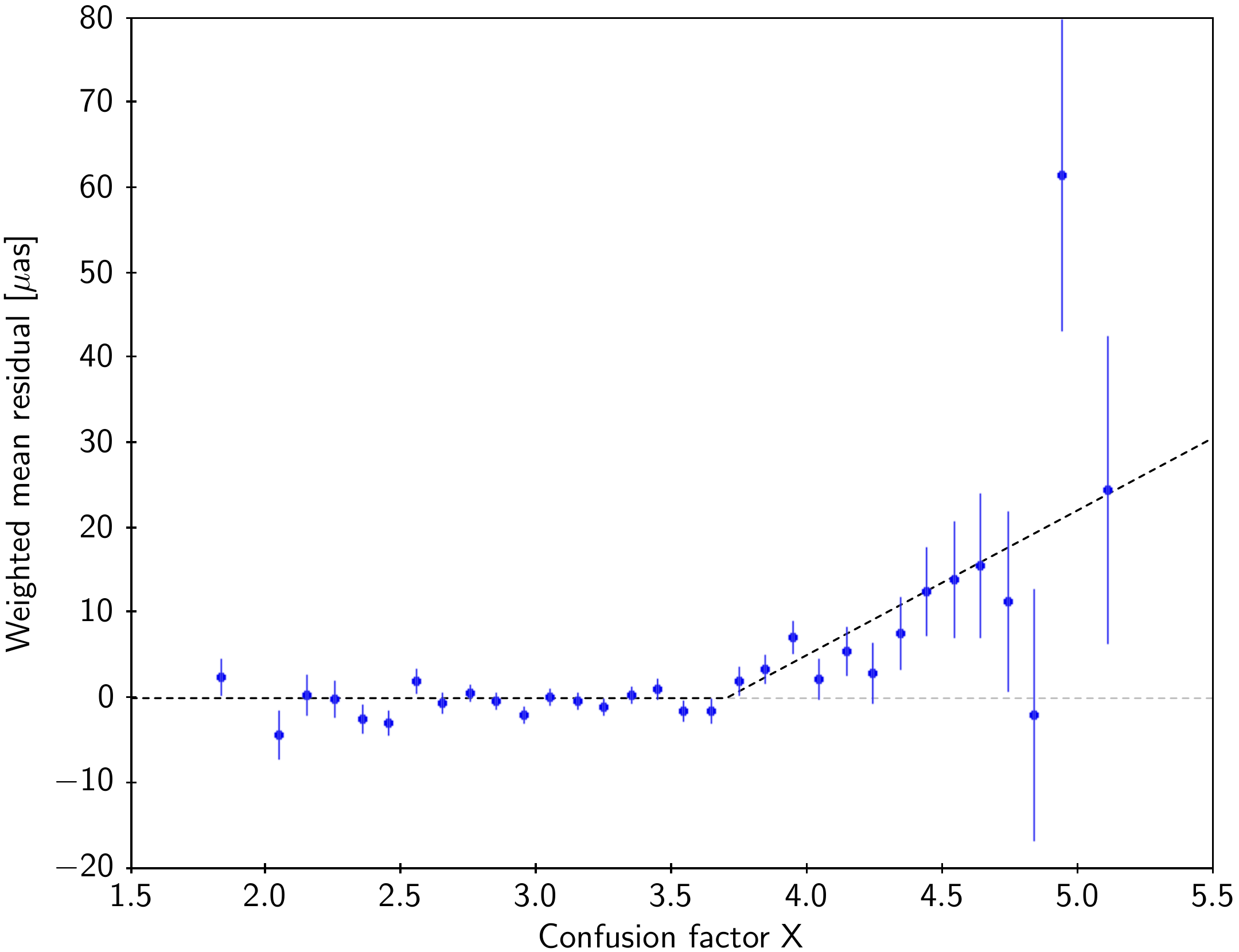}
  \caption{Mean residual in quasar parallax after a regression on $G$, $\nu_\text{eff}$,
  and $\beta$ (see text), plotted against the confusion factor from Eq.~(\ref{eq:X1}).
  In each bin the dot is the mean residual weighted by $\sigma_\varpi^{-2}$, with
  error bars indicating the estimated standard deviation of the weighted mean. The
  broken dashed line is the dependence modelled by Eq.~(\ref{eq:X2}).}
  \label{fig:X}
\end{figure}

At this point it is necessary to consider to what extent the fit is affected by a possible 
contamination of the quasar sample by Galactic stars. The contaminating stars will on
average have higher measured parallaxes than the quasars of similar magnitude, thus 
biasing the fitted function towards more positive values. The effect is only expected to be 
important at the faint end and where the star density is high. In order to explore this 
we introduce the `confusion factor'
\begin{equation}\label{eq:X1}
X = \log_{10} D_{21} + 0.3(G-21)\, ,
\end{equation}
where $D_{21}$ is the mean density of sources brighter than $G=21$ in the vicinity 
of the quasar, expressed in deg$^{-2}$. Densities are calculated by counting the total 
number of sources in EDR3 in pixels of solid angle 0.8393~deg$^2$ (healpix level 6),
divided by the solid angle. Since the density of faint sources in EDR3 roughly doubles 
with each magnitude ($\text{d}\log_{10}D_G/\text{d}G\simeq 0.3$), the second term in 
Eq.~(\ref{eq:X1}) is the approximate change in density with $G$. Thus $X$ is simply a 
convenient proxy for $\log_{10} D_G$, the density of sources brighter than the quasar. For the 
present quasar sample, $X$ ranges from about 1.25 to 5.5, with a median at 3.4. Less 
than 1\% of the quasars have $X>4.5$. 

If the mean quasar parallax is plotted versus $X$, there is a strong increase over practically 
the whole range of $X$ values. This is not primarily caused by contamination,
but rather by the positive trend of parallaxes versus $G$ shown in Fig.~\ref{figA10}a, 
transferred to $X$ via the second term in Eq.~(\ref{eq:X1}). 
In Fig.~\ref{fig:X} we plot instead the residual in parallax, from the regression in
Table~\ref{tab:qsz0}, versus the confusion factor. Here, the mean residual is close to zero 
for $X\lesssim 3.7$, after which it increases roughly linearly with $X$ as suggested
by the dashed line. Based on this plot, we assume that the contamination bias at a 
particular position is proportional to
\begin{equation}\label{eq:X2}
f(X) = \max(0,~X-3.7)\, ,
\end{equation}
with $X$ given by Eq.~(\ref{eq:X1}). A dependence on position is expected not only 
from the varying star density, as encoded in $X$, but also from variations in the quality of
\textit{Gaia} astrometry created by the scanning law (see, for example, several plots
in \citealt{EDR3-DPACP-128} and \citealt{EDR3-DPACP-126}). Globally, the precision and
number of observations improve towards the ecliptic poles, which to first order can be
described as a linear dependence on $\sin^2\beta$.
We consequently model the contamination bias in the mean quasar parallax by adding
the nuisance terms 
\begin{equation}\label{eq:X3}
\left( r_0(G)\cos^2\beta + r_1(G)\sin^2\beta\right) f(X)\, 
\end{equation}
to the fitted model. Here, $r_0(G)$ and $r_1(G)$ represent the contamination bias 
at $\beta=0$ and $\beta=\pm 90^\circ$, respectively; both functions are piecewise 
linear functions of $G$ using the basis functions $g_i(G)$ in Eq.~(\ref{eq:g}). Only the 
last two basis functions ($i=11$ and $12$) are used, as the $r$-coefficients are quite 
insignificant for $G\lesssim 20$. 

\begin{figure}[t]
\centering
  \includegraphics[width=0.95\hsize]{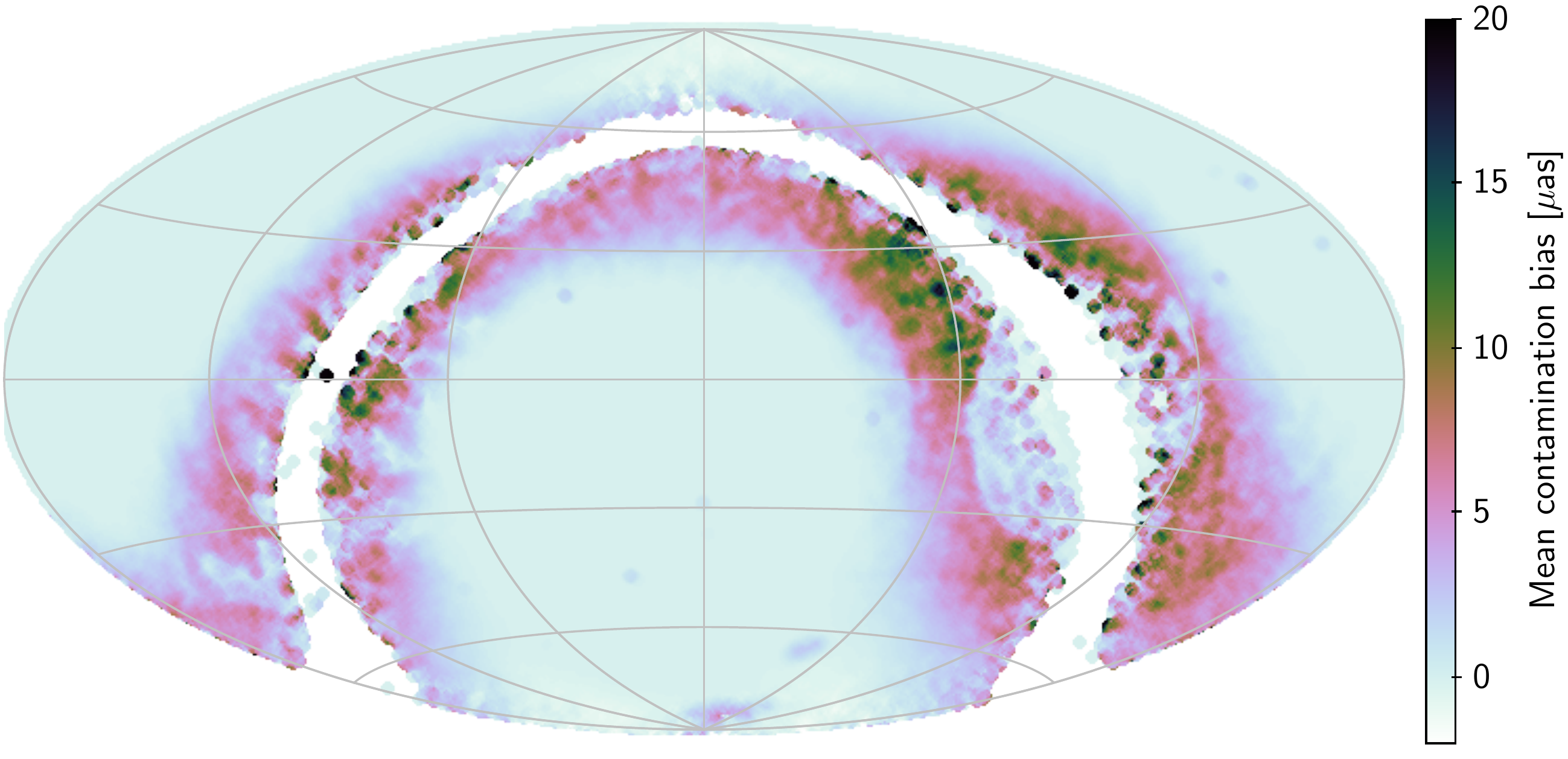}
  \caption{Celestial map of the mean contamination bias of the quasars, as given by
  Eq.~(\ref{eq:X3}) and Table~\ref{tab:qsz1}. The map uses the same projection in
  ecliptic coordinates as in Fig.~\ref{figA20}.}
  \label{fig:qsoX}
\end{figure}

The resulting fit, including the contamination terms $r_0$ and $r_1$, is given in 
Table~\ref{tab:qsz1}. Comparing with Table~\ref{tab:qsz0}, where the fit did not include
these terms, we conclude that the global bias (as shown by the difference in $q_{00}$) 
is about $+4~\mu$as at $G=21.0$ and below $1~\mu$as at $G=20.0$. Figure~\ref{fig:qsoX} 
is map of the bias, calculated from Eq.~(\ref{eq:X3}) and averaged over the quasars at 
each location. The expected increase in bias towards the Galactic plane is very evident, 
but also several features related to the different surveys contributing to the sample.
These features probably reflect variations in the magnitude completeness, made visible
through the steep increase in estimated bias towards $G=21.0$.

\begin{table*}
\caption{Coefficients for the function $Z_5(G,\,\nu_\text{eff},\,\beta)$ fitted to the
quasar parallaxes.  
\label{tab:qsz0}}
\footnotesize\setlength{\tabcolsep}{4pt}
\begin{tabular}{crrrrrrrrrrrrrrr}
\hline\hline
\noalign{\smallskip}
 $G$ & \multicolumn{1}{c}{$q_{00}$} & \multicolumn{1}{c}{$q_{01}$} & \multicolumn{1}{c}{$q_{02}$} & \multicolumn{1}{c}{$q_{10}$} & \multicolumn{1}{c}{$q_{11}$} \\
\noalign{\smallskip}\hline\noalign{\smallskip}
13.1--15.9 & $-30.90\pm   4.51$ & $ +8.50\pm   6.54$ & $ -2.44\pm   4.35$ & $ -15.4\pm   31.3$ & $ +31.1\pm   45.3$ \\
16.1--17.5 & $-27.04\pm   1.73$ & $ -0.76\pm   2.66$ & $ -1.63\pm   1.77$ & $ -17.2\pm   12.0$ & $ +40.4\pm   18.4$ \\
19.0 & $-16.39\pm   1.54$ & $ +3.82\pm   2.38$ & $ -3.93\pm   1.79$ & $ -15.1\pm   11.7$ & $ +14.6\pm   17.9$ \\
20.0 & $-10.21\pm   1.91$ & $ -4.05\pm   2.92$ & $ -9.91\pm   2.64$ & $ -12.9\pm   16.2$ & $ +97.0\pm   25.0$ \\
21.0 & $-10.58\pm   5.16$ & $-15.57\pm   7.64$ & $-26.32\pm   8.83$ & $ -37.0\pm   47.6$ & $+124.3\pm   71.8$ \\
\noalign{\smallskip}
\hline
\end{tabular}
\tablefoot{The table gives $q_{jk}(G)$ at the values of $G$ in the first column. For other values of $G$, 
linear interpolation should be used. Units are: $\mu$as (for $q_{0k}$) and $\mu$as~$\mu$m ($q_{1k}$).} 
\end{table*}

\begin{table*}
\caption{Coefficients for the extended function $Z_5(G,\,\nu_\text{eff},\,\beta,\, X)$ fitted to the
quasar parallaxes. The function $Z_5$ corrected for contamination bias is obtained by setting $r_0=r_1=0$.
\label{tab:qsz1}}
\footnotesize\setlength{\tabcolsep}{4pt}
\begin{tabular}{crrrrrrrrrrrrrrr}
\hline\hline
\noalign{\smallskip}
 $G$ & \multicolumn{1}{c}{$q_{00}$} & \multicolumn{1}{c}{$q_{01}$} & \multicolumn{1}{c}{$q_{02}$} & \multicolumn{1}{c}{$q_{10}$} & \multicolumn{1}{c}{$q_{11}$} & \multicolumn{1}{c}{$r_{0}$} & \multicolumn{1}{c}{$r_{1}$} \\
\noalign{\smallskip}\hline\noalign{\smallskip}
13.1--15.9 & $-30.90\pm   4.51$ & $ +8.50\pm   6.54$ & $ -2.44\pm   4.35$ & $ -15.4\pm   31.3$ & $ +31.1\pm   45.3$ & $-$ & $-$ \\
16.1--17.5 & $-27.04\pm   1.73$ & $ -0.76\pm   2.66$ & $ -1.63\pm   1.77$ & $ -17.2\pm   12.0$ & $ +40.4\pm   18.4$ & $-$ & $-$ \\
19.0 & $-16.39\pm   1.54$ & $ +3.82\pm   2.38$ & $ -3.94\pm   1.79$ & $ -15.1\pm   11.7$ & $ +14.6\pm   17.9$ & $-$ & $-$ \\
20.0 & $-10.57\pm   1.97$ & $ -4.04\pm   2.92$ & $-10.78\pm   2.82$ & $ -11.2\pm   16.3$ & $ +97.4\pm   25.0$ & $ -1.66\pm   8.61$ & $+10.82\pm   9.89$ \\
21.0 & $-14.62\pm   5.33$ & $-15.58\pm   7.64$ & $-18.43\pm   9.43$ & $ -29.0\pm   47.8$ & $+125.4\pm   71.8$ & $+111.02\pm  31.48$ & $-31.38\pm  35.50$ \\
\noalign{\smallskip}
\hline
\end{tabular}
\tablefoot{The table gives $q_{jk}(G)$ and $r_k(G)$ at the values of $G$ in the first column. For other values of $G$, 
linear interpolation should be used. A dash (--) indicates that the coefficient should be ignored (taken as zero). 
Units are: $\mu$as (for $q_{0k}$), $\mu$as~$\mu$m (for $q_{1k})$, and $\mu$as~dex$^{-1}$
(for $r_k$).
.} 
\end{table*}

\begin{figure*}
\centering
  \includegraphics[width=\hsize]{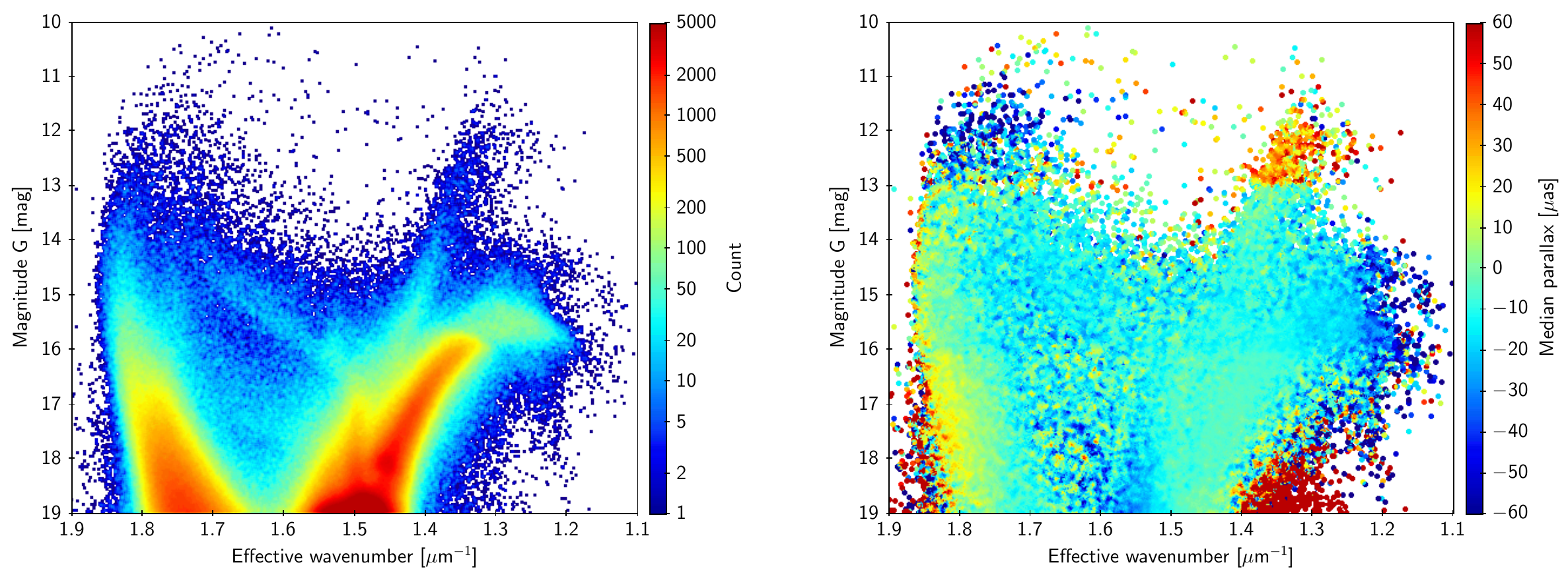}
    \caption{Colour-magnitude diagrams of the LMC sample. \textit{Left:} colour-coded by 
    the number of sources at a given point in the diagram. \textit{Right:} colour-coded by 
    the median parallax in EDR3.}
    \label{figA02ab}
\end{figure*}

Our best estimate of $Z_5(G,\,\nu_\text{eff},\,\beta)$ from the quasar sample is therefore 
given by the coefficients $q_{jk}$ in Table~\ref{tab:qsz1}. The coefficients $r_0$ and $r_1$
must be ignored, as they represent the contamination bias which should not be included in $Z_5$.
 
Several interesting observations can be made concerning the results in Table~\ref{tab:qsz1}.
The coefficient $q_{00}$ represents the mean quasar parallax at the given magnitudes, 
averaged over the celestial sphere and reduced to the reference wavenumber,
$\nu_\text{eff}=1.48~\mu$m$^{-1}$. As described in Appendix~\ref{sec:num}, linear
interpolation between the tabulated values should be used for other magnitudes. 
Its values agree rather well with the mean relation displayed in Fig.~\ref{figA10}\textit{a}. 
Of the remaining coefficients, the most important ones (in terms of how much they
lessen the chi-square of the fit) are $q_{02}$ and $q_{11}$. $q_{02}$ describes a
quadratic dependence on $\sin\beta$; the consistently negative sign means that the
parallax bias is more negative towards the ecliptic poles, and this effect is strongest at the
faint end. The interaction coefficient $q_{11}$ is consistently positive, and increasing with 
magnitude; in terms of the chi-square it is much more important than the corresponding 
simple coefficients $q_{01}$ and $q_{10}$. At the median quasar colour, 
$\nu_\text{eff}=1.59~\mu$m$^{-1}$, $q_{11}$ describes a clear north--south asymmetry 
of the parallaxes, which is what is seen in Fig.~\ref{figA10}\textit{c}. At the reference wavenumber 
$1.48~\mu$m$^{-1}$ the asymmetry, given by $q_{01}$, is smaller and less consistent. 
The coefficients $q_{10}$ represent the colour gradient averaged over the celestial sphere.
They are all consistent with a mean value of $-15~\mu$as~$\mu$m, which is much 
smaller than the slope $\simeq -55~\mu$as~$\mu$m indicated by Fig.~\ref{figA10}\textit{b}.
This apparent contradiction is explained by a correlation between magnitude
and colour in the quasar sample: the faint quasars are, on average, redder than the
brighter ones; together with the overall trend in Fig.~\ref{figA10}\textit{a} this creates a stronger 
variation with colour in the sample as a whole than is present at a fixed magnitude.
This, as well as the strong variation of $q_{02}$ and $q_{11}$ with magnitude, illustrates
the many complex dependencies in the data and the difficulty to determine a unique 
function $Z_5$ based on a limited sample with intrinsic correlations. It also shows the 
danger in using simple plots of the quasar parallaxes versus a single quantity such as 
colour for inferences on the parallax bias.

While Table~\ref{tab:qsz1} (ignoring $r_0$ and $r_1$) in principle defines $Z_5$ 
for any combination of arguments
$G$, $\nu_\text{eff}$, and $\beta$, it is in practice only valid in the subspace of the 
arguments that is well populated by the quasars. Most importantly, this does not include
sources that are brighter than $G\simeq 14$, redder than $\nu_\text{eff}\simeq 1.48$, or
bluer than $\nu_\text{eff}\simeq 1.72$. In order to extend $Z_5$ in these directions, we resort
to differential methods using physical pairs and sources in the LMC.

\subsection{Large Magellanic Cloud (LMC)}
\label{sec:LMC}

The distance modulus of the LMC, $(m-M)_0 = 18.49\pm 0.09$~mag 
\citepads{2014AJ....147..122D}, corresponds to a parallax in the range 
19.2 to 20.9~$\mu$as. The depth and inclination of the system means that
the parallaxes of individual stars have a true dispersion (and gradient) of the order
of one $\mu$as. For the present analysis we do not assume any specific mean
distance to the LMC, only that the selected member sources have the same 
(but unknown) parallax, independent of their colours and magnitudes. 
The LMC data can therefore be used to map the bias function 
$Z_5(G,\nu_\text{eff},\beta)$ at the position of the LMC, $\beta\simeq -85^\circ$, 
up to an unknown additive constant.

The selection of sources in the LMC area for the analysis in this section is described 
in Appendix~\ref{app:LMC}. The sample consists of more than 2~million sources from 
EDR3, brighter than $G=19$ and located within a $5^\circ$ radius of the LMC centre.
As discussed in the appendix, it is believed to be reasonably clean at least down to 
$G\simeq 18$. A colour-magnitude diagram (CMD) of the sample is shown in 
the left panel of Fig.~\ref{figA02ab}.

In the right panel of Fig.~\ref{figA02ab} the same CMD is colour-coded by the median 
parallax at each point of the diagram. The predominantly greenish colour shows that the 
overall median parallax is close to zero, roughly consistent with the expected true parallax 
of 20~$\mu$as and a global parallax bias around $-20~\mu$as. The dark red patch at 
$G>18$, $\nu_\text{eff}<1.4$ is the only part of the diagram dominated by foreground stars.
Several systematic deviations from the overall median are clearly visible, and cannot 
be attributed to foreground stars. These include the rather sharp divisions at
$G\simeq 13.0$ and $G\simeq 16.0$, coinciding with the magnitude boundaries of 
window classes WC0b, WC1, and WC2 (Sect.~\ref{sec:wcgates} and Fig.~\ref{fig:gClass}); 
a strong difference (or gradient) in the bias for $G<13.0$ between the main-sequence 
(at $\nu_\text{eff}\gtrsim 1.7$) and the red
supergiants (at $\nu_\text{eff}\lesssim 1.4$); an up-turn of the bias for the bluest
stars, which is stronger for $G>16.0$ than for the brighter stars; and a down-turn of the
bias for the reddest stars, at least for $G\simeq 14$ to 16. For the faintest stars there
appears to be a depression of the bias at intermediate colours ($\nu_\text{eff}\simeq 1.55$).
The cause of this depression is not known. 
  
\begin{figure*}
\centering
  \includegraphics[height=190mm]{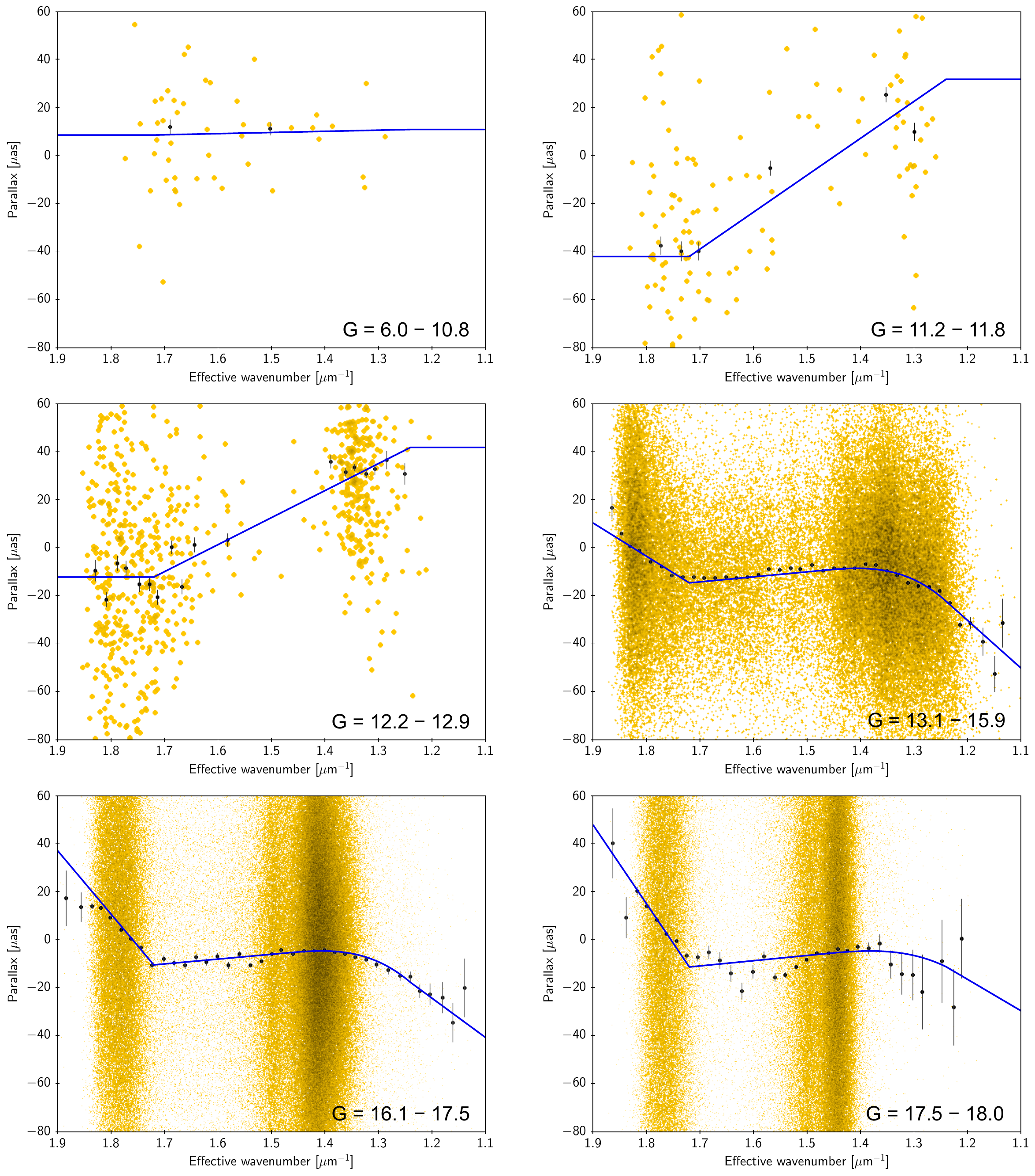}   
    \caption{Median parallaxes for the LMC sample in Fig.~\ref{figA02ab} plotted against 
    $\nu_\text{eff}$ in six magnitude intervals, as indicated in the diagrams.
	The orange points show the parallaxes of individual sources and give an impression of
	the coverage in $\nu_\text{eff}$ and scatter in parallax. The black dots, with error
	bars, show the median parallax, and its uncertainty, in bins of $\nu_\text{eff}$ with at least
	20 sources per bin. The solid blue curves show the fitted combination of basis functions 
	in Eq.~(\ref{eq:lmc1}) evaluated for a representative magnitude in the interval (respectively
	$G=10.0$, 11.5, 12.5, 15.0, 17.0, and 18.0.)}
    \label{figA03}
\end{figure*}

The systematic variation of parallax with colour and magnitude 
is further explored in Fig.~\ref{figA03}, 
where each panel displays a different magnitude interval. The black dots show the 
median parallax binned by $\nu_\text{eff}$.  As seen in the three bottom panels, 
the up-turn of parallax values for the bluest stars sets in abruptly around 
$\nu_\text{eff}=1.72$~$\mu$m$^{-1}$, which is clearly related to the restriction in the 
range of wavenumbers to $1.24\le\nu_\text{eff}\le 1.72~\mu$m$^{-1}$ for the LSF/PSF
calibration, as discussed in Sect.~\ref{sec:nueff}. At the other (red) end of the interval no 
clear break is seen, although rather few stars in the LMC are redder than 1.24~$\mu$m$^{-1}$. 
Between 1.24 and 1.72~$\mu$m$^{-1}$ the relation is approximately linear in the left 
(bluer) half of the interval, but clearly non-linear in the right (redder) half, at least for 
$G>13$.

\begin{table*}
\caption{Function $\varpi_\text{EDR3}(G,\,\nu_\text{eff})$ fitted to the LMC sample.  
\label{tab:lmc}}
\footnotesize\setlength{\tabcolsep}{4pt}
\begin{tabular}{cccccc}
\hline\hline
\noalign{\smallskip}
 $G$ & \multicolumn{1}{c}{$p_{0}$} & \multicolumn{1}{c}{$p_{1}$} & \multicolumn{1}{c}{$p_{2}$} & \multicolumn{1}{c}{$p_{3}$} & \multicolumn{1}{c}{$p_{4}$} \\
\noalign{\smallskip}\hline\noalign{\smallskip}
 6.0--10.8 & $ +9.76\pm   2.83$ & $  -5.3\pm   17.3$ & $-$ & $-$ & $-$ \\
11.2--11.8 & $ -4.35\pm   1.45$ & $-147.7\pm    8.1$ & $-$ & $-$ & $-$ \\
12.2 & $ +9.42\pm   1.23$ & $-147.6\pm    6.8$ & $-$ & $-$ & $-$ \\
12.9 & $+18.14\pm   1.03$ & $ -85.1\pm    5.9$ & $-$ & $-$ & $-$ \\
13.1 & $ -5.08\pm   0.47$ & $ -24.7\pm    3.0$ & $ -1450\pm    192$ & $+103.2\pm  141.7$ & $+101.3\pm   10.8$ \\
15.9 & $-11.74\pm   0.22$ & $ -19.1\pm    1.9$ & $ -1197\pm     70$ & $+244.2\pm   48.0$ & $+156.2\pm    7.5$ \\
16.1 & $ -6.18\pm   0.23$ & $ -15.6\pm    2.5$ & $ -1409\pm    136$ & $+103.8\pm   82.7$ & $+171.9\pm   10.3$ \\
17.5 & $ -5.31\pm   0.20$ & $ -22.5\pm    2.5$ & $ -1122\pm    325$ & $+196.6\pm  231.7$ & $+315.9\pm   11.3$ \\
19.0 & $ -7.61\pm   1.24$ & $ -20.3\pm   15.3$ & $-$ & $-$ & $+354.3\pm   76.6$ \\
\noalign{\smallskip}
\hline
\end{tabular}
\tablefoot{The table gives $p_{j}(G)$ in Eq.~(\ref{eq:lmc1}) at the values of $G$ in the first column. 
For other values of $G$, linear interpolation should be used.
A dash (--) indicates that the coefficient should be ignored (taken as zero).
Units are: $\mu$as (for $p_0$), $\mu$as~$\mu$m ($p_1$, $p_3$, and $p_4$), and 
$\mu$as~$\mu$m$^2$ ($p_2$).}
\end{table*}

These variations can be described by the general model in Appendix~\ref{sec:num}, although
several simplifications are needed for a well-determined fit. Most importantly, because the
LMC sample only covers a small area near the south ecliptic pole, the dependence of
parallax on $\beta$ cannot be determined; the resulting fit is valid at the mean position of 
the LMC, $\beta\simeq -85^\circ$ or $\sin\beta\simeq -0.996$, but not necessarily at other 
locations. Modelled on Eqs.~(\ref{eq:Z1}) and (\ref{eq:Z2}), but using only the basis functions
in $G$ (Eq.~\ref{eq:g}) and $\nu_\text{eff}$ (Eq.~\ref{eq:c}), the function fitted to the
EDR3 parallaxes of the LMC sample is
\begin{equation}\label{eq:lmc1}
\varpi_\text{EDR3}(G,\,\nu_\text{eff}) = \sum_{j=0}^{4} p_j(G) \, c_j(\nu_\text{eff})\, , 
\end{equation}
where
\begin{equation}\label{eq:lmc2}
p_j(G) = \sum_{i=0}^{10} y_{i\!j} g_i(G)\,  
\end{equation}
are piecewise linear functions in $G$, and $y_{i\!j}$ are free parameters of the model,
estimated by a robust weighted least-squares procedure.%
\footnote{Here, and elsewhere in the paper, whenever (weighted) least-squares estimation 
is used, it is made robust against outliers by iteratively removing points that deviate by more 
than four times a robust estimate of the (weighted) RMS residual among all the data points.} 
Owing to the small number 
of sources at the bright end and their limited coverage in $\nu_\text{eff}$, the functions 
$p_j(G)$ are forced to be constant in each of the magnitude intervals 6.0--10.8 and 
11.2--11.8 (using the device described in footnote~\ref{fn2}), and the parameters with 
$j\ge 2$ are set to zero for $i\le 6$. At the faint end, the fit is limited to sources brighter 
than $G=18$ in order to minimise contamination effects, and the basis functions $g_i(G)$
for $i=11$ and 12, which lack support for $G<18$ (see Fig.~\ref{fig:basisG}), are omitted 
in Eq.~(\ref{eq:lmc2}). The resulting fit is given in Table~\ref{tab:lmc}, where
coefficients assumed to be zero are indicated by a dash (--). The fitted function, 
evaluated at representative magnitudes, is shown by the blue curves in Fig.~\ref{figA03}.

In Table~\ref{tab:lmc} the coefficients $p_0$ give, at the different magnitudes, the mean parallax 
of the LMC sample reduced to the reference wavenumber $\nu_\text{eff}=1.48~\mu$m$^{-1}$. 
As they refer to the location of the LMC it is not useful to compare them with the coefficients 
$q_{00}$ from the quasars (Table~\ref{tab:qsz1}), which are averages over the celestial sphere.
Indeed, as we do not want the present analysis to depend in any way on an assumed distance
modulus to the LMC, the fitted coefficients $p_0$ are not further used in the determination of $Z_5$.
The remaining coefficients $p_1$ through $p_4$ map the variation of the parallax bias
at the LMC location as a function of $\nu_\text{eff}$. For WC0 ($G\le 12.9$) only the mean
gradient in wavenumber ($p_2$) is determined, and exhibits very significant variations with 
magnitude; the major breaks at $G\simeq 11$ and 13 are clearly visible in Fig.~\ref{figA02ab} (right).
For WC1 and WC2 ($G\ge 13.1$), the most striking feature is the relative constancy of 
$p_1$, $p_2$, and $p_3$ with magnitude: the tabulated values all agree, to within 
$\pm 2\sigma$, with their weighted mean values, which are
$p_1=-20.0\pm 1.2~\mu$as~$\mu$m, 
$p_2=-1257\pm 58~\mu$as~$\mu$m$^3$, and 
$p_3=+200\pm 39~\mu$as~$\mu$m. On the other hand, the coefficients $p_4$, describing
the colour gradient in the blue end ($\nu_\text{eff}>1.72~\mu$m$^{-1}$), show a very clear
progression with magnitude.

\subsection{Combined fit using quasars and LMC sources (G~>\,13)}
\label{sec:qsolmc}

The quasar sample contains few sources redder than  
$\nu_\text{eff}\simeq 1.44$~$\mu$m$^{-1}$ or bluer than 1.69~$\mu$m$^{-1}$,
and therefore cannot be used to estimate $q_{jk}$ for $j=2$, 3, and 4. The LMC 
sample, on the other hand, gives a good determination of $p_j$ for $j=1\dots 4$
in the magnitude range 13 to 18, but only for the specific location of the LMC. In this
section we attempt to combine the two datasets in order to extend the model to
the full range of colours for $G>13$. This is not possible without certain additional 
assumptions, detailed below; however, the distance to the LMC is still left as a free
parameter.
  
From Eq.~(\ref{eq:b}), using $\sin\beta_\text{LMC}\simeq -0.996$, it can be seen 
that the coefficients in a combined model must satisfy
\begin{equation}\label{eq:conx}
p_j = q_{j,0} - 0.996\, q_{j,1} + 0.659\, q_{j,2}\, ,\quad j=1\dots 4\, .
\end{equation}
The case $j=1$ is the only one for which all the coefficients $p_j$ and $q_{jk}$ are
available in Tables~\ref{tab:qsz1} and \ref{tab:lmc}, and for which a direct comparison
is therefore possible; this is shown in Table~\ref{tab:A4} for $G\ge 13.1$. 
Although the agreement between $\smash{p_1^\text{LMC}}$ and $\smash{p_1^\text{QSO}}$ is not
impressive, the differences are roughly compatible with the uncertainties. 
The main conclusion from 
the comparison is that the LMC data are vastly superior to the quasars for estimating the 
gradient of the bias with colour at the location of the LMC. 

\begin{table}
\caption{Comparison of the parallax gradient in colour, $p_1$, at the position of the LMC, 
as estimated from LMC data, quasars (QSO), and physical pairs (PP).
\label{tab:A4}}
\small
\begin{tabular}{lccc}
\hline\hline
\noalign{\smallskip}
$G$ & $p_1^\text{LMC}$ & $p_1^\text{QSO}$ & $p_1^\text{PP}$ \\
\noalign{\smallskip}
\hline
\noalign{\smallskip}
10.8 & $-5.3\pm 17.3$ & & $+23.3$\\
11.2 & $-147.7\pm 8.1$ & & $-101.0$\\
11.8 & $-147.7\pm 8.1$ & & $-163.1$\\
12.2 & $-147.6\pm 6.8$ & & $-143.8$\\
12.9 & $-85.1\pm 5.9$ & & $-93.4$\\
13.1 & $-24.7\pm 3.0$ & $-46.5\pm 63.5$ & $-19.6$ \\
15.9 & $-19.1\pm 1.9$ & $-46.5\pm 63.5$ \\
16.1 & $-15.6\pm 2.5$ & $-57.5\pm 22.7$ \\
17.5 & $-22.5\pm 2.5$ & $-57.5\pm 22.7$ \\
19.0 & $-20.3\pm 15.3$ & $-29.7\pm 22.1$ \\
\noalign{\smallskip}
\hline
\end{tabular}
\tablefoot{$p_1^\text{LMC}$ is taken from Table~\ref{tab:lmc}. 
$p_1^\text{QSO}=q_{10}-0.996\,q_{11}+0.659\,q_{12}$ with coefficients from Table~\ref{tab:qsz1}.
$\smash{p_1^\text{PP}=q_{10}-0.996\,q_{11}}$ with coefficients from Table~\ref{tab:pairs10_14}.
Values are expressed in $\mu$as~$\mu$m and refer to the mean gradient in the interval 
$\nu_\text{eff}=1.48$ to 1.72~$\mu$m$^{-1}$. Where applicable, uncertainties are $\pm 1\sigma$
from the covariance matrix of the respective solution.}
\end{table}

\begin{table*}
\caption{Coefficients for the extended function $Z_5(G,\,\nu_\text{eff},\,\beta,\, X)$ as 
obtained in a combined fit to the quasar and LMC samples. 
The function $Z_5$ corrected for contamination bias is obtained by setting $r_0=r_1=0$.
\label{tab:qsolmcNoJoin}}
\tiny\setlength{\tabcolsep}{3pt}
\begin{tabular}{ccccccccccccccc}
\hline\hline
\noalign{\smallskip}
 $G$ & \multicolumn{1}{c}{$q_{00}$} & \multicolumn{1}{c}{$q_{01}$} & \multicolumn{1}{c}{$q_{02}$} & \multicolumn{1}{c}{$q_{11}$} & \multicolumn{1}{c}{$q_{20}$} & \multicolumn{1}{c}{$q_{30}$} & \multicolumn{1}{c}{$q_{40}$} & \multicolumn{1}{c}{$r_{0}$} & \multicolumn{1}{c}{$r_{1}$} \\
\noalign{\smallskip}\hline\noalign{\smallskip}
13.1 & $-23.69\pm   5.84$ & $ +7.27\pm   5.16$ & $ +5.54\pm  12.11$ & $ +26.5\pm    2.8$ & $ -1475\pm    179$ & $+107.9\pm  132.2$ & $+104.3\pm   10.2$ & $-$ & $-$ \\
15.9 & $-38.33\pm   2.47$ & $ +5.61\pm   2.81$ & $+15.42\pm   5.35$ & $ +18.7\pm    1.8$ & $ -1189\pm     65$ & $+243.8\pm   44.5$ & $+155.2\pm    7.0$ & $-$ & $-$ \\
16.1 & $-31.05\pm   1.78$ & $ +2.83\pm   2.09$ & $ +8.59\pm   3.95$ & $ +15.5\pm    2.3$ & $ -1404\pm    125$ & $+105.5\pm   76.4$ & $+170.7\pm    9.4$ & $-$ & $-$ \\
17.5 & $-29.18\pm   0.80$ & $ -0.09\pm   1.01$ & $ +2.41\pm   1.98$ & $ +24.5\pm    2.1$ & $ -1165\pm    299$ & $+189.7\pm  214.2$ & $+325.0\pm    9.5$ & $-$ & $-$ \\
19.0 & $-18.40\pm   0.64$ & $ +5.98\pm   1.33$ & $ -6.46\pm   1.73$ & $  +5.5\pm   10.0$ & $-$ & $-$ & $+276.6\pm   55.3$ & $-$ & $-$ \\
20.0 & $-12.65\pm   0.99$ & $ -4.57\pm   3.02$ & $ -7.46\pm   2.97$ & $ +97.9\pm   25.8$ & $-$ & $-$ & $-$ & $+41.22\pm  10.57$ & $ -3.13\pm  10.48$ \\
21.0 & $-18.22\pm   3.38$ & $-15.24\pm   8.01$ & $-18.54\pm  10.50$ & $+128.2\pm   74.0$ & $-$ & $-$ & $-$ & $+73.60\pm  30.84$ & $-16.71\pm  31.30$ \\
\noalign{\smallskip}
\hline
\end{tabular}
\tablefoot{The table gives $q_{jk}(G)$ at the values of $G$ in the first column. For other values of $G$, 
linear interpolation should be used. 
$r_0$ and $r_1$ are the fitted coefficients of the contamination terms. 
A dash (--) indicates that the coefficient should be ignored (taken as zero).
Units are: $\mu$as (for $q_{0k}$), $\mu$as~$\mu$m ($q_{1k}$, $q_{30}$, $q_{40}$), $\mu$as~$\mu$m$^3$ ($q_{20}$), and $\mu$as~dex$^{-1}$ ($r_0$, $r_1$).} 
\end{table*}

If a combined solution using both the quasar and LMC data is attempted, retaining all the
free parameters $q_{jk}$ for $G>13$, the result would be a model that fits both datasets
very well, and is very well-determined in the LMC area, but has very large uncertainties 
in most other locations. This is obviously not very useful. To proceed, we need to constrain
the model. We boldly make the following assumptions for WC1 and WC2 ($G\ge 13.1$):
\begin{enumerate}
\item
The gradient with colour in the mid-blue region ($\nu_\text{eff}=1.48$ to 1.72~$\mu$m$^{-1}$)
is fully described by the interaction term $q_{11}$, that is $q_{10}=q_{12}=0$.
This is consistent with the observation in Sect.~\ref{sec:qso} that $q_{12}$ is generally
insignificant and that $q_{11}$ is far more important than $q_{10}$ for the overall chi-square.
\item
The curvature with colour in the mid-red region ($\nu_\text{eff}=1.24$ to 1.48~$\mu$m$^{-1}$)
is fully described by the non-interaction term $q_{20}$, that is $q_{21}=q_{22}=0$. This 
assumption cannot be tested by means of the quasars, but is partially supported by the 
test in Appendix~\ref{sec:rc}, using red clump stars, showing similar curvatures at 
$\sin\beta=\pm 0.86$.
\item
The colour gradient in the red end  ($\nu_\text{eff}<1.24$~$\mu$m$^{-1}$) is fully described by 
$q_{30}$, that is $q_{31}=q_{32}=0$. 
\item
The colour gradient in the blue end  ($\nu_\text{eff}>1.72$~$\mu$m$^{-1}$) is fully described by 
$q_{40}$, that is $q_{41}=q_{42}=0$. 
\end{enumerate}
Indirect support for the third assumption is provided by the EDR3--DR2 differences $\Delta Z$
analysed in Sect.~\ref{sec:dr2}, for example the bottom panels in Fig.~\ref{fig:dPlx3m2vsNuEff}
for $G=13$--16 and 16--18, where the changes in gradient at $\nu_\text{eff}=1.24~\mu$m$^{-1}$
are not distinctly different in the two hemispheres. As DR2 did not use the clamping of 
$\nu_\text{eff}$ at 1.24 and 1.72~$\mu$m$^{-1}$ described in Sect.~\ref{sec:nueff}, 
it is likely that any abrupt changes in the gradient with colour seen in $\Delta Z$ at these 
wavenumbers are caused by the EDR3 data. The same argument may be advanced in support 
of the fourth assumption, although the evidence in Fig.~\ref{fig:dPlx3m2vsNuEff} for breaks 
at $\nu_\text{eff}>1.72~\mu$m$^{-1}$ in the northern hemisphere is very weak.

With these assumptions the coefficients $q_{jk}$ are effectively determined by the LMC data
via the conditions in Eq.~(\ref{eq:conx}) and it is possible to fit both datasets with a single
model. To account for the offset between $p_0$ and the parallax bias at the LMC location,
one additional unknown must be introduced, representing the true mean parallax of the 
LMC; this is constrained to be independent of $G$. Results are given in Table~\ref{tab:qsolmcNoJoin},
except for the fitted LMC parallax, which is $+22.11 \pm 1.10~\mu$as.

\begin{figure*}
\centering
  \includegraphics[height=195mm]{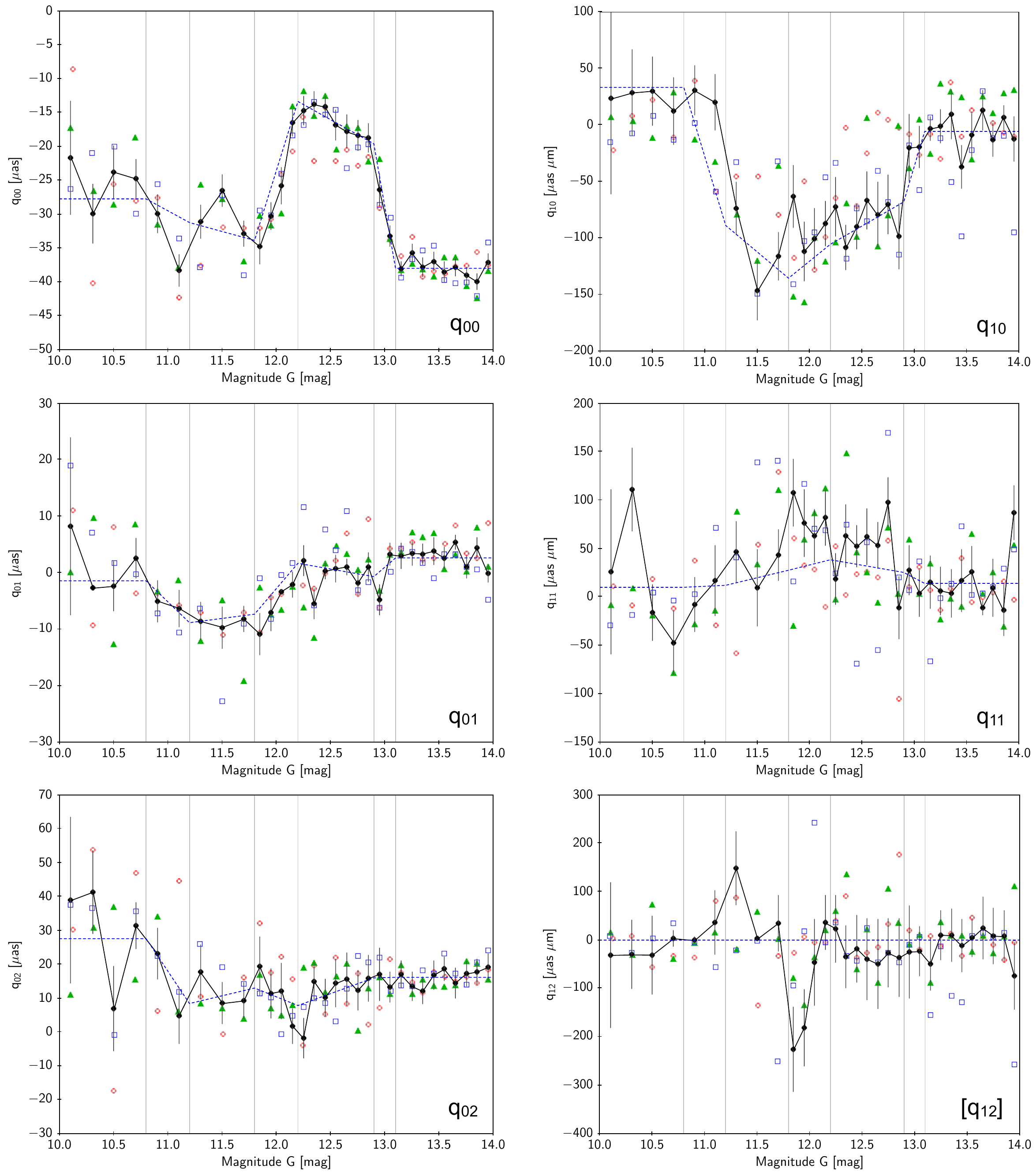}   
    \caption{Coefficients $q_{jk}$ estimated from physical pairs as functions of $G$.
    Results for $q_{12}$ are considered insignificant and set to zero when fitting 
    the other five coefficients.    
    The different symbols represent different selections on $\rho\Delta\mu$: 
    0--0.5~arcsec~mas~yr$^{-1}$ (red circles);
    0.5--1~arcsec~mas~yr$^{-1}$ (green triangles);
    1--2~arcsec~mas~yr$^{-1}$ (blue squares);
    0--2~arcsec~mas~yr$^{-1}$ (filled black circles with lines and error bars).
    The dashed blue line is the global fit in Table~\ref{tab:pairs10_14}.
    The vertical grey lines show the breakpoints for the basis functions in $G$ defined by 
    Eq.~(\ref{eq:g}).}
    \label{figA13}
\end{figure*}

\begin{table}
\caption{Coefficients of $Z_5(G,\,\nu_\text{eff},\,\beta)$ as estimated from 
physical pairs for $G>10$.  
\label{tab:pairs10_14}}
\small
\begin{tabular}{crrrrrrr}
\hline\hline
\noalign{\smallskip}
$G$  & $q_{00}$ & $q_{01}$ & $q_{02}$ & $q_{10}$ & $q_{11}$ & $q_{20}$\\ 
\noalign{\smallskip}
\hline
\noalign{\smallskip}
10.0--10.8 &  $-27.75$ &  $-1.43$ &  $+27.56$ &  $+33.3$ & $+10.0$ & $-1257$ \\
11.2 &  $-31.24$ &  $-8.84$ &  $+8.40$ &  $-88.9$ & $+12.1$ & $-1257$ \\
11.8 & $ -33.87$ & $  -7.33$ & $ +12.98$ & $-135.6$ & $ +27.6$ & $-1257$ \\
12.2 & $ -13.37$ & $  +1.68$ & $  +7.82$ & $-105.6$ & $ +38.4$ & $-1257$ \\
12.9 & $ -19.61$ & $  -0.68$ & $ +15.98$ & $ -68.1$ & $ +25.4$ & $-1257$ \\
13.1--$14.0$ & $ -37.99$ & $  +2.63$ & $ +16.14$ & $  -5.7$ & $ +14.0$ & $-1257$ \\
\noalign{\smallskip}
\hline
\end{tabular}
\tablefoot{The functions $q_{jk}(G)$, obtained by linear interpolation in the table, are expressed 
in $\mu$as ($q_{0k}$), $\mu$as~$\mu$m ($q_{1k}$), and $\mu$as~$\mu$m$^3$ ($q_{20}$).
Values for $q_{20}$ are assumed as described in the text.}
\end{table}

\subsection{Physical pairs (G~<\,13)}
\label{sec:pairs}

Having mapped $Z_5(G,\,\nu_\text{eff},\,\beta)$ for $G>13$ by mean of the quasars 
and the LMC, we now turn to the brighter sources. In the LMC area the
gross variation of the parallax bias with colour was mapped in Sect.~\ref{sec:LMC} 
(Table~\ref{tab:lmc}), but this local result must now be extended to the whole sphere. 
To this end we use binaries (here called `physical pairs'), in which it can be assumed 
that the components have similar true parallaxes, although their magnitudes and colours
may be very different. Using the results from previous sections to `anchor' the parallax 
bias of the fainter component among the quasars, it is then possible to estimate the 
bias of the brighter component. Details of the method are given in 
Appendix~\ref{sec:pairsMethod}, which also describes the selection of data used in
the analysis below.

\begin{table*}
\caption{Coefficients of $Z_5(G,\,\nu_\text{eff},\,\beta)$ obtained by joining the results in 
Tables~\ref{tab:qsolmcNoJoin} and \ref{tab:pairs10_14}. 
\label{tab:qsolmcpairs0}}
\small
\begin{tabular}{crrrrrrrr}
\hline\hline
\noalign{\smallskip}
 $G$ & \multicolumn{1}{r}{$q_{00}$} & \multicolumn{1}{r}{$q_{01}$} & \multicolumn{1}{r}{$q_{02}$} & \multicolumn{1}{r}{$q_{10}$} & \multicolumn{1}{r}{$q_{11}$} & \multicolumn{1}{r}{$q_{20}$} & \multicolumn{1}{r}{$q_{30}$} & \multicolumn{1}{r}{$q_{40}$} \\
\noalign{\smallskip}\hline\noalign{\smallskip}
10.0--10.8 &  $-27.75$ &  $-1.43$ &  $+27.56$ &  $+33.3$ & $+10.0$ & $-1257$ & $-$ & $-$ \\
11.2 &  $-31.24$ &  $-8.84$ &  $+8.40$ &  $-88.9$ & $+12.1$ & $-1257$ & $-$ & $-$ \\
11.8 & $ -33.87$ & $  -7.33$ & $ +12.98$ & $-135.6$ & $ +27.6$ & $-1257$ & $-$ & $-$ \\
12.2 & $ -13.37$ & $  +1.68$ & $  +7.82$ & $-105.6$ & $ +38.4$ & $-1257$ & $-$ & $-$ \\
12.9 & $ -19.61$ & $  -0.68$ & $ +15.98$ & $ -68.1$ & $ +25.4$ & $-1257$ & $-$ & $-$ \\
13.1 & $ -37.99$ & $  +2.63$ & $ +16.14$ & $  -5.7$ & $ +14.0$ & $-1257$ & $+107.9$ & $+104.3$ \\
15.9 & $-38.33$ & $ +5.61$ & $+15.42$ & $-$ & $ +18.7$ & $ -1189$ & $+243.8$ & $+155.2$ \\
16.1 & $-31.05$ & $ +2.83$ & $ +8.59$ & $-$ & $ +15.5$ & $ -1404$ & $+105.5$ & $+170.7$ \\
17.5 & $-29.18$ & $ -0.09$ & $ +2.41$ & $-$ & $ +24.5$ & $ -1165$ & $+189.7$ & $+325.0$ \\
19.0 & $-18.40$ & $ +5.98$ & $ -6.46$ & $-$ & $  +5.5$ & $-$ & $-$ & $+276.6$ \\
20.0 & $-12.65$ & $ -4.57$ & $ -7.46$ & $-$ & $ +97.9$ & $-$ & $-$ & $-$ \\
21.0 & $-18.22$ & $-15.24$ & $-18.54$ & $-$ & $+128.2$ & $-$ & $-$ & $-$ \\
\noalign{\smallskip}
\hline
\end{tabular}
\tablefoot{The table gives $q_{jk}(G)$ at the values of $G$ in the first column. For other values of $G$, 
linear interpolation should be used. A dash (--) indicates that the coefficient should be ignored (taken as zero).
Units are: $\mu$as (for $q_{0k}$), $\mu$as~$\mu$m ($q_{1k}$, $q_{30}$, $q_{40}$), and 
$\mu$as~$\mu$m$^3$ ($q_{20}$).}
\end{table*}

\begin{table*}
\caption{Final coefficients of $Z_5(G,\,\nu_\text{eff},\,\beta)$ obtained by joining the results in 
Table~\ref{tab:qsolmcNoJoin} with the analysis in Sect.~\ref{sec:pairs6}. 
\label{tab:final5}}
\small
\begin{tabular}{crrrrrrrr}
\hline\hline
\noalign{\smallskip}
 $G$ & \multicolumn{1}{r}{$q_{00}$} & \multicolumn{1}{r}{$q_{01}$} & \multicolumn{1}{r}{$q_{02}$} & \multicolumn{1}{r}{$q_{10}$} & \multicolumn{1}{r}{$q_{11}$} & \multicolumn{1}{r}{$q_{20}$} & \multicolumn{1}{r}{$q_{30}$} & \multicolumn{1}{r}{$q_{40}$} \\
\noalign{\smallskip}\hline\noalign{\smallskip}
6.0 &  $-26.98$ &  $-9.62$ &  $+27.40$ &  $-25.1$ & $-0.0$ & $-1257$ & $-$ & $-$ \\
10.8 &  $-27.23$ &  $-3.07$ &  $+23.04$ &  $+35.3$ & $+15.7$ & $-1257$ & $-$ & $-$ \\
11.2 &  $-30.33$ &  $-9.23$ &  $+9.08$ &  $-88.4$ & $-11.8$ & $-1257$ & $-$ & $-$ \\
11.8 & $ -33.54$ & $  -10.08$ & $ +13.28$ & $-126.7$ & $ +11.6$ & $-1257$ & $-$ & $-$ \\
12.2 & $ -13.65$ & $ -0.07$ & $  +9.35$ & $-111.4$ & $ +40.6$ & $-1257$ & $-$ & $-$ \\
12.9 & $ -19.53$ & $  -1.64$ & $ +15.86$ & $ -66.8$ & $ +20.6$ & $-1257$ & $-$ & $-$ \\
13.1 & $ -37.99$ & $  +2.63$ & $ +16.14$ & $  -5.7$ & $ +14.0$ & $-1257$ & $+107.9$ & $+104.3$ \\
15.9 & $-38.33$ & $ +5.61$ & $+15.42$ & $-$ & $ +18.7$ & $ -1189$ & $+243.8$ & $+155.2$ \\
16.1 & $-31.05$ & $ +2.83$ & $ +8.59$ & $-$ & $ +15.5$ & $ -1404$ & $+105.5$ & $+170.7$ \\
17.5 & $-29.18$ & $ -0.09$ & $ +2.41$ & $-$ & $ +24.5$ & $ -1165$ & $+189.7$ & $+325.0$ \\
19.0 & $-18.40$ & $ +5.98$ & $ -6.46$ & $-$ & $  +5.5$ & $-$ & $-$ & $+276.6$ \\
20.0 & $-12.65$ & $ -4.57$ & $ -7.46$ & $-$ & $ +97.9$ & $-$ & $-$ & $-$ \\
21.0 & $-18.22$ & $-15.24$ & $-18.54$ & $-$ & $+128.2$ & $-$ & $-$ & $-$ \\
\noalign{\smallskip}
\hline
\end{tabular}
\tablefoot{The table gives $q_{jk}(G)$ at the values of $G$ in the first column. For other values of $G$, 
linear interpolation should be used. A dash (--) indicates that the coefficient should be ignored (taken as zero).
Results are uncertain for $\nu_\text{eff}>1.72~\mu$m$^{-1}$ ($G_\text{BP}-G_\text{RP}\lesssim 0.15$) and
$\nu_\text{eff}<1.24~\mu$m$^{-1}$ ($G_\text{BP}-G_\text{RP}\gtrsim 3.0$).
Units are: $\mu$as (for $q_{0k}$), $\mu$as~$\mu$m ($q_{1k}$, $q_{30}$, $q_{40}$), 
and $\mu$as~$\mu$m$^3$ ($q_{20}$).}
\end{table*}

\subsubsection{Results for $G>$ 10}
\label{sec:pairs10}

The method outlined in Appendix~\ref{sec:pairsMethod} was applied to pairs where the 
magnitude of the bright component ($G_1$) is in the range 10 to 14~mag, and that of 
the faint component ($G_2$) is in the range $\max(G_1,13.1)$ to $G_1+4$~mag. 
The bias-corrected parallax of the faint component was computed as 
$\varpi_2-Z_5(G_2,\,\nu_\text{eff2},\,\beta)$, where $\varpi_2$ and $\nu_\text{eff2}$ 
are the published EDR3 parallax and effective wavenumber of the faint component,
and $Z_5$ is defined by the coefficients in Table~\ref{tab:qsolmcNoJoin}.
The investigated magnitude interval overlaps with Table~\ref{tab:qsolmcNoJoin} for $G=13.1$
to 14, which provides a consistency check and possibly improved estimates of the
parallax bias in an interval that is poorly covered by the quasars.

The LMC data show that significant variations of the bias as a function of (at least)
$G$ and $\nu_\text{eff}$ exist for sources brighter than $G\simeq 13$. Owing to the 
relatively small number of physical pairs available, the analysis cannot be very complex 
but should still include the main known or expected dependencies. A severe limitation is
that the scarcity of bright sources with $\nu_\text{eff}<1.3$ or $>1.6~\mu$m$^{-1}$ 
makes it practically impossible to determine $q_{20}$, $q_{30}$, and $q_{40}$.
From the LMC data we have $q_{20}\simeq -1257~\mu$as~$\mu$m$^3$ for $G>13.1$,
so even though it cannot be usefully estimated from the physical pairs in the
magnitude interval 13.1 to 14, consistency requires that the corresponding term is
subtracted from the parallax of the brighter component before fitting the remaining
parameters. It turns out that this procedure indeed reduces the sum minimised in 
Eq.~(\ref{eq:L1}), albeit not by a significant amount. A similar improvement is obtained
by applying the same a~priori correction for $G<13.1$. We therefore assume that 
$q_{20}\simeq -1257~\mu$as~$\mu$m$^3$ throughout the range $G<14$. 
Concerning $q_{30}$ and $q_{40}$ we assume that they are zero for $G<13.1$.  
Noting that a fit including the six coefficients $q_{jk}$ ($j\le 1$, $k\le 2$) gives 
insignificant results for $q_{12}$ at all magnitudes (lower right panel of Fig.~\ref{figA13}), 
we also assume $q_{12}=0$.

Figure~\ref{figA13} shows results for the remaining five 
coefficients $q_{00}$, $q_{01}$, $q_{02}$, $q_{10}$, and $q_{11}$. 
The interaction of these terms with magnitude is fully mapped by independent fits
in 35 bins of $G_1$ covering the interval $10<G_1<14$. 
All five coefficients show significant variations with $G$, which in most cases can
be related to the boundaries discussed in Sect.~\ref{sec:wcgates}.

The different symbols in Fig.~\ref{figA13} show the results from using different intervals 
in $\rho\Delta\mu$. The filled black circles are for $\rho\Delta\mu<2$~arcsec~mas~yr$^{-1}$, 
which gives the most precise estimates; the error bars are $\pm 1\sigma$ uncertainties 
obtained by bootstrap resampling. The coloured symbols (open circles, triangles, and 
squares) are for non-overlapping intervals in $\rho\Delta\mu$, and are thus statistically
independent,%
\footnote{They are not completely independent, though: systems with more than two 
components may appear with the same bright component in different intervals of 
$\rho\Delta\mu$.}
which gives an additional indication of the uncertainty. The absence of any obvious trend 
in $q_{00}$ with $\rho\Delta\mu$ suggests that contamination bias is negligible.

More reliable estimates of the coefficients are obtained in a simultaneous fit of all the
parameters, using Eq.~(\ref{eq:L1}). Here, $q_{jk}$ are constrained to be piecewise linear 
functions of $G$ with breakpoints at $G=10.8$, 11.2, 11.8, 12.2, 12.9, and 13.1 (see
Appendix~\ref{sec:num}). For numerical stability, we require that $q_{jk}$ are constant 
for $G<10.8$ and $>13.1$. The resulting fit is given in Table~\ref{tab:pairs10_14}
and shown by the blue dashed lines in Fig.~\ref{figA13}. 

The coefficients in the last row of Table~\ref{tab:pairs10_14} are in excellent agreement with 
the joint quasar and LMC results (Table~\ref{tab:qsolmcNoJoin}) at $G=15.9$, while the 
agreement is less good at $G=13.1$ where the coefficients in Table~\ref{tab:qsolmcNoJoin}
are considerably more uncertain. In Table~\ref{tab:qsolmcpairs0} we have joined the two
datasets by adopting the quasar/LMC results for $G\ge 15.9$ and the results from the 
physical pairs for $G\le 13.1$, but taking $q_{30}$ and $q_{40}$ at $G=13.1$ from 
Table~\ref{tab:qsolmcNoJoin} as they could not be determined from the pairs.

\subsubsection{Extending the analysis to $G>$ 6}
\label{sec:pairs6}

The previous analysis of physical pairs could not go brighter than $G=10$ owing to the
restrictions $G_2-G_1<4$~mag and $G_2>13.1$, where the latter condition came from the
necessity to use the bias function from Sect.~\ref{sec:qsolmc} (valid for $G>13.1$) to 
correct the parallax of the faint component. As a consequence, the number of
available pairs was rapidly decreasing towards the bright end, not only because of the general 
scarcity of bright stars but also because a decreasing fraction of them have faint components
in the required range.

Using the coefficients in Table~\ref{tab:pairs10_14} to define a provisional parallax bias for
$G>10$, it is now possible to extend the analysis to pairs as bright as $G_1\simeq 6$ by removing 
the second constraint, that is, by including all pairs with $G_1< G_2<G_1+4$~mag. Not only 
does this extend the analysis to $G>6$, but the results for $G>10$ are also improved by the
many more pairs included. For example, between $G_1=10.0$ and 10.8, twice as many pairs 
become available for the analysis. This also allows us to remove the constraint that the 
coefficients are constant for $G<10$. Naturally, the resulting estimates are different from those in 
Table~\ref{tab:pairs10_14}, with the most important differences seen towards the bright end.
Repeating the analysis, using the improved coefficients for the biases of the faint 
components, results in yet another, slightly different set of coefficients. However, after a
few more iterations, the coefficients were found to be completely stable, and hence internally
consistent with the data for all useful pairs. The end result is shown in Table~\ref{tab:final5}, 
which is our final estimate of the bias function $Z_5(G,\,\nu_\text{eff},\,\beta)$ 
for sources with five-parameter solutions in EDR3. Figure~\ref{figA14} is a visualisation of
this function. The top panel of Fig.~\ref{figA23} shows values of the bias function for a
representative selection of sources, taking into account the actual distribution
of effective wavenumbers at a given magnitude.

\begin{figure}
\centering
  \includegraphics[width=0.9\hsize]{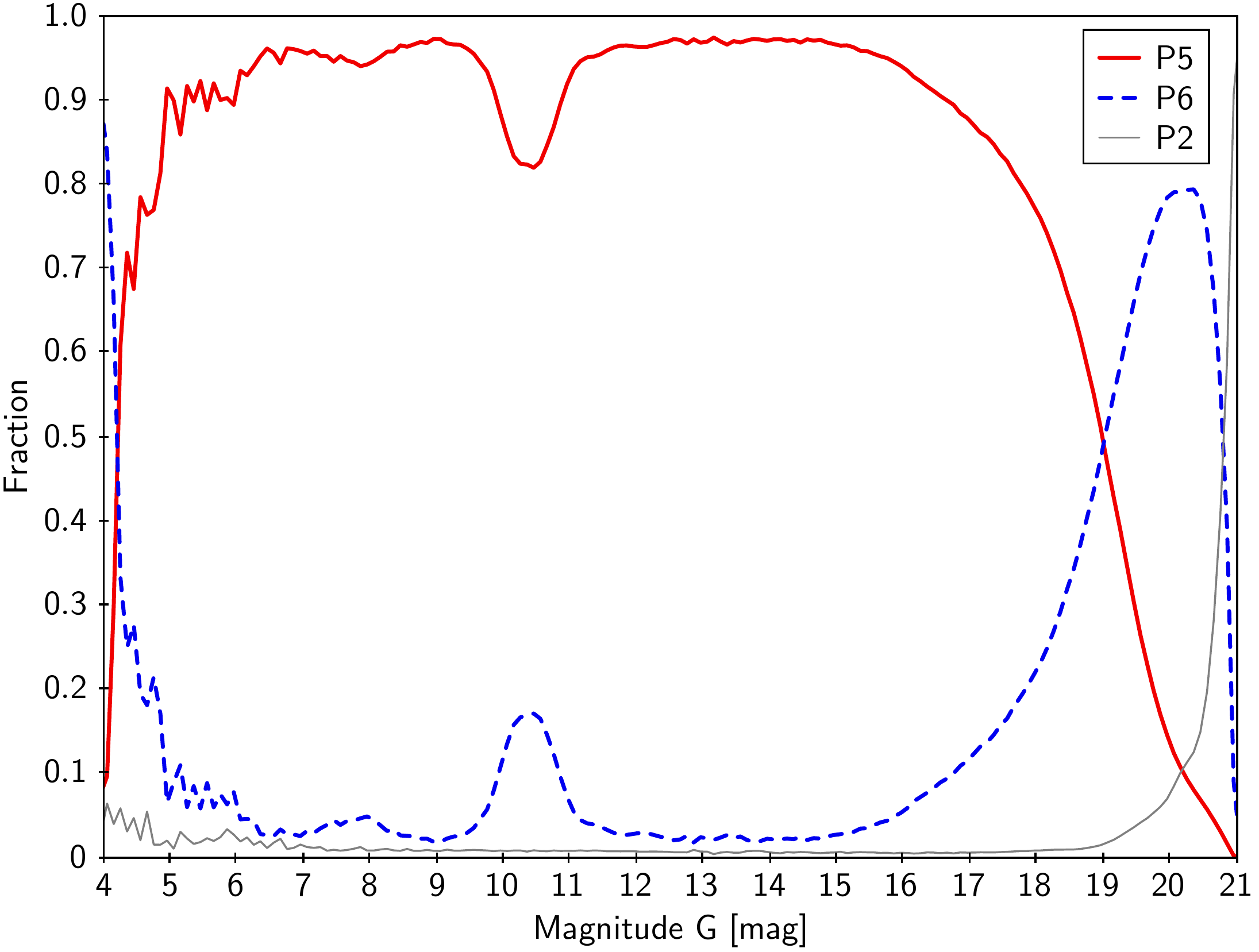}
    \caption{Fraction of sources in \textit{Gaia} EDR3 with different kinds of solutions: 
    five-parameter solutions (P5 = red solid curve), six-parameter solutions (P6 = blue dashed curve), 
    and two-parameter solutions (P2 = thin grey curve). The two-parameter solutions are ignored
    in this paper as they do not have parallaxes.}
    \label{fig56}
\end{figure}

\section{Six-parameter solutions}
\label{sec:p6}

Because of their different treatments in the image parameter determination and astrometroc
solution, the five- and six-parameter solutions have different systematics and it is necessary
to consider their parallax biases separately. In principle the same methodology as was used for 
the five-parameter solutions could be applied to the six-parameter solutions, but in practice
this in not possible owing to the much smaller number of six-parameter solutions at
all magnitudes, except for $G\gtrsim 19$ and in the very bright end (Fig.~\ref{fig56}).
Recalling (Sect.~\ref{sec:nueff}) that six-parameter solutions are used for sources that did 
not have reliable colour information from the BP and RP photometers in \textit{Gaia} DR2, 
we may also note that their observations are more often disturbed by other sources than 
the five-parameter solutions, and therefore generally more problematic. 

\begin{figure}
\centering
  \includegraphics[height=195mm]{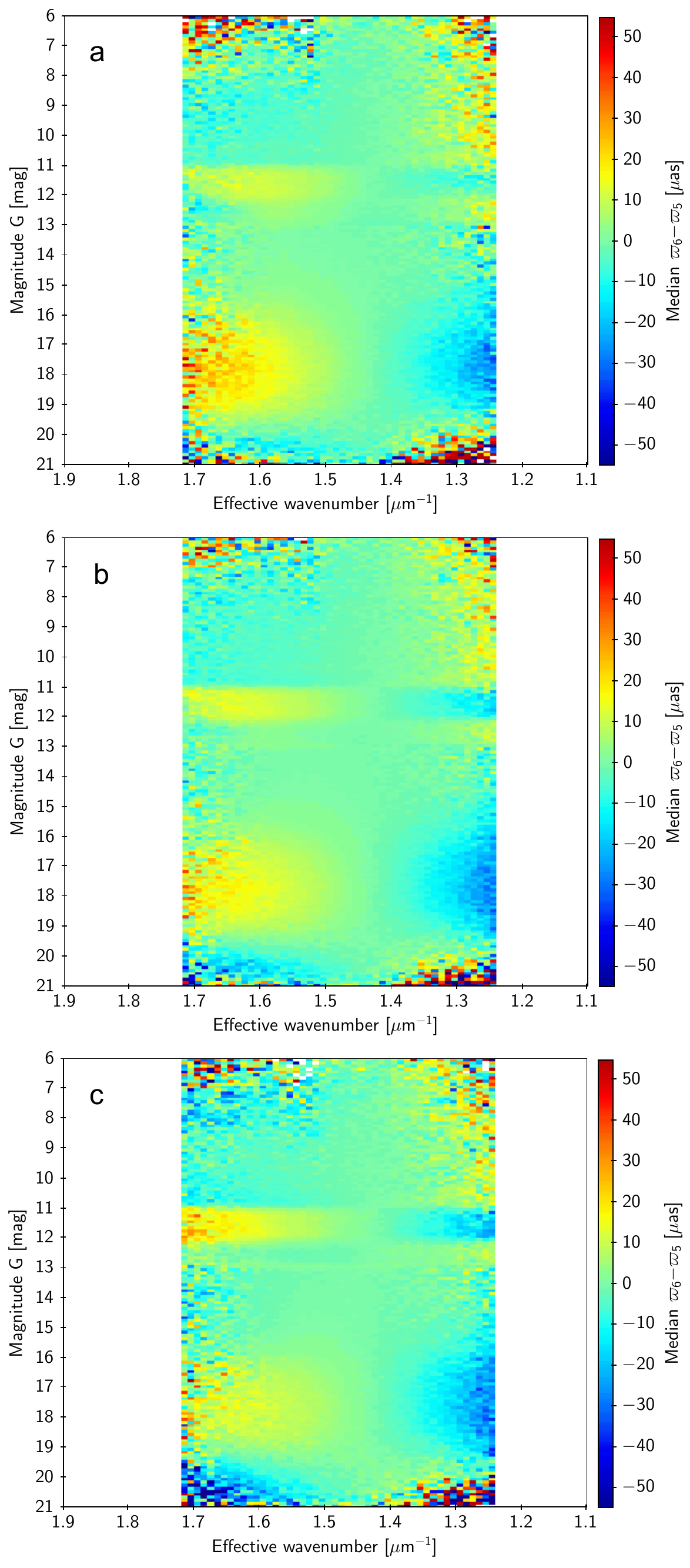}   
    \caption{Median parallax difference between six- and five-parameter solutions as a function
    of $\nu_\text{eff}$ and $G$ for the 8~million sources with both kinds of solutions.
    The panels show selections depending on ecliptic latitude $\beta$:
    southern hemisphere (\textit{a}), all latitudes (\textit{b}), and northern hemisphere (\textit{c}).
    In this sample there are no sources outside of the interval $1.24<\nu_\text{eff}<1.72~\mu$m$^{-1}$.}
    \label{figA17nu}
\end{figure}

To circumvent the scarcity of suitable six-parameter solutions we bootstrap the estimation
of $Z_6$ in the following way on the already determined $Z_5$.
For 8.16~million of the primary sources we have both five- and six-parameter solutions 
(see Sect~\ref{sec:nueff}), and this sample is used here to map the systematic 
differences in parallax between the two kinds
of solution. Figure~\ref{figA17nu} shows the median of $\varpi_6-\varpi_5$ versus $G$ and
$\nu_\text{eff}$, where $\varpi_5$ and $\varpi_6$ are the parallaxes of a given source as
obtained in the five- and six-parameter solutions. The most prominent features in the plots 
are the positive gradient in $\nu_\text{eff}$ for $16\lesssim G\lesssim 19.5$ and 
$11\lesssim G\lesssim 12$, and an opposite gradient for the faintest stars. There are
clear differences between the southern and northern ecliptic hemispheres. The systematics 
of $\varpi_6-\varpi_5$ thus depend in a complex way on at least $G$, $\nu_\text{eff}$, 
and $\beta$. Owing to the restrictions in the selection of primary sources for the
astrometric solution, the colour region outside of the interval 1.24 to 1.72~$\mu$m$^{-1}$
in $\nu_\text{eff}$ cannot be mapped with this sample.
 
\begin{figure}
\centering
  \includegraphics[height=195mm]{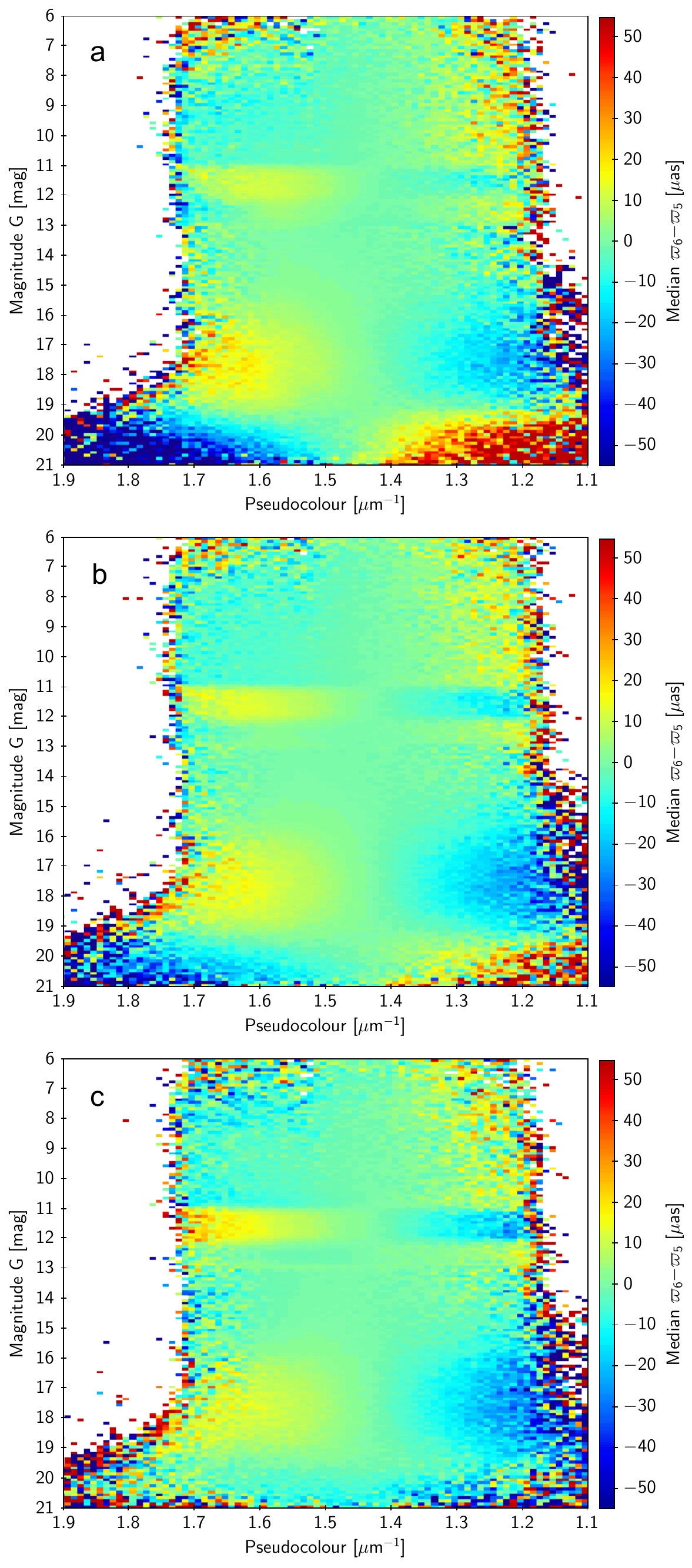}  
    \caption{Same data as in Fig.~\ref{figA17nu} but plotted against the astrometrically 
    determined pseudocolour ($\hat{\nu}_\text{eff}$). Uncertainties in the pseudocolour 
    scatter some points outside of the interval 1.24 to 1.72~$\mu$m$^{-1}$.}
    \label{figA17pc}
\end{figure}

In the \textit{Gaia} Archive, no effective wavenumber derived from photometry is provided 
for sources with six-parameter solutions, and many of them also lack a colour index such as
$G_\text{BP}-G_\text{RP}$. The parallax bias function $Z_6$ must therefore 
be expressed in terms of the pseudocolour $\hat{\nu}_\text{eff}$ instead of $\nu_\text{eff}$.
Figure~\ref{figA17pc} shows the same difference $\varpi_6-\varpi_5$ as in Fig.~\ref{figA17nu},
but plotted versus $\hat{\nu}_\text{eff}$. The main trends are the same, only amplified for the
faintest sources. Uncertainties in $\hat{\nu}_\text{eff}$ scatter some points outside of the
interval 1.24 to 1.72~$\mu$m$^{-1}$ especially in the faint end, where the 
pseudocolour becomes rather uncertain. In the southern hemisphere, the strong gradient versus 
$\hat{\nu}_\text{eff}$ in the faint end is produced by the predominantly negative correlation
between parallax and pseudocolour seen in Fig.~\ref{figA20} (panel \textit{b}).

\begin{figure}
\centering
  \includegraphics[height=195mm]{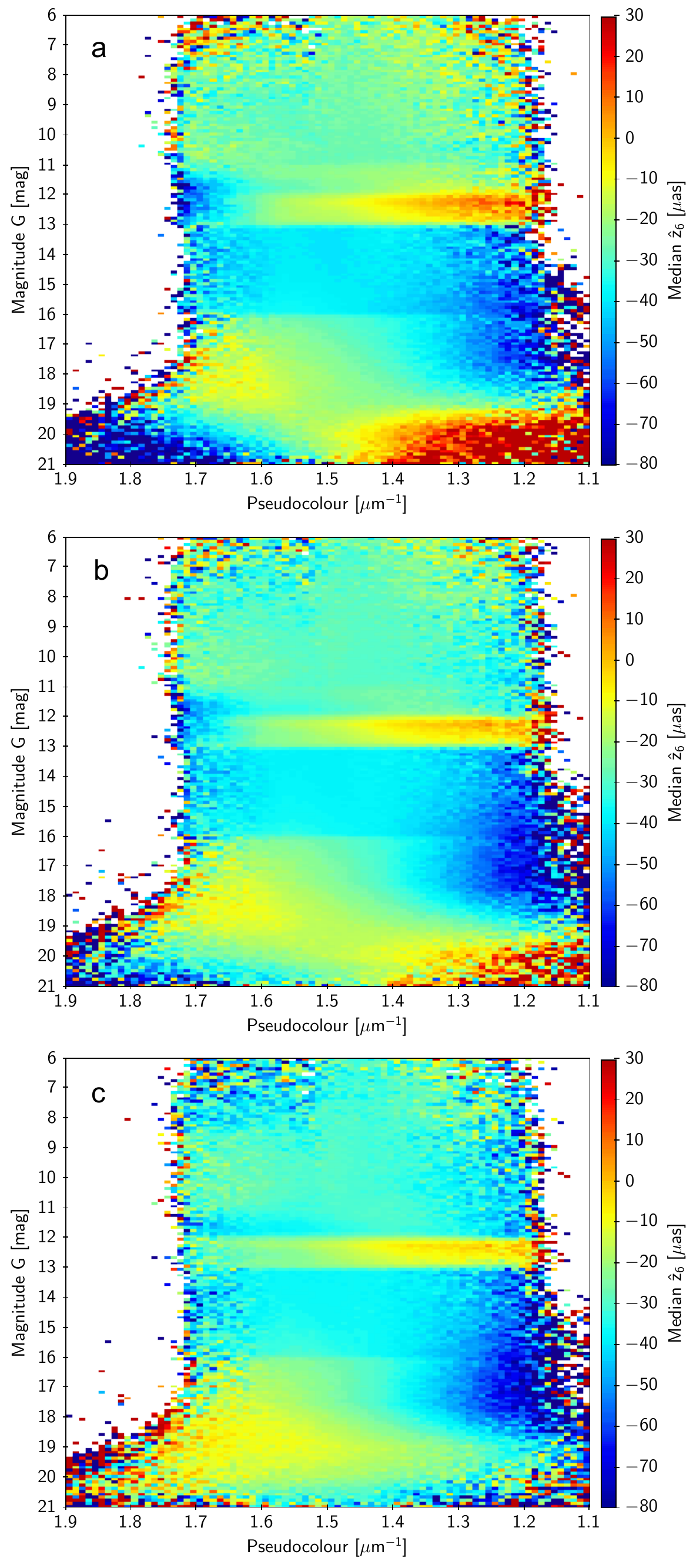}  
    \caption{Median $\hat{z}_6$ as a function
    of $\nu_\text{eff}$ and $G$ for the sample in Figs.~\ref{figA17nu} and \ref{figA17pc}.
    The panels show selections depending on ecliptic latitude:
    southern hemisphere (\textit{a}), all latitudes (\textit{b}), and northern hemisphere (\textit{c}).}
    \label{figA18}
\end{figure}

To estimate the parallax bias function for six-parameter solutions, 
$Z_6(G,\,\hat{\nu}_\text{eff},\,\beta)$, the following procedure was used:
\begin{enumerate}
\item 
For each of the $\sim$8~million primary sources with both kinds of solution, and for which 
both $\nu_\text{eff}$ and $\hat{\nu}_\text{eff}$ are available, 
a corrected five-parameter parallax is calculated as
\begin{equation}\label{eq:plx5corr}
\varpi_5^\text{corr} = \varpi_5 - Z_5(G,\,\nu_\text{eff},\,\beta)\, ,
\end{equation}
using the coefficients in Table~\ref{tab:final5}.
\item 
From this, an estimate of the parallax bias in the six-parameter solutions is obtained as
\begin{equation}\label{eq:plx6zero}
\hat{z}_6=\varpi_6-\varpi_5^\text{corr}\, .
\end{equation}
\item 
Finally, a robust weighted least-squares fit of the general model in Eqs.~(\ref{eq:Z1})
and (\ref{eq:Z2}) to $\hat{z}_6$ is made, with $\hat{\nu}_\text{eff}$ replacing 
$\nu_\text{eff}$ in Eq.~(\ref{eq:c}). In the fit, the data are weighted by the inverse 
variance, calculated as the sum of the formal variances of $\varpi_5$ and $\varpi_6$. 
This overestimates the random errors (because the five- and six-parameter solutions 
are positively correlated), but neglects the uncertainty of the correction $Z_5$.
\end{enumerate}
Figure~\ref{figA18} shows the median values of $\hat{z}_6$ from step~2, plotted
versus $G$ and $\hat{\nu}_\text{eff}$ using the same divisions by ecliptic latitude as in 
Figs.~\ref{figA17nu} and \ref{figA17pc}.
From Fig.~\ref{figA18} it is clear that $Z_6$ cannot be usefully determined for 
$\hat{\nu}_\text{eff}>1.72~\mu$m$^{-1}$, from the available sample of common five- and 
six-parameter solutions, at least not for $G\lesssim 18$ (and even for fainter sources it 
would be very dubious, given the correlation mentioned above). For a similar reason, the results 
in the red end
$\hat{\nu}_\text{eff}<1.24~\mu$m$^{-1}$ are also highly uncertain. Since the boundaries
at 1.24 and 1.72~$\mu$m$^{-1}$ have no special significance in the six-parameter solutions
(except that the chromaticity calibration does not go beyond these limits), it could well be that
an extrapolation of the fit gives reasonable results at the more extreme colours. However, rather
than relying on this we have chosen to assume $q_{3k}=q_{4k}=0$ in the fitted model; effectively
this means that $Z_6$ is `clamped' to its value at 1.24 or 1.72~$\mu$m$^{-1}$ for more
extreme pseudocolours. This has the added benefit of restricting $Z_6$ to a finite interval
even when the pseudocolour is completely wrong. In line with the analysis of $Z_5$ we also
assume $q_{21}=q_{22}=0$. 

\begin{figure}
\centering
  \includegraphics[height=195mm]{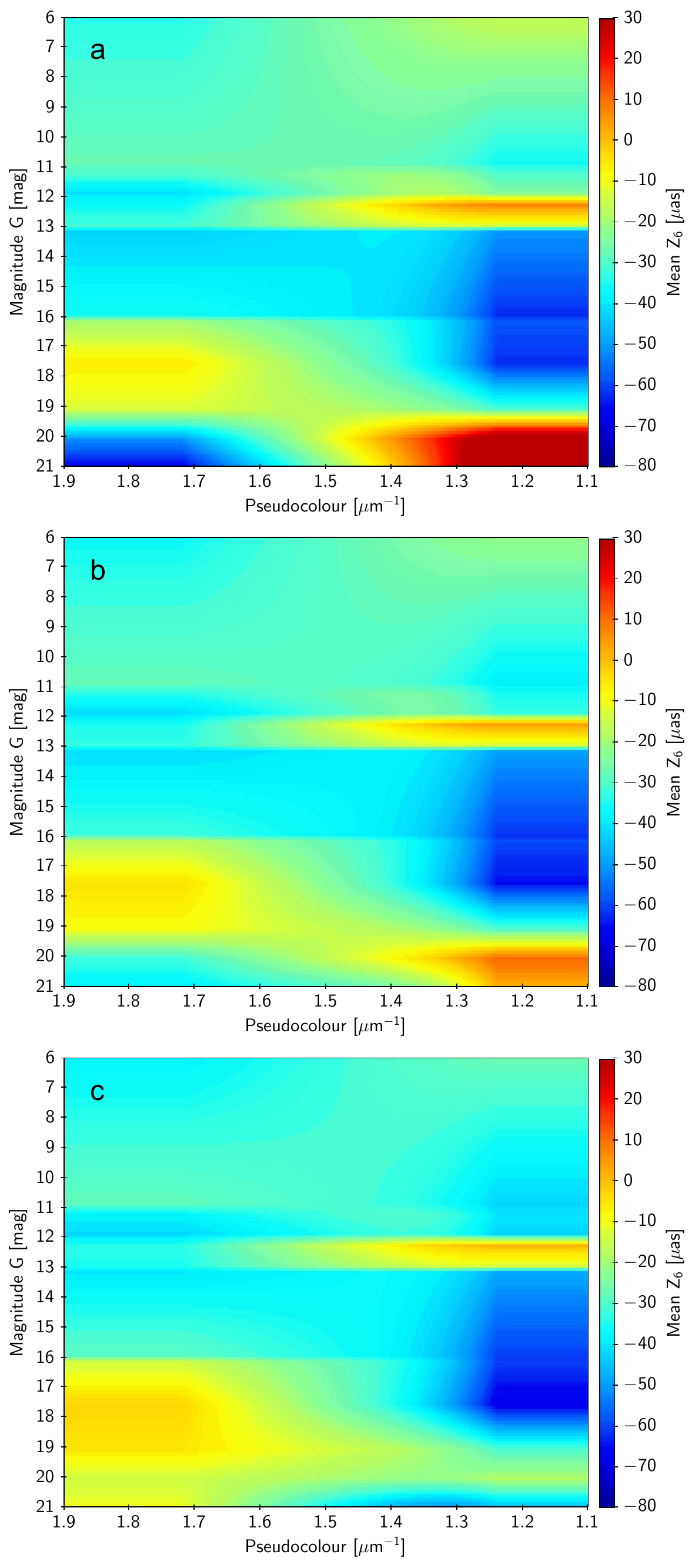}  
    \caption{Parallax bias $Z_6$ according to Table~\ref{tab:final6}. 
    The panels show mean values for the southern hemisphere (\textit{a}), 
    all latitudes (\textit{b}), and northern hemisphere (\textit{c}).}
    \label{figA19mean}
\end{figure}

The resulting coefficients for $Z_6$ are given in Table~\ref{tab:final6}. 
The function $Z_6(G,\,\hat{\nu}_\text{eff},\,\beta)$ is visualised in Figs.~\ref{figA19mean} and
\ref{figA19}. To facilitate comparison with the sample data in Fig.~\ref{figA18}, the data
shown in Fig.~\ref{figA19mean} have been averaged over all latitudes in the 
middle panel, and over each hemisphere in the top and bottom panels.
(To calculate the average over the whole celestial sphere one only needs to 
include the coefficients $q_{j0}$; similarly, the averages in the two hemispheres are obtained 
from $q_{j0}\pm q_{j1}/2$.)

\section{Validation of the bias corrections}
\label{sec:val}

In this section the bias functions $Z_5$ and $Z_6$ defined in Tables~\ref{tab:final5} and \ref{tab:final6}
are applied to EDR3 parallaxes in order to check the validity of the corrections. To some extent the 
same data are used as for deriving the corrections in Sects.~\ref{sec:qso} and \ref{sec:LMC}, in which
case the tests are not a true validation of the functions but rather a consistency check of the procedures
used to derive them. Exceptions are
the tests on the quasar sample with six-parameter solutions and the bright ($G<13$) LMC sample,
neither of which were used in the derivation, and the mean parallax of the LMC, which was a free
parameter in the fit. Additional tests are described in \citet{EDR3-DPACP-126}
and \citet{EDR3-DPACP-113}.

\begin{figure}
\centering
  \includegraphics[height=180mm]{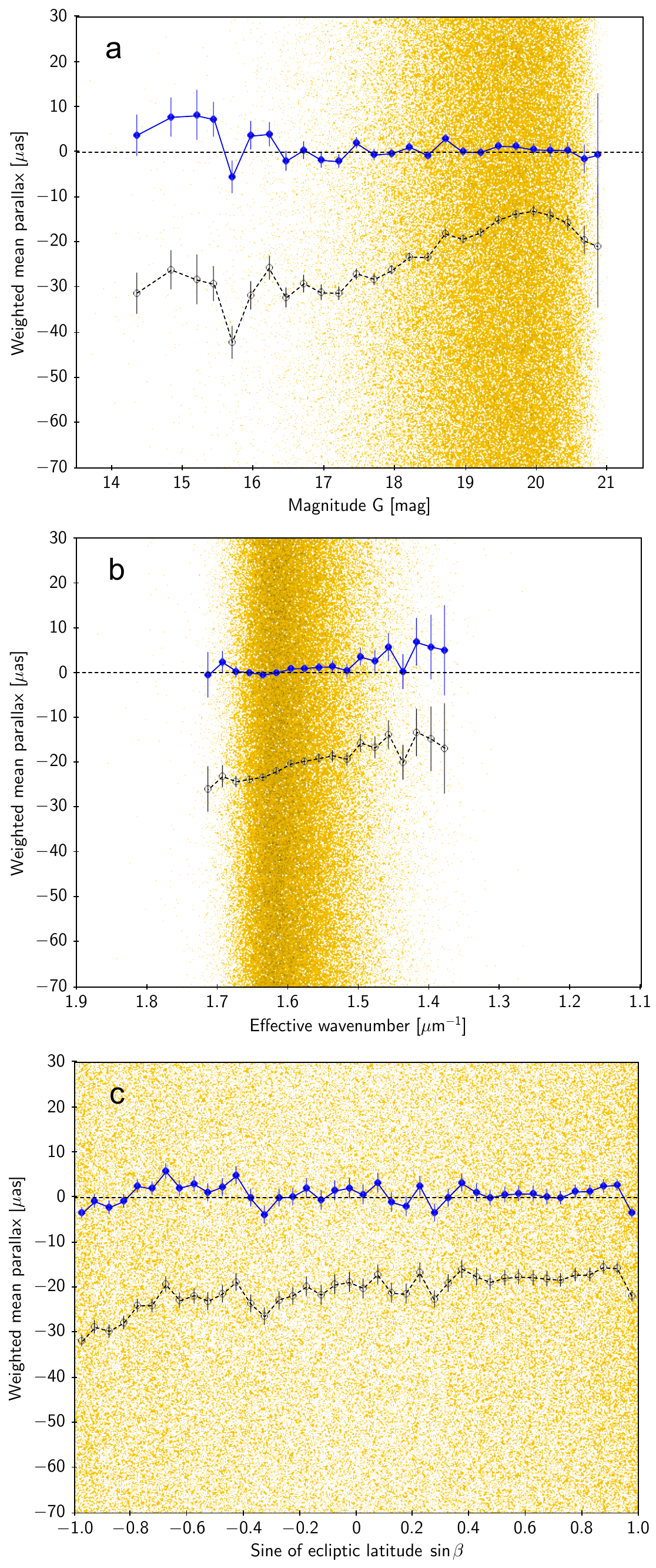} 
    \caption{Parallaxes for 1.2~million quasars with five-parameter solutions in \textit{Gaia} EDR3.
    Yellow dots show the individual values plotted versus magnitude (\textit{a}), 
    effective wavenumber (\textit{b}), and sine of ecliptic latitude (\textit{c}). Open black circles
    show mean values of the uncorrected parallaxes ($\varpi$) in bins of magnitude etc.; 
    filled blue circles show mean values of the corrected parallaxes ($\varpi-Z_5$) in the same bins. 
    Mean values are calculated using weights $\sigma_\varpi^{-2}$. Error bars indicate the estimated 
    standard deviation of the weighted mean in each bin.}
    \label{figValP5Qso}
\end{figure}

\begin{figure}
\centering
  \includegraphics[height=180mm]{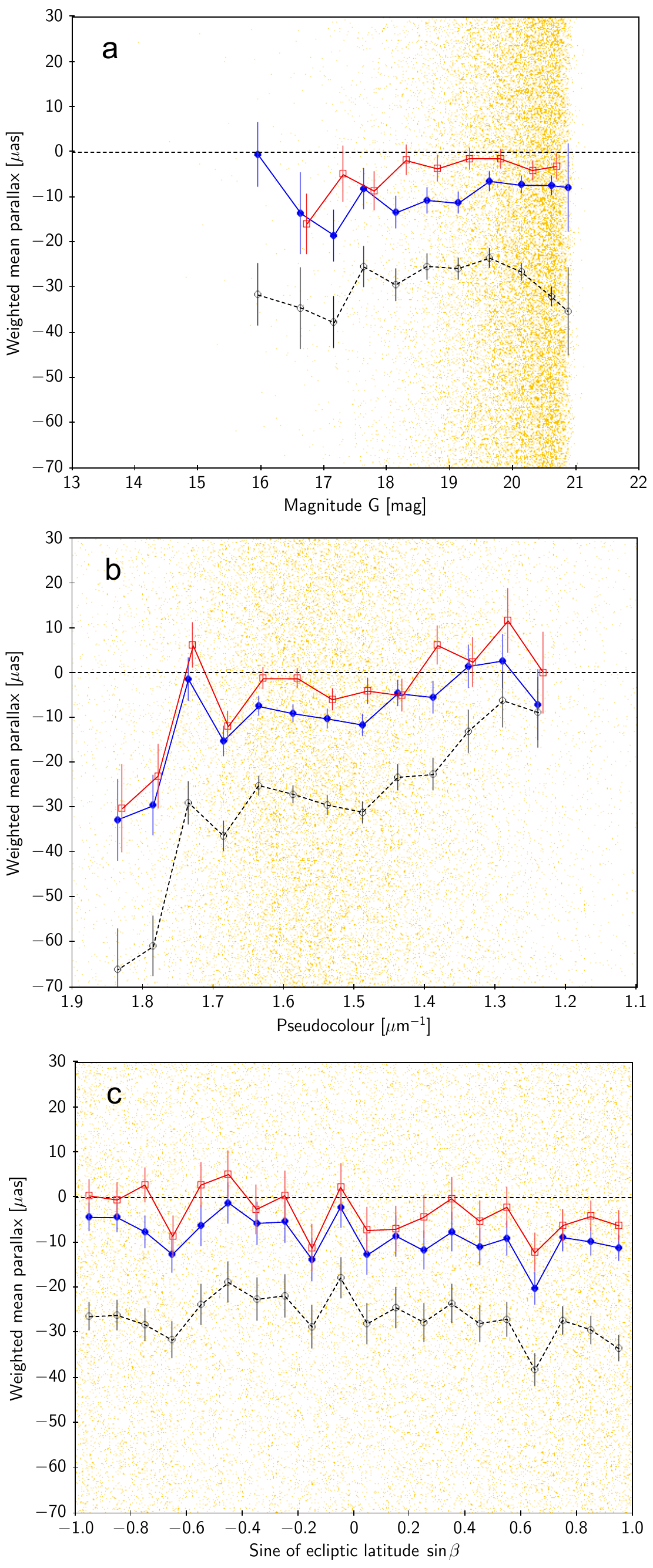} 
    \caption{Parallaxes for 0.4~million quasars with six-parameter solutions in \textit{Gaia} EDR3.
    Yellow dots show the individual values plotted versus magnitude (\textit{a}), 
    effective wavenumber (\textit{b}), and sine of ecliptic latitude (\textit{c}). Open black circles
    show mean values of the uncorrected parallaxes ($\varpi$) in bins of magnitude etc.; 
    filled blue circles show mean values of the corrected parallaxes ($\varpi-Z_6$) in the same bins. 
    Red open squares show mean values of the corrected parallaxes for the 78\% of the 
    quasars that have insignificant excess source noise, using a slightly different binning. 
    Mean values are calculated using weights $\sigma_\varpi^{-2}$. Error bars indicate the estimated 
    standard deviation of the weighted mean in each bin.}
    \label{figValP6Qso}
\end{figure}

\begin{figure}
\centering
  \includegraphics[width=\hsize]{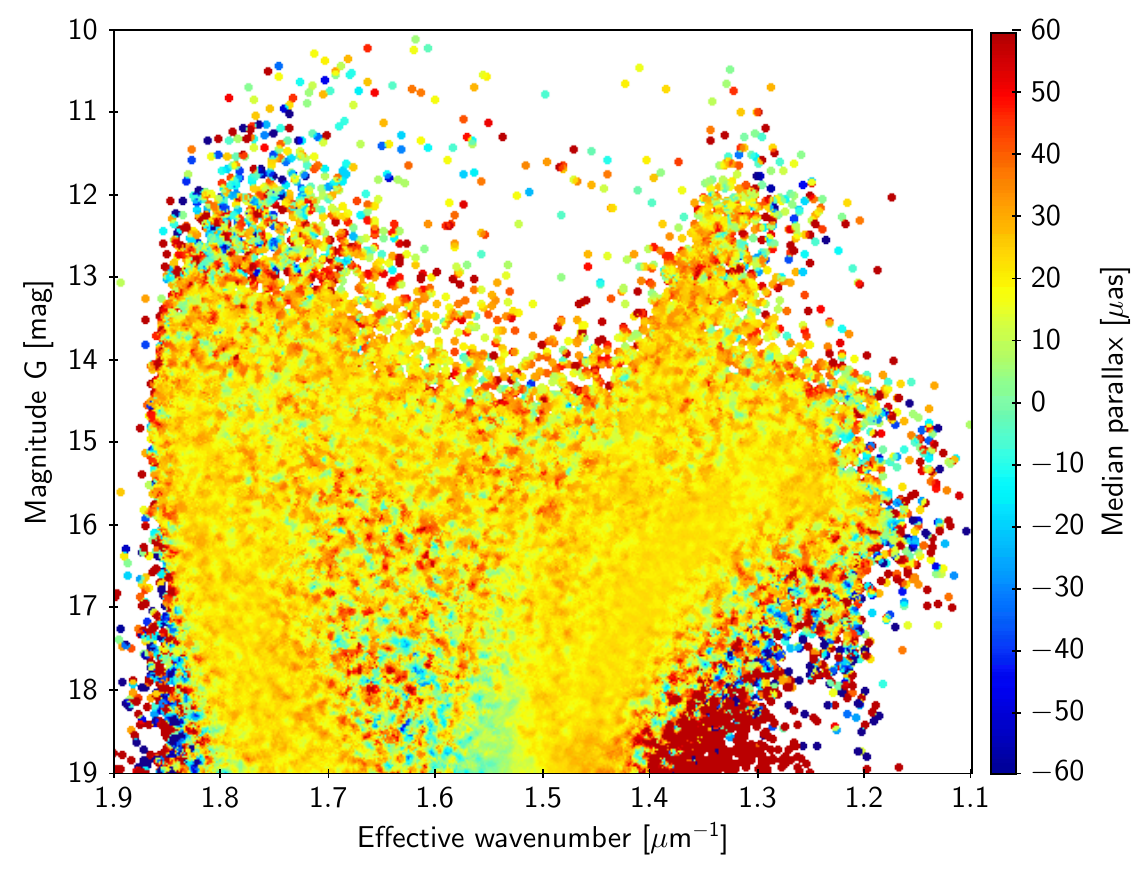}
    \caption{Colour-magnitude diagram of the LMC sample color-coded by the median of 
    the EDR3 parallax after subtracting $Z_5$ as given by the cofficients in Table~\ref{tab:final5}.}
    \label{figA02c}
\end{figure}

\subsection{Using quasars}
\label{sec:valQso}

We apply the bias 
corrections to the quasar sample in EDR3, expecting the mean corrected parallax to be close to 
zero independent of magnitude, colour, etc. The EDR3 table \gacs{agn\_cross\_id} contains
1\,215\,942~sources with five-parameter solutions and 398\,231 with six-parameter
solutions. In the five-parameter case, this sample is nearly the same as used in 
Sect.~\ref{sec:p5} to derive the faint part of $Z_5$. By contrast, the six-parameter sample was 
not previously used: as detailed in Sect.~\ref{sec:p6}, $Z_6$ was estimated differentially with 
respect to $Z_5$, using a sample of sources for which both kinds of solution were available.

Figure~\ref{figValP5Qso} shows the results for the five-parameters solutions, divided according 
to magnitude, pseudocolour, and ecliptic latitude. Mean values of the uncorrected parallaxes 
($\varpi$) are shown as open black circles, those of the corrected values ($\varpi-Z_5$) as
filled blue circles. The yellow dots, showing individual uncorrected values, are mainly intended 
to give an impression of the distributions in $G$, $\hat{\nu}_\text{eff}$, and $\sin\beta$ of the 
sources. Although the corrected parallaxes are not perfectly centred on zero, especially for 
the brightest and reddest sources, the overall improvement is fairly satisfactory. 
 
Figure~\ref{figValP6Qso} shows the corresponding results for the six-parameters solutions.
Although the corrected values ($\varpi-Z_6$) are clearly better than the uncorrected ones, it appears 
that a slightly larger correction than $Z_6$ might often be required. A peculiar effect noted in this
sample is that the mean corrected parallax is a strong function of various goodness-of-fit 
measures such as the RUWE and excess source noise. For example, the mean corrected parallax 
is generally closer to zero for the subset of quasars with insignificant excess source noise 
($\gacs{astrometric\_excess\_noise\_sig}<2$), as shown by the red squares in 
Fig.~\ref{figValP6Qso}. This trend is not present in the five-parameter sample, and we have
at the present time no explanation for it.

\subsection{Using the LMC sample}
\label{sec:valLmc}

The LMC data in Table~\ref{tab:lmc} for $G<13$ were not used in any of the analysis leading
to the bias estimates $Z_5$ in Table~\ref{tab:final5}. For example, the faint components of the 
physical pairs were anchored in the combined quasar--LMC solution, which was derived using
only sources with $G>13$. We can therefore
use the bright LMC data for a partial validation of $Z_5$. Of particular interest is
the conspicuous difference in parallax bias between the blue upper main sequence and the red giants, 
seen in the right panel of Fig.~\ref{figA02ab} and in Fig.~\ref{figA03} for $G=11.2$--11.8 and
12.2--12.9. Figure~\ref{figA02c} is a CMD of the LMC sample, similar to Fig.~\ref{figA02ab}, 
but colour coded by the median value of $\varpi-Z_5$ at each point, where $Z_5$ 
is defined by the coefficients in Table~\ref{tab:final5}. It is clear from the diagram that $Z_5$
provides a reasonable correction in most parts of the CMD, including the bright part, although 
there is a suggestion that the bias is more negative (by $\simeq 10~\mu$as) than 
as given by $Z_5$ for the blue branch at $G<13$. This could indicate that the sizes of 
$q_{10}$ and/or $q_{11}$, as obtained from the physical pairs, are underestimated in 
Table~\ref{tab:final5}. As remarked in Sect.~\ref{sec:LMC}, the dark red patch in the lower 
right part of the diagram (roughly $G>18$, $\nu_\text{eff}<1.4$) is caused by foreground 
stars dominating this part of the CMD. 
The unmodelled depression of the bias for the faintest stars at intermediate colours, 
mentioned in Sect.~\ref{sec:LMC}, is seen as a greenish patch in Fig.~\ref{figA02c}. 

In Sect.~\ref{sec:qsolmc} the distance the LMC was a free parameter in a fit of $Z_5$
to the combined quasar and LMC data, yielding a mean parallax of $+22.11 \pm 1.10~\mu$as.
This is about two standard deviations higher than commonly accepted values, e.g.\
$+20.17 \pm 0.25~\mu$as (\citeads{2019Natur.567..200P}; including systematic uncertainty).
In absolute measure, however, the difference of about 2~$\mu$as is small compared with 
the regional variations of the quasar parallaxes shown for example in Fig.~\ref{figA20}\textit{c}.
The angular power spectrum of quasar parallaxes in \textit{Gaia} EDR3 is discussed in
\citet{EDR3-DPACP-128}, where it is estimated that the RMS variation of the parallax systematics
(excluding the global offset) is about 10~$\mu$as on angular scales $\gtrsim 10^\circ$.
The uncertainty of $\pm 1.1~\mu$as from the fit in Sect.~\ref{sec:qsolmc} does not 
take these variations into account, as the LMC only probes a smaller area.
We can therefore conclude that the mean corrected parallax of the LMC is in remarkably good 
agreement with the accepted value.

\section{The way forward}
\label{sec:future}

The bias functions $Z_5$ and $Z_6$ provide a recipe for the systematic correction of the EDR3 
parallaxes based on the particular choices of data, methodology, and bias models described in
the preceding sections. These choices contain considerable elements of uncertainty and 
arbitrariness, and are no doubt also coloured by our preconceived notions. The results should 
therefore in no way be regarded as definitive. On the contrary, it is vitally important that alternative 
routes are explored towards getting a better handle on the systematics in \textit{Gaia} data. 
It should be noted that \textit{Gaia} Data Release~3 expected in 2022 will be a superset of EDR3, 
so investigations made on EDR3 parallaxes will not be superseded until Data Release~4 (DR4)
much later. Although DR4 is expected to be significantly better in terms of systematics, it will not 
be unbiased. Thus methodological developments made using the current data will remain
applicable for a long time. Here we point out a few possible directions for this work.

Global analysis: The approach taken here is highly heuristic, where models are built up 
gradually in interaction with the data analysis, using a range of different analysis tools. 
This is often a good way to arrive at a reasonable approximation quickly, and it naturally 
incorporates the already existing knowledge on the structures and difficulties inherent in the 
data. But it probably does not give the optimal result. Once a general model has been established, 
it would be advantageous to make a singe global fit to all the data. Using standard statistical 
techniques (e.g.\ \citealt{burnham+anderson2002}) the appropriate submodel can be selected by
objective criteria, and confidence intervals evaluated. This should give more precise estimates 
by combing datasets optimally and avoiding the accumulation of errors in the current 
multi-step fits, and could also reveal new dependencies and interactions.

More and different data: Increasing the amount of data included in a global analysis 
could improve the precision of the bias estimation and perhaps extend its validity in 
magnitude--colour space. Examples of datasets that should be explored are the stars in
open and globular clusters, which have the potential to cover a wide range in magnitudes
and different environments, e.g.\ crowded areas. They could for example provide a more
direct link between $Z_5$ (for the brighter stars) and $Z_6$ (for the faint members) in
each cluster.

Data-driven modelling: The access to high-quality astrometric, photometric, and 
radial-velocity data for very large stellar samples makes it possible to construct and
fit statistical models that do not depend on the physical models of specific types of
stars (e.g.\ \citeads{2019MNRAS.487.3568S}; \citeads{2019AJ....158..147H}).
These methods depend on having a very large number of objects to beat down 
statistical uncertainties, and will therefore not provide a high resolution in several 
variables, but rather an independent overall validity check.

Bias modelling for specific applications: General bias models of the kind developed here
are not necessarily the best way to handle parallax bias in specific applications. It might for
example be better to include it as a free parameter directly in the physical model, e.g.\ for
luminosity calibrations. This approach was taken by many researchers using \textit{Gaia} DR2
data (see Sect.~\ref{sec:intro} for some examples), and it remains a valid alternative to
correcting the data. 

\begin{figure}
\center
  \includegraphics[width=0.98\hsize]{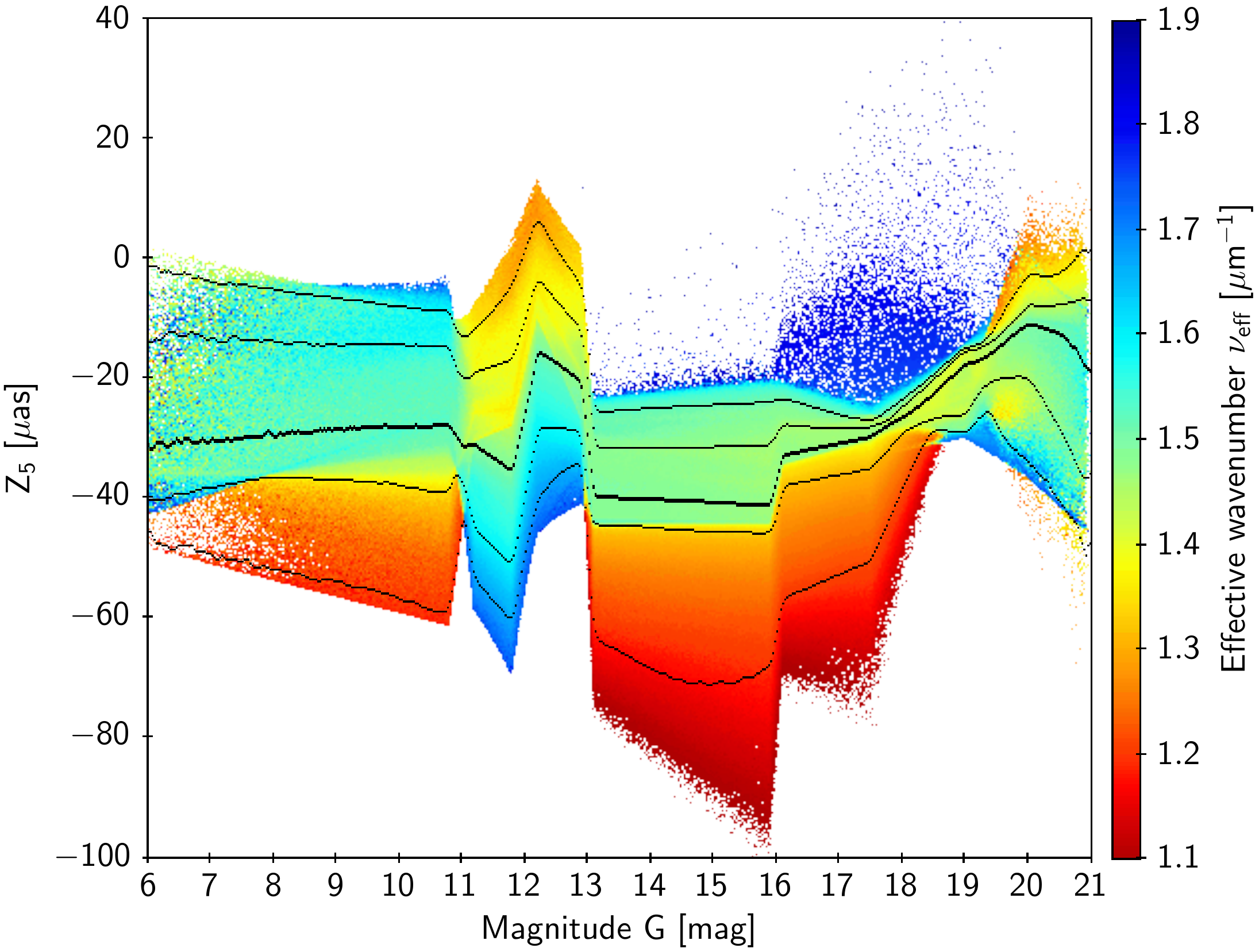}\\
  \includegraphics[width=0.98\hsize]{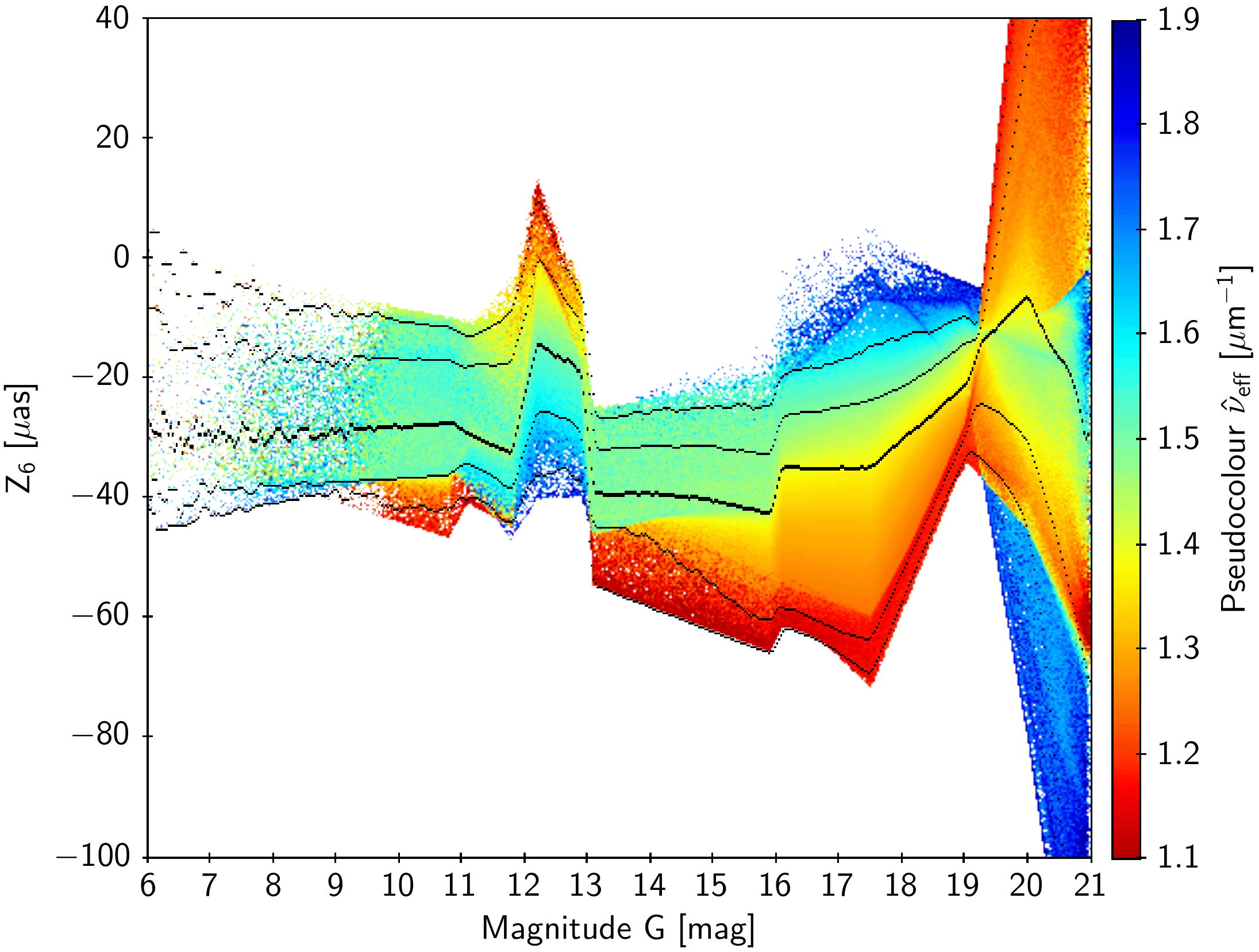}
    \caption{Parallax bias $Z_5$ (\textit{top}) and $Z_6$ (\textit{bottom}) computed 
    according to Tables~\ref{tab:final5} and \ref{tab:final6} for a representative sample of
    sources with five- and six-parameter solutions in EDR3. The dots show values for
    the complete sample of sources (for $G<11.5$) or for a random selection ($G>11.5$).
    The colour scale indicates the mean $\nu_\text{eff}$ or $\hat{\nu}_\text{eff}$ at a given 
    point, thus giving an impression of the mean colour dependence of the bias.
    The black curves show the 1st, 10th, 50th, 90th, and 99th percentiles. The thick curve
    is the 50th percentile or median.}
    \label{figA23}
\end{figure}

Systematics in proper motions: A~priori there is no reason to expect that the proper
motions in EDR3 are less affected by systematics than the parallaxes -- after all, they
are jointly determined from the same observations. 
\citet{EDR3-DPACP-126} show an example (their Fig.~25) where a discontinuity of about
$25~\mu$as~yr$^{-1}$ is seen at $G\simeq 13$ in the EDR3 proper motions $\mu_{\alpha*}$ 
of cluster stars. It is therefore clearly interesting to extend the present analysis to the 
components of proper motion. Although less crucial in most astrophysical and 
galactic astronomy applications of \textit{Gaia} data than the parallax bias, 
proper motion biases are relevant for a number of the most exacting applications,
such as the search for exoplanets. Mapping these biases in clusters may however 
be non-trivial in view of internal motions, which may include systematic patterns from 
expansion, rotation, etc.

Feedback to \textit{Gaia} calibration models: Several distinct features in $Z_5$ and $Z_6$,
such as the abrupt changes and reversal of gradients in colour at $G\simeq 11$ and 13
(Fig.~\ref{figA23}), can be traced back to specific elements of the instrument calibration 
in the data processing chain for EDR3. This can be used in a kind of reversed engineering
to understand how the calibration models need to be improved in order to bring down
systematics in future releases. This is part of the normal cyclic development work in the 
\textit{Gaia} data processing consortium, and is much helped by having at our disposal 
tools to evaluate systematics in the astrometric solution.

\begin{figure}
\centering
  \includegraphics[height=180mm]{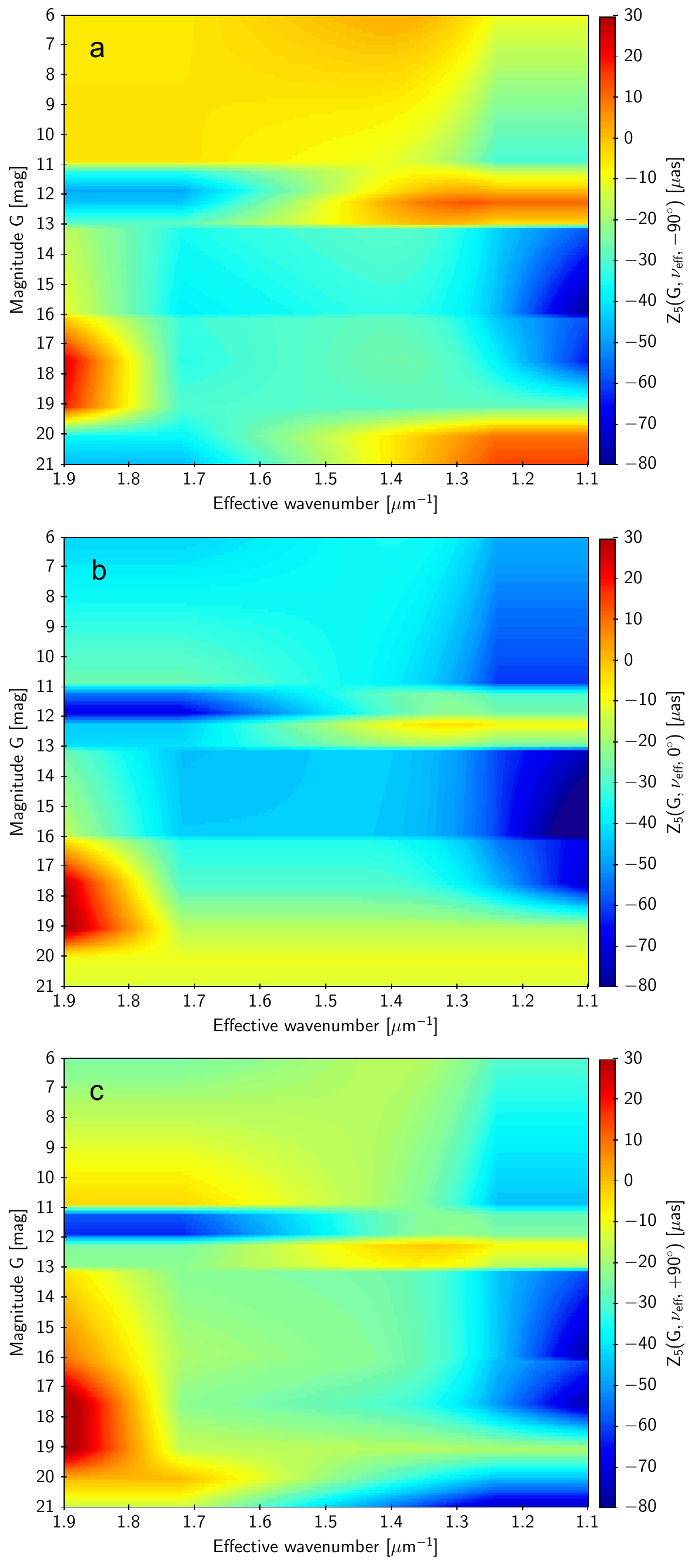}  
    \caption{Parallax bias $Z_5(G,\,\nu_\text{eff},\,\beta)$ according to Table~\ref{tab:final5}. 
    The panels show cuts at (\textit{a}) ecliptic latitude $\beta=-90^\circ$, 
    (\textit{b}) $\beta=0^\circ$, and (\textit{c}) $\beta=+90^\circ$.}
    \label{figA14}
\end{figure}

\begin{figure}
\centering
  \includegraphics[height=180mm]{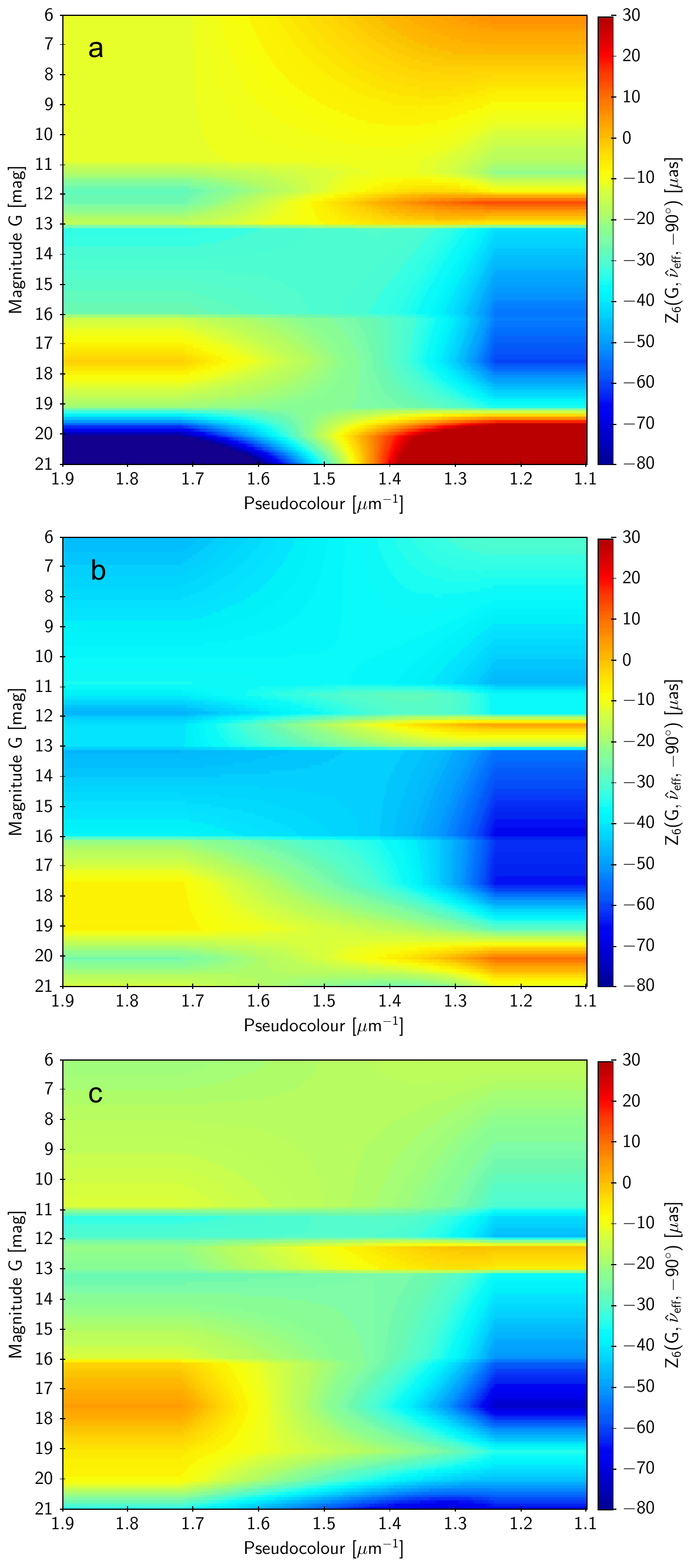}  
    \caption{Parallax bias $Z_6(G,\,\hat{\nu}_\text{eff},\,\beta)$ according to Table~\ref{tab:final6}. 
    The panels show cuts at (\textit{a}) ecliptic latitude $\beta=-90^\circ$, 
    (\textit{b}) $\beta=0^\circ$, and (\textit{c}) $\beta=+90^\circ$.}
    \label{figA19}
\end{figure}

\section{Summary and conclusions}
\label{sec:sum}

We have investigated the parallax bias in \textit{Gaia} EDR3 and the variation of this bias 
with magnitude, colour, and ecliptic latitude. The direct estimation of the bias using quasars 
is complemented by indirect methods using physical binaries and stars in the LMC. 
The indirect methods are strictly differential with respect to the quasars; in particular, no
particular value is assumed for the distance to the LMC.

The expected functional form of the dependencies on magnitude, colour, and ecliptic
latitude was derived partly from a consideration of the known properties of the instrument 
and data processing, partly from a mapping of the systematic difference between parallaxes
in \textit{Gaia} EDR3 and DR2 (Table~\ref{tab:dZ}).

Complex dependencies on all three variables are evident in all the data, and the dependencies 
are not the same for sources that have five- and six-parameter solutions. For sources with 
five-parameter solutions in \textit{Gaia} EDR3 ($\gacs{astrometric\_params\_solved}=31$)
the fitted bias function, $Z_5(G,\,\nu_\text{eff},\,\beta)$,
is given by Eqs.~(\ref{eq:c})--(\ref{eq:Z1}), using the functions $q_{jk}(G)$ obtained by linear 
interpolation in Table~\ref{tab:final5}. Here, $G$ is the mean magnitude of the source in the
broad-band \textit{Gaia} photometric system (\gacs{phot\_g\_mean\_mag}), $\nu_\text{eff}$
is the effective wavenumber (\gacs{nu\_eff\_used\_in\_astrometry}), and $\beta$ the ecliptic
latitude (\gacs{ecl\_lat}). A dash ($-$) in the table should be interpreted as zero.
The function $Z_5$ is illustrated in Fig.~\ref{figA14} and the top panel of Fig.~\ref{figA23}. 
Nominally, the bias ranges from
$-94$ to $+36~\mu$as, with an RMS scatter of 18~$\mu$as when the full range of colours
and magnitudes is considered; weighted by the actual distribution of colours per
magnitude in EDR3, as in Fig.~\ref{figA23} (top), the RMS scatter is about 13~$\mu$as.

For sources with six-parameter solutions in \textit{Gaia} EDR3 ($\gacs{astrometric\_params\_solved}=95$) 
the parallax bias function, $Z_6(G,\,\hat{\nu}_\text{eff},\,\beta)$, is similarly given by the functions $q_{jk}(G)$
obtained by interpolation in Table~\ref{tab:final6}. Here, $\hat{\nu}_\text{eff}$ is the 
astrometrically estimated effective wavenumber, known as pseudocolour (\gacs{pseudocolour}). 
The function $Z_6$ is illustrated in Fig.~\ref{figA19} and in the bottom panel of Fig.~\ref{figA23}. 
It is very uncertain when $\hat{\nu}_\text{eff}$ is $<1.24~\mu$m$^{-1}$ or $>1.72~\mu$m$^{-1}$. 
The bias ranges from $-151~\mu$as to $+130~\mu$as, with an RMS scatter of 21~$\mu$as for the
full range of colours, or 15~$\mu$as if weighted by the actual distribution of colours.

\begin{table}
\caption{Coefficients for the function $Z_6(G,\,\hat{\nu}_\text{eff},\,\beta)$.  
\label{tab:final6}}
\footnotesize\setlength{\tabcolsep}{4pt}
\begin{tabular}{crrrrrrrrrrrrrrr}
\hline\hline
\noalign{\smallskip}
 $G$ & $q_{00}$ & $q_{01}$ & $q_{02}$ & $q_{10}$ & $q_{11}$ & $q_{12}$ & $q_{20}$ \\
\noalign{\smallskip}\hline\noalign{\smallskip}
 6.0 & $-27.85$  & $ -7.78$  & $+27.47$  & $ -32.1$  & $ +14.4$  & $  +9.5$  & $   -67$\rlap{:}  \\
10.8 & $-28.91$  & $ -3.57$  & $+22.92$  & $  +7.7$  & $ +12.6$  & $  +1.6$\rlap{:} & $  -572$   \\
11.2 & $-26.72$  & $ -8.74$  & $ +9.36$  & $ -30.3$  & $  +5.6$  & $ +17.2$  & $ -1104$   \\
11.8 & $-29.04$  & $ -9.69$  & $+13.63$  & $ -49.4$  & $ +36.3$  & $ +17.7$  & $ -1129$   \\
12.2 & $-12.39$  & $ -2.16$  & $+10.23$  & $ -92.6$  & $ +19.8$  & $ +27.6$  & $  -365$   \\
12.9 & $-18.99$  & $ -1.93$  & $+15.90$  & $ -57.2$  & $  -8.0$  & $ +19.9$  & $  -554$   \\
13.1 & $-38.29$  & $ +2.59$  & $+16.20$  & $ -10.5$  & $  +1.4$  & $  +0.4$\rlap{:} & $  -960$   \\
15.9 & $-36.83$  & $ +4.20$  & $+15.76$  & $ +22.3$  & $ +11.1$  & $ +10.0$  & $ -1367$   \\
16.1 & $-28.37$  & $ +1.99$  & $ +9.28$  & $ +50.4$  & $ +17.2$  & $ +13.7$  & $ -1351$   \\
17.5 & $-24.68$  & $ -1.37$  & $ +3.52$  & $ +86.8$  & $ +19.8$  & $ +21.3$  & $ -1380$   \\
19.0 & $-15.32$  & $ +4.01$  & $ -6.03$  & $ +29.2$  & $ +14.1$  & $  +0.4$\rlap{:} & $  -563$   \\
20.0 & $-13.73$  & $-10.92$  & $ -8.30$  & $ -74.4$  & $+196.4$  & $ -42.0$\rlap{:} & $  +536$\rlap{:}  \\
21.0 & $-29.53$  & $-20.34$  & $-18.74$\rlap{:} & $ -39.5$\rlap{:} & $+326.8$  & $-262.3$  & $ +1598$\rlap{:}  \\
\noalign{\smallskip}
\hline
\end{tabular}
\tablefoot{The table gives $q_{jk}(G)$ at the values of $G$ in the first column. For other values of $G$, 
linear interpolation should be used.
A colon (:) after the coefficient indicates that it is not significant at the $3\sigma$ level.
Results are very uncertain for $\hat{\nu}_\text{eff}>1.72~\mu$m$^{-1}$ (nominally corresponding to
$G_\text{BP}-G_\text{RP}\lesssim 0.15$) and $\hat{\nu}_\text{eff}<1.24~\mu$m$^{-1}$ (nominally
$G_\text{BP}-G_\text{RP}\gtrsim 3.0$).
Units are: $\mu$as (for $q_{0k}$), $\mu$as~$\mu$m ($q_{1k}$), and $\mu$as~$\mu$m$^3$ ($q_{20}$).}
\end{table}

The derived relations are only applicable to the parallaxes in \textit{Gaia} EDR3.
Regarded as a systematic correction to the parallax, the bias function 
$Z_5$ or $Z_6$ should be \textit{subtracted} from the value (\gacs{parallax}) given in 
the Archive. Python implementations of both functions are available via the \textit{Gaia}
web pages at 
\url{https://www.cosmos.esa.int/web/gaia/edr3-code}.

As recipes for the systematic correction of the EDR3 parallaxes, these functions should 
be regarded as provisional and indicative. While we have reason to believe that
their application will in general reduce systematics in the parallaxes, this 
may not always be the case. Users are urged to make their own judgement concerning 
the relevance of the indicated bias correction for their specific applications. 
Whenever possible, depending on the type and number of sources under consideration, 
users of EDR3 should try to derive more targeted bias estimates for their specific use cases. 

It is difficult to quantify uncertainties in $Z_5$ and $Z_6$. 
In the region of parameter space well populated by the quasars (essentially $G\gtrsim 16$ and
$1.4\lesssim\nu_\text{eff}\lesssim 1.7~\mu$m$^{-1}$), they may be as small as a few $\mu$as, but beyond  
that region uncertainties are bound to be greater because of the indirect methods used. For redder
sources ($\nu_\text{eff}\lesssim 1.4~\mu$m$^{-1}$, corresponding to $G_\text{BP}{-}G_\text{RP}\gtrsim 1.6$),
both $Z_5$ and $Z_6$ depend critically on the assumption (item~2 in Sect.~\ref{sec:qsolmc})
that the curvature in colour, seen in the LMC, is the same over the whole sky. If this turns out 
to be wrong, it could mean that the bias function for very red sources in the northern hemisphere 
is wrong by several tens of $\mu$as. As discussed in Appendix~\ref{sec:rc} we do not think this 
is likely, but the possibility should be kept in mind. Beyond the non-clamped interval 
$1.72 > \nu_\text{eff} > 1.24~\mu$m$^{-1}$ (corresponding to 
$0.15 \lesssim G_\text{BP}-G_\text{RP}\lesssim 3.0$) all results are in any case quite uncertain    
owing to the way colour information is handled in the LSF/PSF calibration and astrometric 
solutions for EDR3 (Sect.~\ref{sec:nueff}).

Two unmodelled features mentioned in the paper could merit further investigation. 
One is the depression of the bias for $G\gtrsim 18$, $\nu_\text{eff}\simeq 1.55~\mu$m$^{-1}$
seen as a greenish patch in Fig.~\ref{figA02c}. As this region of colour--magnitude space is well
covered by the quasars, the feature is probably not generally present but could be particularly strong 
in the LMC area. The other unmodelled feature is the dependence of the parallax bias for six-parameter
solutions on excess source noise illustrated in Fig.~\ref{figValP6Qso}. It is possible that both features
are caused by to a hitherto unexplored tendency, at faint magnitudes, for the bias to become more 
negative in crowded areas, where the source excess noise or RUWE also tend to be higher.

As discussed in connection with Fig.~\ref{figA20} (panel \textit{c}) and more extensively 
elsewhere \citep{EDR3-DPACP-128}, additional systematics in the EDR3 
parallaxes have been identified, for example depending on position on small and intermediate 
angular scales. These cannot easily be mapped to any useful precision, and should rather be 
modelled as correlated random errors. The angular power spectrum of quasar parallaxes
presented in Sect.~5.6 of \citet{EDR3-DPACP-128} may be used to that end.
Depending on the angular scale considered, the estimated RMS variation with position
ranges from 5 to 26~$\mu$as, that is of the same order of magnitude as the RMS variations 
in $Z_5$ and $Z_6$ as functions of colour or magnitude.

While it is easy enough to demonstrate that the EDR3 parallaxes contain significant 
systematics, it is extremely difficult to obtain accurate estimates of the bias 
beyond the limited region of parameter space well populated by the quasars. This paper 
does not claim to give a definitive answer but merely a rough characterisation of what we 
have found to be the main dependencies. It is likely that better, and possibly quite different 
estimates can be obtained in the future by means of more refined and comprehensive 
analysis methods. Continued exploration of the systematics is important not least in order
to gain a better understanding of their causes. In the end, this will hopefully lead to much
lower levels of systematics in future \textit{Gaia} data releases.

\begin{acknowledgements}

This work has made use of data from the European Space Agency (ESA) mission
{\it Gaia} (\url{https://www.cosmos.esa.int/gaia}), processed by the {\it Gaia}
Data Processing and Analysis Consortium (DPAC,
\url{https://www.cosmos.esa.int/web/gaia/dpac/consortium}). Funding for the DPAC
has been provided by national institutions, in particular the institutions
participating in the {\it Gaia} Multilateral Agreement.
This work was financially supported by 
the Swedish National Space Agency (SNSA/Rymdstyrelsen),
the European Space Agency (ESA) in the framework of the \textit{Gaia} project, and
the German Aerospace Agency (Deutsches Zentrum für Luft- und Raumfahrt e.V., DLR) through grants 50QG0501, 50QG0601, 50QG0901, 50QG1401 and 50QG1402.
We thank the Centre for Information Services and High Performance Computing (ZIH) at the Technische Universität (TU) Dresden for generous allocations of computer time.

Diagrams were produced using the astronomy-oriented data handling and visualisation software TOPCAT
\citepads{2005ASPC..347...29T}. 

\end{acknowledgements}

\bibliographystyle{aa} 
\bibliography{refsZ} 

\appendix

\section{Parametrised functions}
\label{sec:num}

The various absolute or differential bias functions discussed in the paper are written as linear 
combinations of a finite set of three-dimensional basis functions, with $G$, $\nu_\text{eff}$ 
(or $\hat{\nu}_\text{eff}$),
and $\beta$ as independent variables. To allow interactions among the variables, the set of 
three-dimensional basis functions should in the most general case be the outer product of the 
one-dimensional basis functions on each axis. The generic function is therefore
\begin{equation}\label{eq:Z}
Z(G,\,\nu_\text{eff},\,\beta)=\sum_i\sum_j \sum_k z_{i\!jk}\,g_i(G)\,c_j(\nu_\text{eff})\,b_k(\beta) \, ,
\end{equation}
where $g_i(G)$ are the basis functions in magnitude, $c_j(\nu_\text{eff})$ the basis functions in 
colour, and $b_k(\beta)$ the basis functions in ecliptic latitude. The coefficients $z_{i\!jk}$ are the
free parameters used to fit $Z$ to the given data.

\begin{figure}
\center
  \includegraphics[width=\hsize]{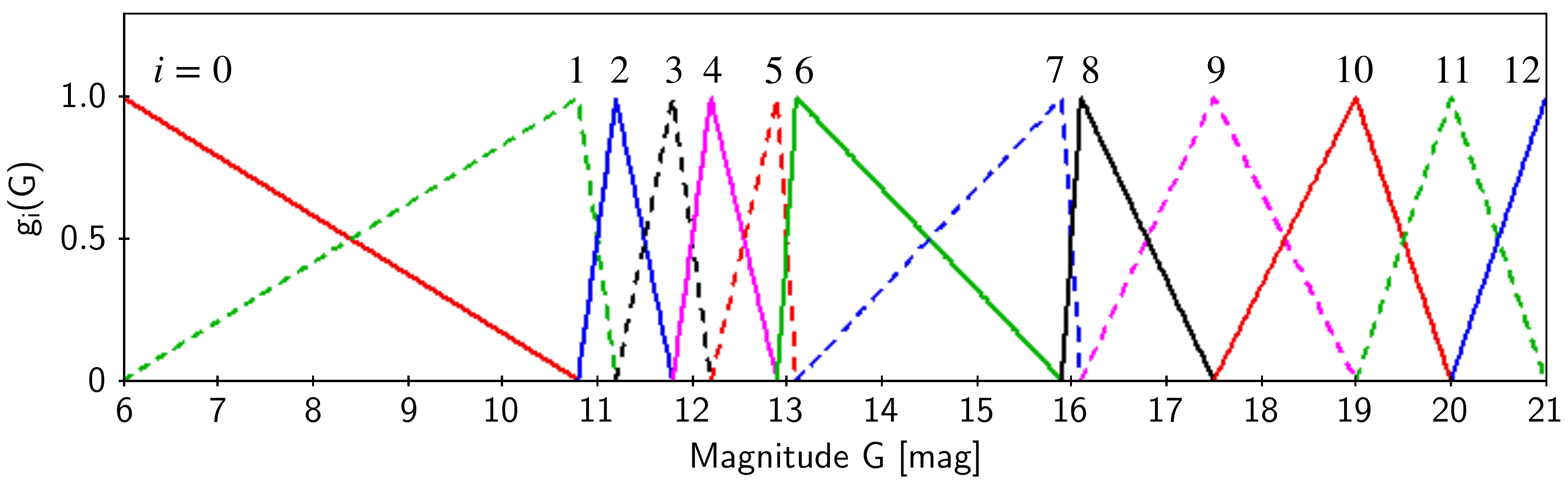}
    \caption{Basis functions $g_i(G)$, $i=0\dots 12$ according to Eq.~(\ref{eq:g}).}
    \label{fig:basisG}
\end{figure}

The magnitude dependence is modelled as a continuous piecewise linear function with breakpoints (knots) at
$\gamma_{0\dots 12}=6.0$, 10.8, 11.2, 11.8, 12.2, 12.9, 13.1, 15.9, 16.1, 17.5, 19.0, 20.0, and 21.0~mag. 
The corresponding basis functions are
\begin{equation}\label{eq:g}
\left.\begin{aligned}
g_0(G) &= \begin{cases} 1 & \text{if $G\le\gamma_0$,}\\ 
(\gamma_1-G)/(\gamma_1-\gamma_0)\hspace{1.2em} & \text{if $\gamma_0<G\le\gamma_1$,}\\
0 & \text{if $\gamma_1<G$,} \end{cases}\\[12pt]
g_i(G) &= \begin{cases} 0 & \text{if $G\le\gamma_{i-1}$,}\\
(G-\gamma_{i-1})/(\gamma_i-\gamma_{i-1}) & \text{if $\gamma_{i-1}<G\le\gamma_i$,}\\
(\gamma_{i+1}-G)/(\gamma_{i+1}-\gamma_i) & \text{if $\gamma_i<G\le\gamma_{i+1}$,}\\
0 & \text{if $\gamma_{i+1}<G$,} \end{cases}\\[-2pt]
& \quad \text{for $i=1\dots 11$,}\\[12pt]
g_{12}(G) &= \begin{cases}  0 & \text{if $G\le\gamma_{11}$,}\\
(G-\gamma_{11})/(\gamma_{12}-\gamma_{11}) & \text{if $\gamma_{11}<G\le\gamma_{12}$,}\\
1 & \text{if $\gamma_{12}<G$} \end{cases}
\end{aligned}\quad\right\}
\end{equation}
(Fig.~\ref{fig:basisG}). A linear combination of these functions may provide a reasonable 
approximation of variations such as those seen in the top panel of Fig.~\ref{fig:dPlx3m2vsGB}.
In particular, the knots at $G=11.0\pm 0.2$, $12.0\pm 0.2$, $13.0\pm 0.1$, and $16.0\pm 0.1$
correspond to the transitions suggested in Fig.~\ref{fig:gClass}. An important property of this
basis set is that, at every breakpoint, exactly one basis function is $=1$ and the rest are $=0$.
This means that, for arbitrary coefficients $q_i$, the function $q(G)=\sum_j q_i g_i(G)$
can be evaluated by linear interpolation among the coefficients, since $q(\gamma_i)=q_i$. This 
property is useful in connection with the alternative form in Eq.~(\ref{eq:Z1}).  

\begin{figure}
\center
  \includegraphics[width=0.9\hsize]{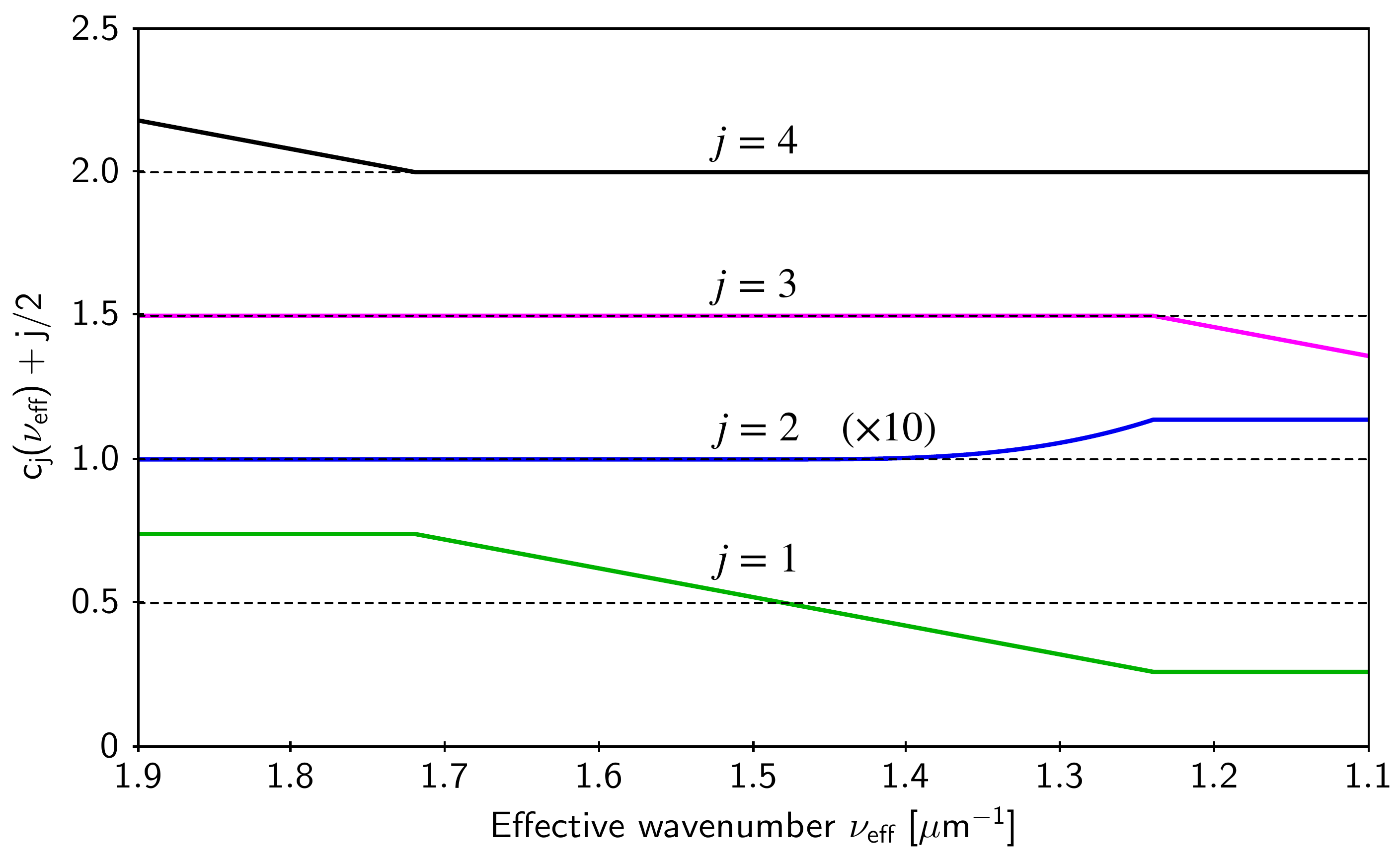}
    \caption{Basis functions $c_j(\nu_\text{eff})$, $j=1\dots 4$ according to Eq.~(\ref{eq:c}). 
    The (constant) function
    $c_0=1$ is not shown. For better visibility the functions are vertically displaced by 
    $j/2$, and the amplitude of $c_2(\nu_\text{eff})$ is increased by a factor 10.}
    \label{fig:basisC}
\end{figure}

The dependence on colour is also modelled as a continuous piecewise polynomial, but of a more
specific form inspired by plots like Figs.~\ref{fig:dPlx3m2vsNuEff} and \ref{figA03}. Interior breakpoints 
are placed at $\nu_\text{eff}=1.24$, 1.48, and 1.72~$\mu$m$^{-1}$. The segments below 1.24, between 
1.48 and 1.72, and above 1.72 are linear, while between 1.24 and 1.48 it is cubic with continuous 
first and second derivatives at 1.48~$\mu$m$^{-1}$. The basis functions are
\begin{equation}\label{eq:c}
\left.\begin{aligned}
c_0(\nu_\text{eff}) &= 1,\\[12pt]
c_1(\nu_\text{eff}) &= \begin{cases} -0.24 & \text{if $\nu_\text{eff}\le 1.24$},\\
\nu_\text{eff}-1.48\hspace{1.2em} & \text{if $1.24<\nu_\text{eff}\le 1.72$,}\\
+0.24 & \text{if $1.72<\nu_\text{eff}$,}\end{cases}\\[12pt]
c_2(\nu_\text{eff}) &= \begin{cases} (+0.24)^3 & \text{if $\nu_\text{eff}\le 1.24$,}\\
(1.48-\nu_\text{eff})^3 & \text{if $1.24<\nu_\text{eff}\le 1.48$,}\\
0 & \text{if $1.48<\nu_\text{eff}$,}\end{cases}\\[12pt]
c_3(\nu_\text{eff}) &= \begin{cases} \nu_\text{eff}-1.24\hspace{1.0em} & \text{if $\nu_\text{eff}\le 1.24$,}\\
0 & \text{if $1.24<\nu_\text{eff}$,}\end{cases}\\[12pt]
c_4(\nu_\text{eff}) &= \begin{cases} 0 & \text{if $\nu_\text{eff}\le 1.72$,}\\
\nu_\text{eff}-1.72\hspace{1.0em} & \text{if $1.72<\nu_\text{eff}$,}\end{cases}
\end{aligned}\quad\right\}
\end{equation}
(Fig.~\ref{fig:basisC}). Their coefficients thus represent, respectively, the value at 
$\nu_\text{eff}=1.48~\mu$m$^{-1}$,
the linear gradient between 1.24 and $1.72~\mu$m$^{-1}$, the added cubic trend in the middle red interval 
(from 1.24 to $1.48~\mu$m$^{-1}$), and the linear gradients at the extreme colours (below 1.24 and above 
1.72~$\mu$m$^{-1}$).

\begin{figure}
\center
  \includegraphics[width=0.9\hsize]{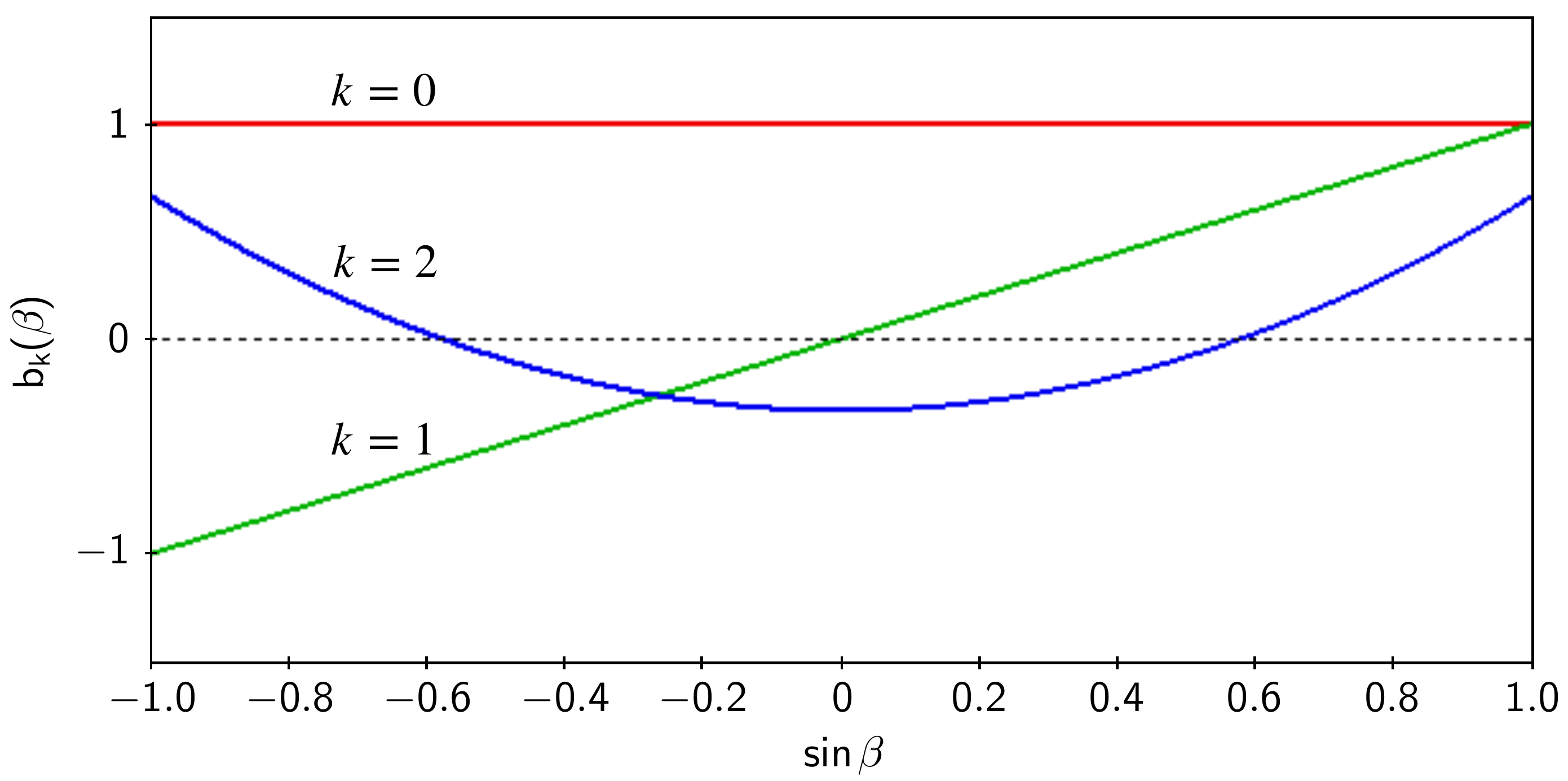}
    \caption{Basis functions $b_k(\beta)$, $k=0\dots 2$ according to Eq.~(\ref{eq:b}).}
    \label{fig:basisB}
\end{figure}

Finally, the basis functions in ecliptic latitude,
\begin{equation}\label{eq:b}
\left.\begin{aligned}
b_0(\beta) &= 1,\\[6pt]
b_1(\beta) &= \sin\beta,\\[6pt]
b_2(\beta) &= \sin^2\beta - {\textstyle\frac{1}{3}}\, 
\end{aligned}\quad\right\}
\end{equation}
(Fig.~\ref{fig:basisB}) describe an arbitrary quadratic dependence on $\sin\beta$. The term 
$-{\textstyle\frac{1}{3}}$ in $b_2$ makes the three functions orthogonal for a uniform distribution 
of sources on the celestial sphere (which is also uniform in $\sin\beta$).

An equivalent form of Eq.~(\ref{eq:Z}) is
\begin{equation}\label{eq:Z1}
Z(G,\,\nu_\text{eff},\,\beta)=\sum_j \sum_k q_{jk}(G)\,c_j(\nu_\text{eff})\,b_k(\beta) \, ,
\end{equation}
where the functions
\begin{equation}\label{eq:Z2}
q_{jk}(G)=\sum_i z_{i\!jk}\,g_i(G) \, , \quad j=0\dots 4\, , \quad k=0\dots 2\, 
\end{equation}
are piecewise linear in $G$. Equation~(\ref{eq:Z1}) is useful because the functions 
$q_{jk}(G)$ can be evaluated by linear interpolation among the coefficients $z_{i\!jk}$,
which allows $Z(G,\,\nu_\text{eff},\,\beta)$ to be given in the compact tabular form
extensively used in this paper.

The coefficients $z_{i\!jk}$ may be determined by standard curve fitting techniques.
The problem is simple in the sense that it is linear in all the coefficients, but in practice
it is complicated by the presence of outliers and the often very incomplete coverage
of the three-dimensional space $(G,\,\nu_\text{eff},\,\beta)$. Robust techniques
such as $L_1$-norm minimisation can be used to cope with outliers. Data coverage is
more problematic and may require some judicious modification of the set of basis functions. 
A simple remedy could be to remove basis functions without support, for example $c_3$ 
and $c_4$ if there are too few sources of extreme colours; this is equivalent to putting 
the corresponding coefficients ($z_{i3k}$ and $z_{i4k}$) equal to 0. The dependence 
on $G$ can be simplified by adding constraints to the fit; for example, the constraint
$z_{i,jk}=z_{i+1,jk}$ will force the function $q_{jk}(G)$ to be constant for 
$\gamma_i\le G\le \gamma_{i+1}$.

\section{Construction of the LMC sample}
\label{app:LMC}

\begin{figure*}
\centering
  \includegraphics[height=180mm]{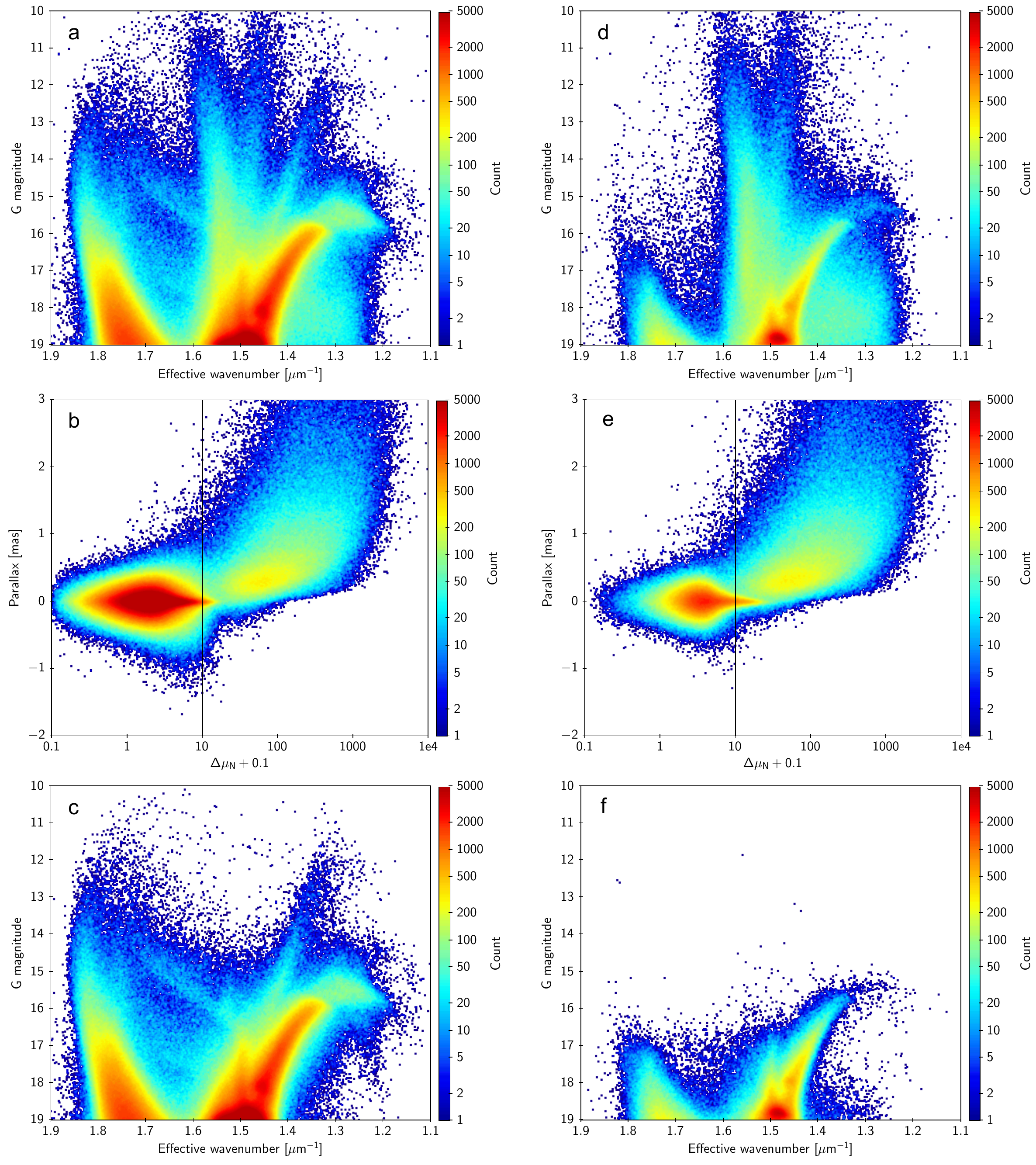}  
    \caption{Illustrating the selection and validation of the LMC sample. 
    \textit{a}: colour-magnitude diagram for the original sample centred on $(l_\text{C},b_\text{C})$.
    \textit{b}: joint distribution for the original sample of parallax and 
    $\Delta\mu_\text{N}$, the normalised proper-motion difference to the fitted model.
    The vertical line marks the cut-off used  for the selection based on proper motions.
    \textit{c}: colour-magnitude diagram for the sub-sample with $\Delta\mu_\text{N}<10$.
    \textit{d--f}: same as \textit{a--c} but for a sample centred on $(l_\text{C}-10^\circ,b_\text{C})$,
    containing much fewer LMC stars but roughly the same number of Galactic foreground
    stars as in \textit{a--c}.}
    \label{figA01}
\end{figure*}

This appendix describes the selection of sources in the LMC area, used for the 
analysis in Sect.~\ref{sec:LMC}, and some tests on the purity of the sample.

Adopting the LMC centre $(\alpha_\text{C},\delta_\text{C})=(78.77^\circ,-69.01^\circ)$
as in \citetads{2018A&A...616A..12G}, we extracted all sources in \textit{Gaia} EDR3
within a radius of $5^\circ$ with five-parameter solutions, $G< 19$, and 
$\gacs{ruwe}< 1.4$. A colour-magnitude (Hess) diagram of the
resulting sample is shown in Fig.~\ref{figA01}\textit{a}. The effective wavenumber
$\nu_\text{eff}$ is used instead of a colour index on the horizontal axis, and 
the direction has been reversed so that bluer stars are to the left and 
redder to the right. Several prominent features in this diagram are produced 
by Galactic foreground stars, and some additional filtering is clearly required. 
Since the purpose here is to study biases in parallax, it is essential that no filtering 
uses the actual parallax values, while the filtering already done based on position 
and \gacs{ruwe} cannot introduce a selection bias on the parallaxes. For the further
selection we use the residuals in proper motion relative to a fitted, very simple 
kinematic model.

The positions and proper motions (including uncertainties and correlations of the
latter) were converted to Galactic coordinates,%
\footnote{The selection described here could equally well have been made using 
the original equatorial values, but for the corresponding selection 
in the comparison areas, described in the next paragraph, it is essential that Galactic 
coordinates are used in order to preserve the orientation of the $x$ and $y$ axes 
relative to the Galactic plane.}
and then to rectangular orthographic 
components $(x,y,\mu_x,\mu_y)$ using Eq.~(2) in \citetads{2018A&A...616A..12G}, 
but replacing everywhere $\alpha$, $\delta$, $\mu_{\alpha*}$, and $\mu_\delta$ by 
$l$, $b$, $\mu_{l*}$, $\mu_b$, and $(\alpha_\text{C},\delta_\text{C})$ by 
$(l_\text{C},b_\text{C})\simeq (279.77^\circ,-33.77^\circ)$. 
Using a robust ($L_1$-norm minimisation) algorithm, the following linear relation
was obtained in a fit including only stars brighter than $G=18$:
\begin{equation}\label{eq:lmcfit}
\left.
\begin{aligned}
\mu_x &\simeq -0.602 -0.409\, x -4.755\, y\, \\
\mu_y &\simeq +1.760 + 4.180\, x -1.464\, y\, 
\end{aligned}\quad\right\}\quad [\text{mas~yr}^{-1}]\, .
\end{equation} 
For each star in the sample, deviations in $(\mu_x,\mu_y)$ from this model were 
transformed back to proper-motion residuals $(\Delta\mu_{l*},\Delta\mu_b)$, from
which the normalised deviations
\begin{equation}\label{eq:deltaMuN}
\Delta\mu_\text{N}=\left(\frac{\left(\frac{\Delta\mu_{l*}}{\sigma_{\mu l*}}\right)^2 +
\left(\frac{\Delta\mu_{b}}{\sigma_{\mu b}}\right)^2 - 
2\rho(\mu_{l*},\mu_b)\left(\frac{\Delta\mu_{l*}}{\sigma_{\mu l*}}\right)
\left(\frac{\Delta\mu_{b}}{\sigma_{\mu b}}\right)}%
{1-\rho(\mu_{l*},\mu_b)^2}\right)^{1/2}
\end{equation}
were computed. Figure~\ref{figA01}\textit{b} shows the joint distribution
of parallax and $\Delta\mu_\text{N}$ for the sample in Fig.~\ref{figA01}\textit{a}. It is seen that
the selection $\Delta\mu_\text{N}<10$ is likely to produce a relatively clean sample of
LMC stars. The colour-magnitude diagram of the filtered sample is shown in Fig.~\ref{figA01}\textit{c}.
This sample, which is the one used for the analysis in Sect.~\ref{sec:LMC}, contains 1457 sources
with $G<13.0$, 88\,285 with $G<16.0$, 519\,203 with $G<17.5$, and 2\,371\,761 with $G<19.0$.

To validate and further quantify the cleanliness of the resulting sample, we performed the 
analogous selection and filtering of sources in two adjacent areas of the same size, but 
centred on $(l_\text{C}\pm 10^\circ,b_\text{C})$. Their positions and proper motions were
transformed to $x$, $y$, $\mu_x$, $\mu_y$ relative to the centre of the respective offset 
area, while residuals and $\Delta\mu_\text{N}$ were computed relative to the fixed values 
in Eq.~(\ref{eq:lmcfit}). This should give approximately the same number and kinematic
selection of Galactic stars in the three areas, thanks to their equal latitude and limited spread 
in longitude. The right panels (\textit{d})--(\textit{f}) of Fig.~\ref{figA01} show the results for 
one of the offset areas; plots for the other offset area are not shown but qualitatively similar. 
The $\varpi$--$\Delta\mu_\text{N}$ plots in the middle panels confirm that the number and 
distribution of Galactic stars (the structure extending towards positive $\varpi$ and high 
$\Delta\mu_\text{N}$) are quite similar in the LMC and offset areas. Further statistical analysis 
shows that the sample in Fig.~\ref{figA01}\textit{c} has negligible contamination by foreground stars 
down to $G\simeq 18$ at all colours, and down to $G=19$ for the bluer sources 
($\nu_\text{eff}\gtrsim 1.6$).

\section{Test using red clump (RC) stars}
\label{sec:rc}

\begin{figure}
\sidecaption
  \includegraphics[height=130mm]{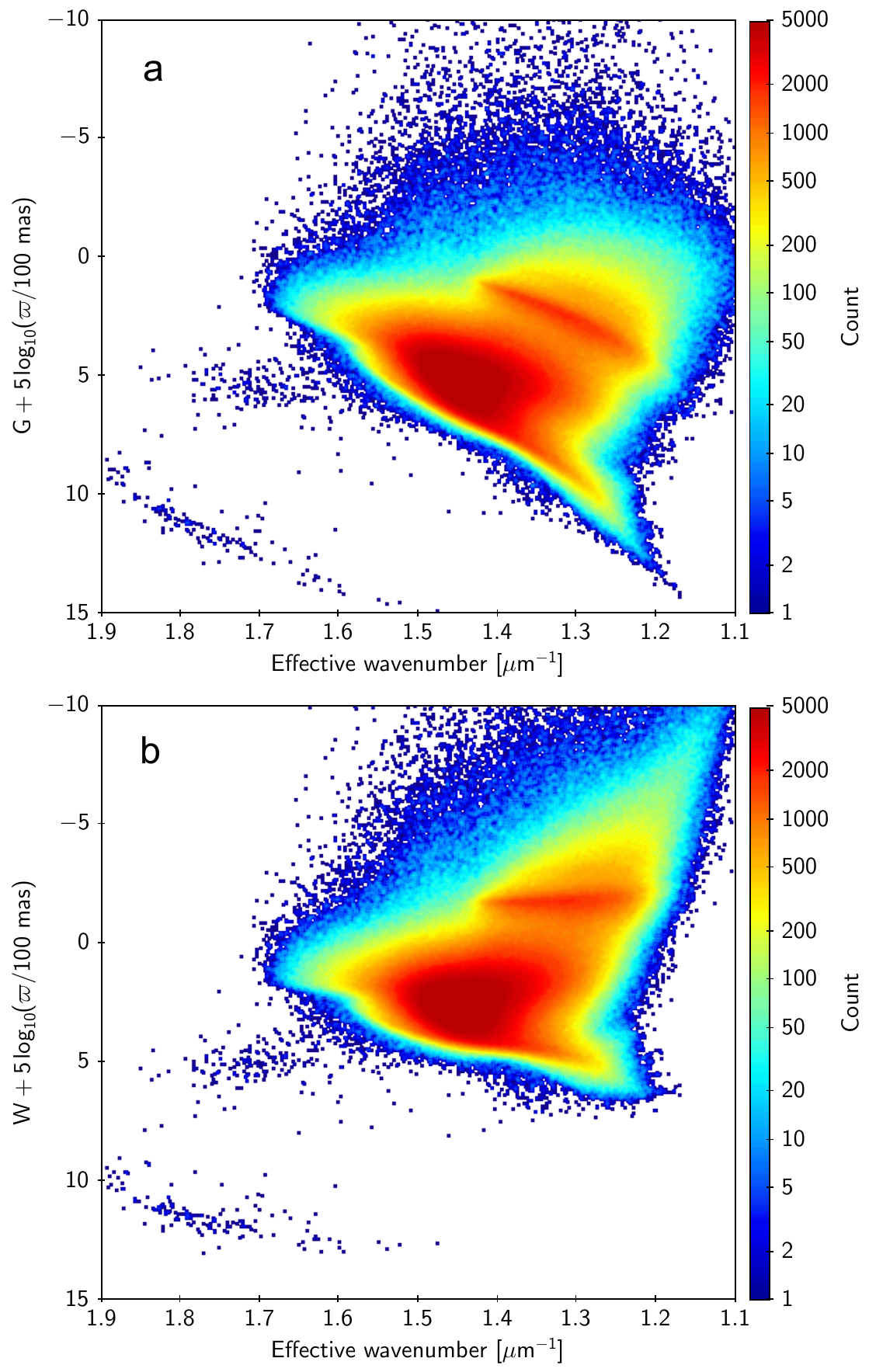}   
    \caption{Hertzsprung--Russell diagrams for the union of samples N and S.
    Top (\textit{a}): absolute magnitudes in the $G$ band.
    Bottom (\textit{b}): absolute magnitudes from $W$ in Eq.~(\ref{eq:W}).}
    \label{fig:rcHR}
\end{figure}

\begin{figure}
\centering
  \includegraphics[height=130mm]{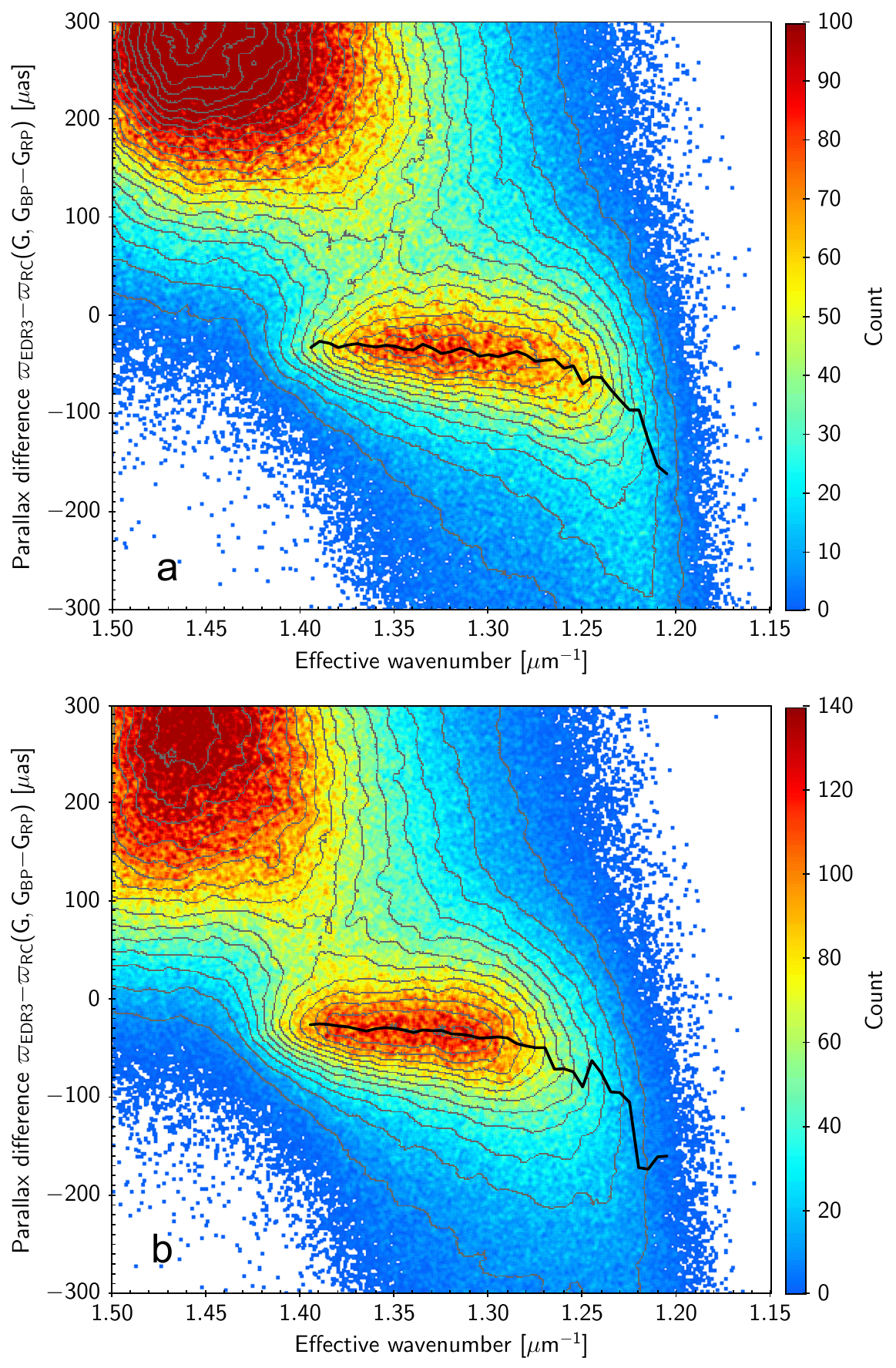}   
    \caption{Differential parallax bias estimated by means of red clump stars.
    The plots show differences between the measured parallaxes $\varpi_\text{EDR3}$
    the photometric parallaxes $\varpi_\text{RC}$ from Eq.~(\ref{eq:rcPlx}) in
    area N (\textit{a}) and S (\textit{b}). Contours of constant density are shown in thin grey lines. 
    The thick black curve traces the ridge of the feature created by the red clump stars.}
    \label{figA11ab}
\end{figure}

\begin{figure}
\centering
  \includegraphics[width=0.9\hsize]{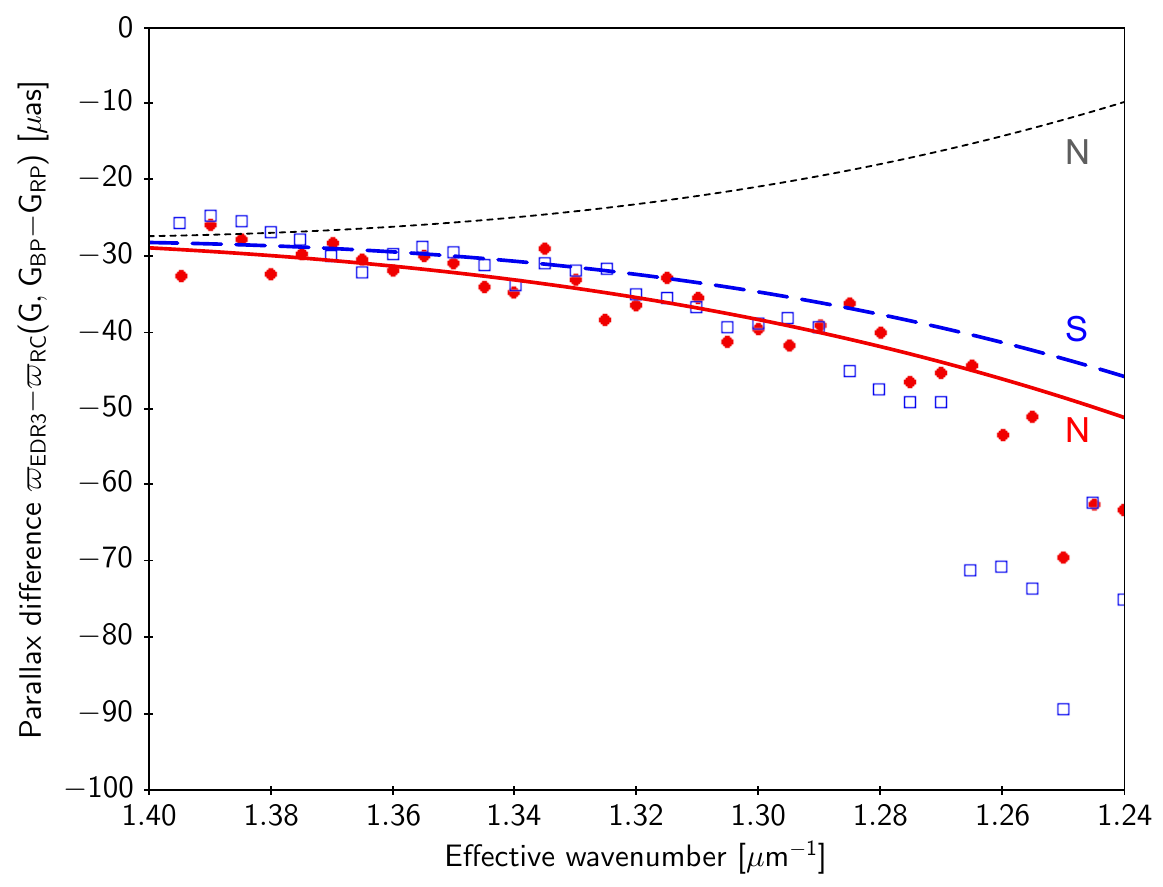}   
    \caption{Differential parallax bias estimated by means of red clump stars.
    The points show the locations of the ridges in Fig.~\ref{figA11ab} for area N
    (red filled circles) and S (blue open squares). The curves show the expected locations
    at $G=17$ according to Table~\ref{tab:qsolmcNoJoin} for area N (solid red) and S 
    (long-dashed blue), assuming that the curvature with colour ($j=2$) is the same in both 
    areas. Had the curvature flipped sign with $\beta$, the ridge location for area N 
    should instead follow the up-bending, short-dashed black curve.}
    \label{figA11c}
\end{figure}

A crucial assumption for the joint quasar and LMC model in
Sect.~\ref{sec:qsolmc} is that the curvature in colour, represented by the 
basis function $c_2(\nu_\text{eff})$ in Eq.~(\ref{eq:c}), is the same over the whole
celestial sphere; in other words, that interaction terms between $\nu_\text{eff}$ 
and $\beta$ are negligible. This assumption is needed because 
there are too few quasars that are red enough ($\nu_\text{eff}\lesssim 1.35$) 
to reliably determine the curvature at any point; instead it is derived entirely from 
the LMC sample, as illustrated by the curved segments in the three panels of 
Fig.~\ref{figA03} for $G>13$. In Table~\ref{tab:qsolmcNoJoin} such interactions,
if they existed, would be represented by non-zero coefficients 
$q_{21}$ and $q_{22}$. The assumed invariance of the colour curvature with 
position rests mainly on the following analysis of the parallaxes of red clump 
(RC) stars in EDR3. 
  
In the observational Hertzsprung--Russell diagram (HRD), the red clump is a 
prominent feature made up of low-mass stars in the core helium-burning stage of 
their evolution. RC stars are widely used as standard candles thanks to their 
relatively small scatter in absolute magnitudes, especially at near-infrared and 
infrared wavelengths \citep[for a general overview, see][]{2016ARA&A..54...95G}. 
Differences in the absolute magnitudes of RC stars are nevertheless expected,
depending on factors such as their ages, effective temperatures, metallicities ([Fe/H]), 
and alpha-element abundances ([$\alpha$/Fe]). 

In the HRD of nearby giants \citep[e.g.\ Fig.~10 in][]{2018A&A...616A..10G}, for which 
extinction is negligible, the RC stars occupy a narrow range of colours, approximately 
$G_\text{RP}-G_\text{BP}=1.2\pm 0.1$, corresponding to 
$\nu_\text{eff}=(1.48\pm 0.02)~\mu$m$^{-1}$. Intrinsically, therefore, the RC stars 
are not nearly red enough to determine the curvature of the parallax bias
versus colour, which only becomes significant for $\nu_\text{eff}\lesssim 1.3~\mu$m$^{-1}$.
In the Galactic plane and bulge it is however possible to find many RC stars that are 
sufficiently reddened by interstellar extinction. 

Recognising that the absolute magnitudes of the RC stars are uncertain and may 
depend on many unknown factors, as mentioned above, our analysis of the RC parallaxes 
must be strictly differential, comparing only samples with similar population 
characteristics. At the same time we need locations at widely different ecliptic latitudes
in order to see a possible variation in curvature with $\beta$. These conditions rule out 
the use of the inner part of the Galaxy, which only covers a limited range of ecliptic 
latitudes. A better strategy is to compare two areas in the Galactic plane, symmetrically 
placed on either side of the Galactic centre, and therefore probing similar ranges of 
galactocentric distances. The difference in $\sin\beta$ between the areas is 
maximised for Galactic longitudes $l=\pm 90^\circ$, so we should choose areas
around $(l,b)=(90^\circ,0)$ or $(\alpha,\delta)=(318.00^\circ,+48.33^\circ)$ (hereafter
called `N'), and $(270^\circ,0)$ or $(\alpha,\delta)=(138.00^\circ,-48.33^\circ)$ ( `S').
Within $5^\circ$ radius of these points, we extracted from EDR3 all sources with
five-parameter solutions satisfying $13<G<17.5$ and $\text{RUWE}<1.4$. 
This gave 1.022~million sources in N and 0.686~million in S. The reason for this 
strong asymmetry seems to be the presence of a nearby ($<1$~kpc) complex of
dust clouds in the middle of the S area, possibly part of the Vela Molecular Ridge 
\citep{1991A&A...247..202M}. While the high extinction produced by the dust
clouds is desirable for our purpose, their proximity reduces the number of stars 
in the sample and increases their mean parallax, which is unfavourable for a 
precise determination of the bias. We therefore added to the previous 
selection two areas of $5^\circ$ radius, centred on $(l,b)=(85^\circ,0)$ and 
$(275^\circ,0)$. The resulting union sets (1.361~million sources in N; 
1.561~million in S) are more similar in terms of the overall distribution of 
colours and distances, and still approximately symmetric in Galactic longitudes. 
The mean value of $\sin\beta$ is $+0.852$ for the sources in N, and 
$-0.867$ in S.

Figure~\ref{fig:rcHR} is an HRD for the union of the two sets N and S. 
The absolute magnitude is (simplistically) computed using $1/\varpi$ 
for the distance and ignoring extinction. In the top panel (\textit{a}) the RC stars 
are seen as the concentration of points along a diagonal line about five 
magnitudes above the main sequence. In effective wavenumber the feature 
starts at $\nu_\text{eff}\simeq 1.43~\mu$m$^{-1}$, and extends at least to 
$\simeq 1.24~\mu$m$^{-1}$ thanks to the extinction reddening. 
In the bottom panel (\textit{b}) the `reddening-free' Wesenheit magnitude
\begin{equation}\label{eq:W}
W = G - \lambda\left(G_\text{BP}-G_\text{RP}\right)
\end{equation}
is used instead of $G$ to compute the values on the vertical axis; here $\lambda\simeq 1.9$ 
is the slope of the reddening line for the photometric bands of \textit{Gaia} 
\citep{2019A&A...625A..14R}. With this transformation the RC stars have an
absolute magnitude $M_W\simeq -1.7$ that is nearly independent of the colour.%
\footnote{By a lucky coincidence, the RC for unreddened nearby giants has a similar
slope in the \textit{Gaia} HRD, so the use of $W$ instead of $G$ not only eliminates 
the effect of the reddening, but reduces the intrinsic scatter of the absolute 
magnitudes.}
Assuming that the RC stars have a fixed and known $M_W$, their photometric 
parallaxes can be computed as 
\begin{equation}\label{eq:rcPlx}
\varpi_\text{RC}(G,\,G_\text{BP}\!-\!G_\text{RP}) = (100~\text{mas})\times
10^{0.2\left[M_W-G+\lambda (G_\text{BP}-G_\text{RP})\right]} \, .
\end{equation}
In Fig.~\ref{figA11ab} we have plotted the differences 
$\varpi_\text{EDR3}-\varpi_\text{RC}(G,\,G_\text{BP}\!-\!G_\text{RP})$ for
the sources in areas N and S versus $\nu_\text{eff}$, using $\lambda=1.9$ 
and $M_W=-1.68$~mag. If all the sources had absolute magnitude $M_W$,
the points in these diagrams would outline the parallax bias as a function
of $\nu_\text{eff}$. Most of the sources are nearby main-sequence stars with
more positive parallaxes; they are seen in the upper-left corners of the diagrams. 
The RC stars form a concentration of points between 0 and $-100~\mu$as on 
the vertical axis. The ridge locations, shown as the black curves in Fig.~\ref{figA11ab},
were estimated by binning the points in colour, using a bin width of 0.01~$\mu$m$^{-1}$, 
and finding the maximum of the distribution for a Gaussian kernel of $10~\mu$as 
standard deviation. In Fig.~\ref{figA11c} the ridge locations for the two areas are
plotted in the same diagram for easier comparison. 

The ridge locations in Figs.~\ref{figA11ab} and \ref{figA11c} are very sensitive to the 
assumed values
of $\lambda$ and $M_W$. The value $\lambda=1.90$ used here was adopted from
\citet{2019A&A...625A..14R}, while $M_W=-1.68$ was selected to give approximately
the expected bias at $\nu_\text{eff}\simeq 1.4~\mu$m$^{-1}$ according to the
analysis in Sect.~\ref{sec:qsolmc}. The red and blue curves in Fig.~\ref{figA11c} show the 
expected variation of the bias according to 
Table~{\ref{tab:qsolmcNoJoin}, evaluated at $G=17$, where most of the sources in the
RC area are found. By adjusting $M_W$ it is possible 
to obtain agreement at a specific $\nu_\text{eff}$ for any reasonable choice of 
$\lambda$, but the slope and curvature of the relations defined by the points will be 
different. If $\lambda$ is a function of the total extinction, this could also change
the curvature of the relation. Little weight should therefore be given to the 
circumstance that the points in Fig.~\ref{figA11c} roughly follow the curves 
computed using the values in Table~{\ref{tab:qsolmcNoJoin}. What is significant, and 
supports the assumption made in Sect.~\ref{sec:qsolmc}, is that the empirical
relation is similar in the two areas for any reasonable $M_W$ and reddening law. 
If, for example, the curvature instead of being independent of $\beta$ were 
proportional to $\sin\beta$, the expected relation in the N area would rather 
follow the up-bending black curve in the diagram, which is clearly contradicted 
by the RC data.

For $\nu_\text{eff}\lesssim 1.26~\mu$m$^{-1}$ the ridges and points in 
Fig.~\ref{figA11ab} go to much more negative values on the vertical axis than 
expected from Table~\ref{tab:qsolmcNoJoin}.
Although deviations from the simplistic reddening law (constant $\lambda$) could
contribute to this trend, we believe that it is mainly a selection effect, similar to the
Malmquist bias. The strong down-turn sets in for sources redder than approximately 
$1.26~\mu$m$^{-1}$, at which point the total extinction in $V$ is at least 4~mag and 
most of the sources defining the ridge are close to the faint magnitude limit of the 
present sample at 
$G=17.5$. A preferential selection of RC stars that are intrinsically 0.2--0.3~mag 
brighter than the mean population is enough to explain the discrepancy at 
$\nu_\text{eff}=1.24~\mu$m$^{-1}$. For $\nu_\text{eff}>1.30~\mu$m$^{-1}$ this 
selection bias would be much smaller, as the sources are on average at least one 
magnitude brighter.

The conclusion from the analysis is that we see no evidence in the RC data for a difference 
in the curvature of parallax bias versus $\nu_\text{eff}$ between the northern and 
southern hemispheres.

\section{Data and method for physical pairs}
\label{sec:pairsMethod}

\begin{figure*}
\centering
  \includegraphics[height=130mm]{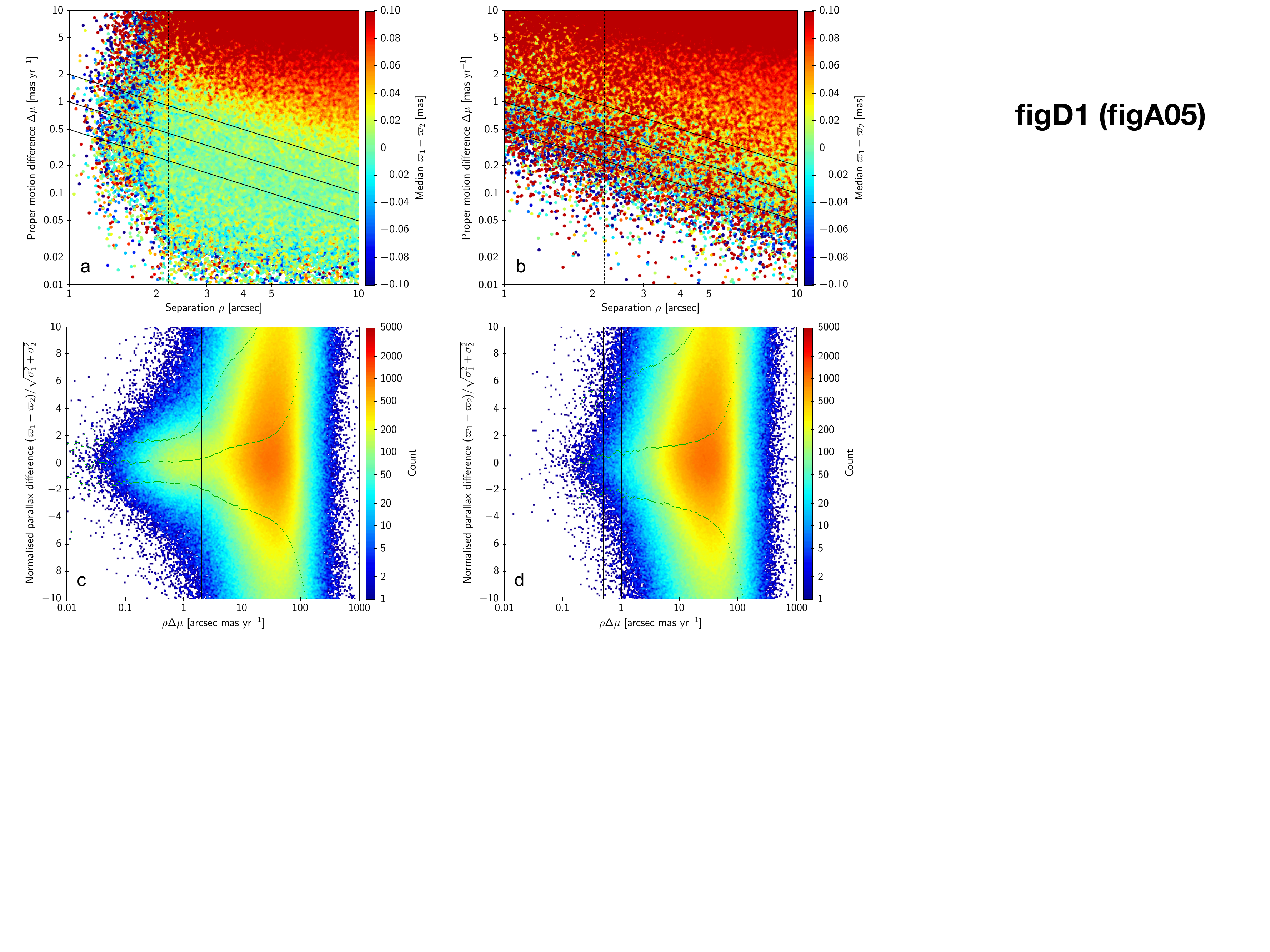}   
    \caption{Illustration of the selection of physical pairs and the contamination by optical pairs. 
    \textit{Top panels}: proper motion difference ($\Delta\mu$) versus separation ($\rho$) for 
    source pairs in sample A (\textit{a}) and B1 (\textit{b}) before making the cut $\rho>2.2$~arcsec
     indicated  by the dashed vertical line. The three solid lines correspond to $\rho\Delta\mu=0.5$, 
     1, and 2~arcsec~mas~yr$^{-1}$. The points are colour-coded by the median parallax 
     difference between the components. In panel \textit{b} the separation of the faint source is 
     measured from the position of the bright source displaced by $+0.1^\circ$ in declination.
     \textit{Bottom panels}: normalised parallax difference versus $\rho\Delta\mu$ for the 
     samples in the top panels, but now after the selection $\rho>2.2$~arcsec. The three 
     vertical lines correspond to the values of $\rho\Delta\mu$ indicated in the top panels. 
     The thin green curves show the 16th, 50th, and 84th percentiles of the distribution in 
     normalised parallax difference.}
    \label{figA05}
\end{figure*}

This appendix describes the selection of data used for the analysis of physical pairs in 
Sect.~\ref{sec:pairs} and the method applied to these data for estimating the parallax 
bias.

We only consider pairs with apparent separation $\rho< 10$~arcsec ($5\times 10^{-5}$~rad).
If their parallax is $\lesssim 20$~mas, as is the case for 99.9\% of the bright stars,
it is not likely that the true parallax difference in the pair exceeds $\pm 1$~$\mu$as. 
In the following we use subscripts 1 and 2 for the bright and faint components in a pair.
If $G_2>13.1$, the recipe derived from the quasars 
can be used to correct the EDR3 parallax 
of the faint component, $\varpi_2$, thus providing an (approximately) unbiased estimate 
of the true parallax of the pair. Considering a sample of physical pairs with similar
$G_1$, $\nu_\text{eff1}$, and $\beta$, the parallax bias can then be estimated as
\begin{equation}\label{eq:bin1}
Z_5(G_1,\nu_\text{eff1},\beta)=\text{med}\bigl[
Z_5(G_2,\nu_\text{eff2},\beta)+\varpi_1-\varpi_2\bigr] \, ,
\end{equation}
where $\text{med}[x]$ is the sample median, which we use for robustness.
A limitation of the method is that the components in a pair always have practically 
the same $\beta$ (which is why $\beta$ is not subscripted in Eq.~\ref{eq:bin1}), 
so the mapping of $Z_5$ with
respect to this parameter for the bright components must rely on the presumed known 
dependence on $\beta$ for the faint components.

A major concern with the method is contamination by `optical pairs', that is, chance 
alignments of physically unconnected sources with different true parallaxes. The use
of the median in Eq.~(\ref{eq:bin1}) provides good protection against outliers, but is
still biased if the outliers tend to fall mainly on one side of the median. This is the case 
for optical pairs, where the fainter component is likely to be more distant than the 
brighter, that is $\varpi_1-\varpi_2>0$, leading to a positive contamination bias in 
Eq.~(\ref{eq:bin1}). Rather than eliminating the contamination bias by using a very clean 
sample, which may then be too small for our purpose, we adopt a heuristic approach, 
where a small amount of contamination is accepted and the calculated median is 
corrected by a statistical procedure. This requires accurate knowledge of the selection 
function of the sample, which in practice precludes using pre-defined catalogues 
such as the Washington Double Star Catalog \citep[][]{2001AJ....122.3466M}.

For the present study, pairs are selected entirely based on EDR3 data. It is imperative 
that the parallax values themselves are not used anywhere in the selection process.
Without risk of introducing selection biases it is possible to define samples in 
terms of angular separation $\rho$, proper motion difference 
$\Delta\mu=[(\mu_{\alpha*1}-\mu_{\alpha*2})^2+(\mu_{\delta 1}-\mu_{\delta 2})^2]^{1/2}$, 
parallax uncertainty and other quality indicators, and various photometric parameters.
The main sample (denoted A) consists of pairs with $G_1<14$, $G_1<G_2<19$,
and $\rho<10$~arcsec. For correcting the contamination bias we need in addition two comparison 
samples (B1 and B2), where faint `components' in the range $G_1<G_2<19$ are selected 
within 10~arcsec    
of positions that are offset in declination by $\pm 0.1^\circ$ from the bright components
in sample A. It is assumed that B1 and B2 contain similar numbers and distributions 
of potential contaminants as A. The use of two comparison fields, rather than one, reduces 
the statistical noise introduced by the correction procedure, and their symmetrical placement
around the bright component reduces the effect of local variations in star density, etc.

We started by extracting sources in \textit{Gaia} EDR3 with five-parameter solutions 
that satisfy 
\begin{equation}\label{eq:bin2}
G<14\quad\&\quad\sigma_\varpi<0.05\,\text{mas}\quad\&\quad\text{RUWE}<2\, ,
\end{equation}
which resulted in 14\,561\,255 sources.%
\footnote{This selection was made on a pre-release version of the EDR3 catalogue, 
which did not yet include the EDR3 photometric data; the $G$ magnitudes (and 
$\nu_\text{eff}$) therefore come from DR2 and the resulting number of sources may 
be slightly different in the published release.} These are the bright components in sample A.
Next, within a radius of 10~arcsec around each such source, we extracted faint 
components (physical and optical) with five-parameter solutions and magnitudes $G_2$ 
in the range $[G_1,\,19]$. This selection is very incomplete for $\rho\lesssim 2$~arcsec, 
mainly because such sources usually do not have a reliable $\nu_\text{eff}$ from the
BP and RP photometry, and therefore no five-parameter solution. For 
$G_2-G_1\gtrsim 5$~mag the selection is incomplete also at much larger separations. 
For a well-defined selection we therefore required, in addition, $\rho>2.2$~arcsec and 
$G_2-G_1<4$~mag, which gave 3\,336\,571 faint components in sample A. No condition 
was imposed on $\sigma_\varpi$ or RUWE for the faint components.

To obtain the faint components in B1 and B2, the same criteria were used as in A (that
is $G_1<G_2<19$, $G_2-G_1<4$, and $2.2<\rho<10$~arcsec), only with $\rho$ measured 
from the offset positions. This gave 3\,309\,468 and 3\,311\,657 sources, respectively. 
Since B1 and B2 probably contain some faraway ($\rho\simeq 0.1^\circ$) physical companions, 
or members of clusters that include a brighter component in A, it can be inferred that 
sample A contains at least 26\,000 physical pairs. 

The further selection of physical pairs is based on separation ($\rho$) and 
proper motion difference ($\Delta\mu$). Small values of $\rho$ and/or $\Delta\mu$ are 
clearly much more likely in physical pairs than in optical. However, orbital motion may give 
a significant proper motion difference in a physical pair, especially if the separation is small. 
On the other hand, the risk of selecting an optical pair is proportional to $\rho^2$, so we can 
afford to increase the tolerance on $\Delta\mu$ for small separations. It turns out that the 
product $\rho\Delta\mu$ is convenient for separating optical from physical pairs. This
is illustrated in the top panels of Fig.~\ref{figA05}. Panel (\textit{a}) shows $\Delta\mu$ versus 
$\rho$ for sample A, but for illustration purposes the diagram includes also pairs with 
$\rho<2.2$~arcsec. Panel (\textit{b}) is the corresponding plot for sample B1; the plot for B2 
(not shown) is extremely similar. Both panels are colour-coded by the parallax difference 
in the pair, and it is obvious that A contains mainly physical pairs for 
$\rho\Delta\mu\lesssim 2$~arcsec~mas~yr$^{-1}$, that is, below the topmost diagonal 
line. From a comparison of panels (\textit{a}) and (\textit{b}) it should be clear why a cut like 
$\rho>2.2$~arcsec is needed: without it, the number of optical pairs in the top-left 
corner of the diagram is grossly overestimated. Similar plots of $G_2-G_1$ 
versus $\rho$ motivate the selection $G_2-G_1<4$~mag
(cf.\ Fig.~6 of \citealt{EDR3-DPACP-128}). 

The bottom panels of Fig.~\ref{figA05} show the normalised parallax difference 
($\varpi_1-\varpi_2$ divided by the combined uncertainty) versus $\rho\Delta\mu$ for 
sample A (panel \textit{c}) and B1 (panel \textit{d}), after applying the selection $\rho>2.2$~arcsec. 
For $\rho\Delta\mu\lesssim 1$~arcsec~mas~yr$^{-1}$, the normalised parallax differences 
roughly follow the expected normal distribution in sample A, whereas the distribution 
is much wider and displaced towards positive values in B1, and for $\rho\Delta\mu\gg 1$ in 
both samples. Subsamples of A with a varying degree of contamination can be obtained
by selecting on $\rho\Delta\mu$; specifically, we will use the limits indicated by the three
solid lines in Fig.~\ref{figA05}, corresponding to $\rho\Delta\mu=0.5$, 1, and 2. For
example, $\rho\Delta\mu<0.5$ is the cleanest (and smallest) subsample, while 
$\rho\Delta\mu<2$ gives a larger but more contaminated subsample. Increasing the
upper limit on $\rho\Delta\mu$ reduces the statistical uncertainty of the parallax bias, 
but could instead increase contamination bias. However, with a good procedure to
correct for the contamination, the result should not depend systematically on the 
upper limit used.
 
The procedure to correct for contamination bias is illustrated in Fig.~\ref{figA06}. With the 
selection $\rho\Delta\mu<1$~arcsec~mas~yr$^{-1}$, sample A contains 69\,993 pairs 
and the median of $\varpi_1-\varpi_2$ is $+4.1\pm 0.2~\mu$as (uncertainty estimated
by bootstrapping). The distribution of the parallax differences in sample A is shown by the 
red histogram ($h_\text{A}$) in the top panel of the figure. With the same selection on 
$\rho\Delta\mu$, samples B1 and B2 contain 7226 and 7097 pairs, respectively. The 
mean of the distributions in B1 and B2, $(h_\text{B1}+h_\text{B2})/2$, shown by the blue 
histogram, is clearly skewed towards positive values, as expected from the contaminants; 
its median parallax difference is $+29.5\pm 1.4~\mu$as. On the assumption that the 
contaminants in sample A are similar, in number and distributions, as in B1 or B2, we
obtain an estimate of the distribution for the true pairs in A by taking the difference between
the observed distributions, that is $\Delta h\equiv h_\text{A}-(h_\text{B1}+h_\text{B2})/2$. 
This is shown by the shaded histogram in Fig.~\ref{figA06}, although only the part with 
positive count differences is visible owing to the logarithmic scale. Calculating the median
of the difference histogram $\Delta h$, including negative counts, gives the corrected estimate 
$+3.4\pm 0.2~\mu$as. The contamination correction is less than a $\mu$as in this case, 
but for other subsamples it can be much larger. 

The bottom panel of Fig.~\ref{figA06} shows the 
uncorrected and corrected medians as functions of the upper limit on $\rho\Delta\mu$. 
With increasing limit the samples get larger and the statistical uncertainties smaller (as
shown by the error bars), but the increasing contamination bias is also
apparent in the uncorrected medians. The corrected medians, on the other hand, remain
virtually constant up to a limit of about 2~arcsec~mas~yr$^{-1}$ on $\rho\Delta\mu$.
This result is only meant to illustrate the principle of the contamination
correction, and ignores the complex dependencies of both the parallax bias and
the contamination bias on $G$, $\nu_\text{eff}$, and $\beta$. The important point is that the 
correction makes it possible to benefit from the larger samples obtained with a less strict 
limit on $\rho\Delta\mu$, 
or alternatively to use a finer division in magnitude or colour without sacrificing the accuracy
of the estimated parallax bias.

Since we only want the median of the distribution difference, it is not necessary to compute 
and subtract histograms; a simpler and more exact way is to use a weighted median.
Given an array of values $\{x_i\}$ with (relative) weights $\{w_i\}$, the weighted median is 
the value $\hat{x}$, such that the sum of weights is the same on either side of 
$\hat{x}$. In the present case $x=\varpi_1-\varpi_2$, and the weighted median is 
computed for the union of A, B1, and B2, setting $w=1$ for pairs in sample A, and $w=-0.5$ 
for pairs in B1 and B2.

It is well known that the median minimises the sum of absolute deviations, that is the 
$L_1$ norm of the residuals. An alternative definition of the weighted median is, therefore,
\begin{equation}\label{eq:wmed}
\hat{x} = \underset{x}{\arg\min} \sum_i \left|x_i-x\right| w_i \, . 
\end{equation}
This expression can immediately be generalised to the multi-dimensional case, where an 
arbitrary function of the parameter vector $\vec{z}$ (such as the general function described 
in Appendix~\ref{sec:num}) is fitted to the data by weighted $L_1$-norm minimisation:
\begin{equation}\label{eq:L1}
\hat{\vec{z}} = \underset{\vec{z}}{\arg\min} \sum_i \left|x_i-f(\vec{z})\right| w_i \, . 
\end{equation}
In all our analysis of the pairs (and for the results shown as the black dots in the bottom panel of
Fig.~\ref{figA06}), we use Eq.~(\ref{eq:L1}) with two additional refinements. First, the 
uncertainty of the resulting median or fit is estimated by bootstrap resampling of the 
union set. Secondly, the 
weights in Eq.~(\ref{eq:L1}) are adjusted to take into account the uncertainties of the parallax 
differences, $\sigma_{\Delta\varpi}=(\sigma_1^2+\sigma_2^2)^{1/2}$. This is done by setting 
the weights to $1/\sigma_{\Delta\varpi}$ in sample A, and to $-0.5/\sigma_{\Delta\varpi}$ 
in B1 and B2. Using weights proportional to $\sigma_{\Delta\varpi}^{-1}$
is consistent with the $L_1$ formalism in Eq.~(\ref{eq:L1}). 

\begin{figure}
\centering
  \includegraphics[width=0.82\hsize]{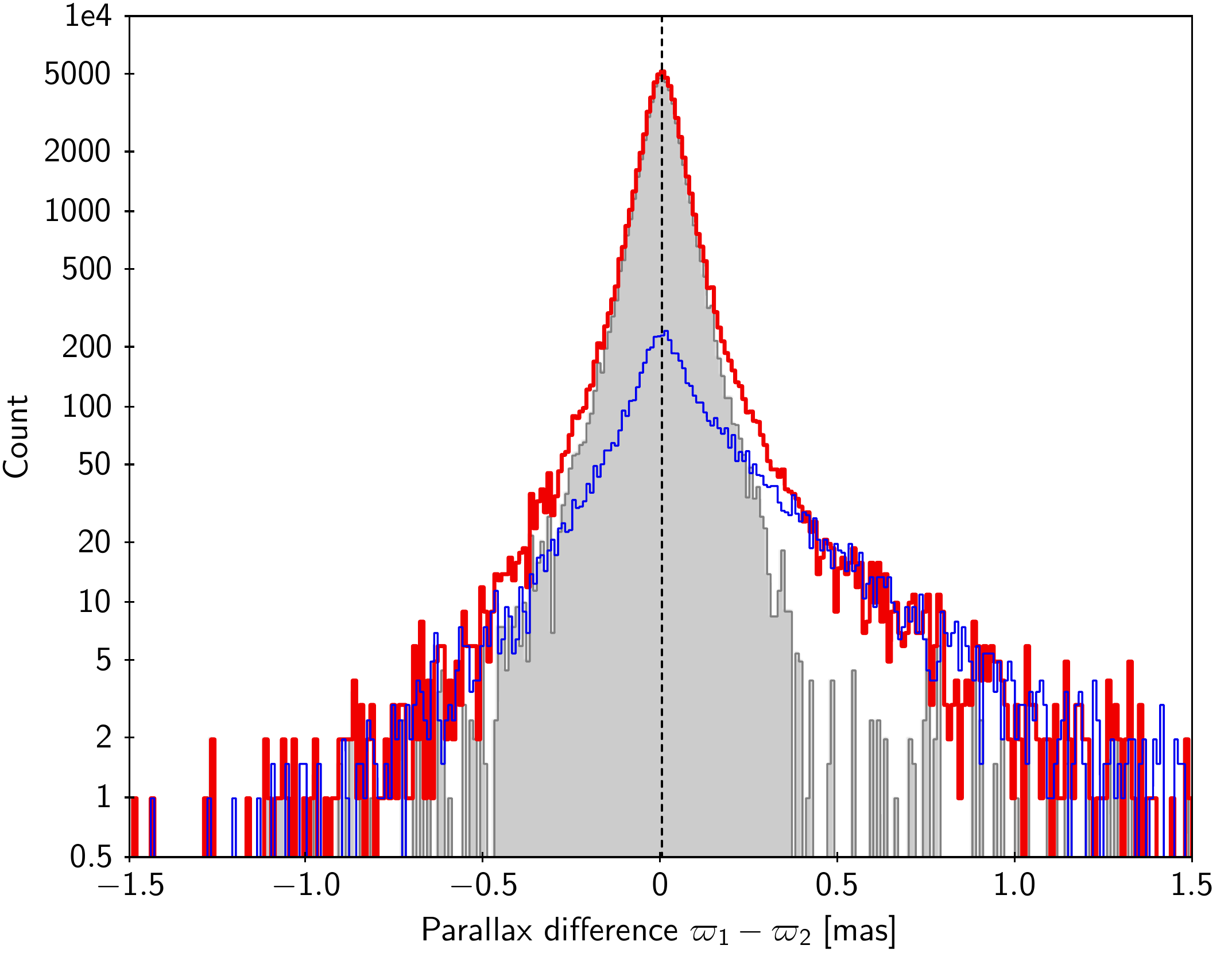}\\
  \includegraphics[width=0.82\hsize]{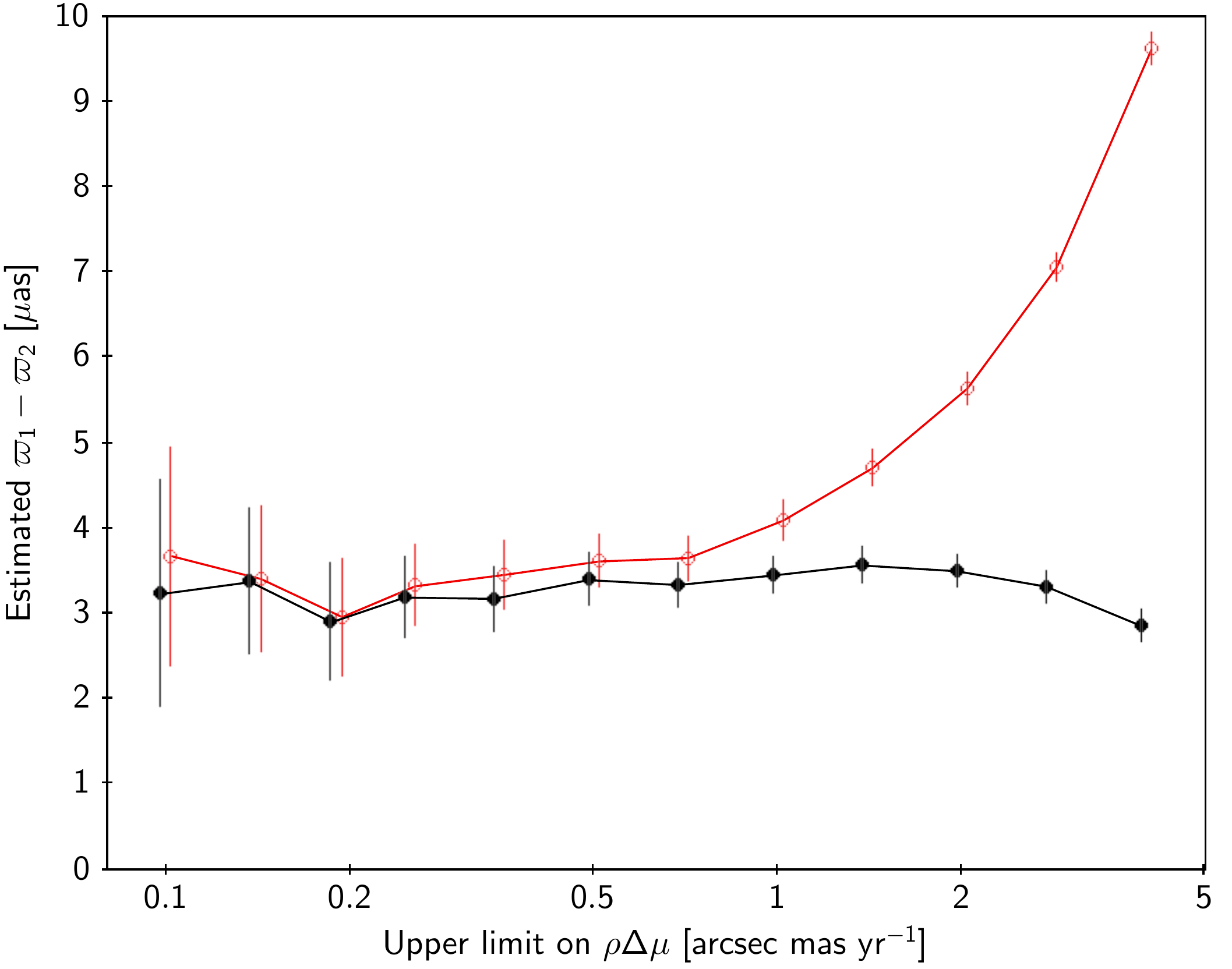}
    \caption{Illustrating the procedure for contamination bias correction. \textit{Top}: 
    distribution of parallax differences in sample A (thick red histogram), and for the
    mean of comparison samples B1 and B2 (thin blue histogram). The shaded grey 
    histogram is the difference between the red and blue histograms. The dashed 
    line is the corrected estimate of the parallax difference, equal to the median of the 
    difference histogram. \textit{Botton}: uncorrected (open red circles) and corrected 
    (filled black) estimates of the mean parallax difference versus the cut in $\rho\Delta\mu$.
    For better visibility, the points have been slightly displaced sideways.}
    \label{figA06}
\end{figure}

\end{document}